\documentclass[aps, prb, twocolumn,amsmath,amssymb,floatfix,reprint,longbibliography, nofootinbib]{revtex4-2}
\usepackage{graphicx}
\usepackage{physics}
\usepackage{times}
\usepackage{dcolumn}
\usepackage{hyperref} 
\hypersetup{colorlinks, allcolors=blue}
\allowdisplaybreaks
\begin{document}

\title{Spin-current rectifier and linear magnon Edelstein effect in $p$-wave antialtermagnets}

\author{Kristian M{\ae}land}
\email[Contact author: ]{kristian.maeland@uni-wuerzburg.de}
\author{Bj{\"o}rn Trauzettel}
\affiliation{Institute for Theoretical Physics and Astrophysics, University of W{\"u}rzburg, D-97074 W{\"u}rzburg, Germany}
\affiliation{Würzburg-Dresden Cluster of Excellence ctd.qmat, D-97074 W{\"u}rzburg, Germany}

\begin{abstract}
We show that $p$-wave magnetic insulators efficiently generate magnetization via a linear thermal magnon Edelstein effect. Furthermore, they can generate perfectly nonreciprocal spin currents through a purely even-order nonlinear response. This makes them promising candidates for spin-current rectification. More generally, we find that odd-parity-wave magnets with coplanar ground states host magnons with zero magnetic moment in the plane. Instead the magnons carry a colinear, out-of-plane magnetic moment, placing coplanar odd-parity-wave magnets within the class of antialtermagnets. We establish these general results using symmetry arguments and demonstrate them explicitly for a two-dimensional model with a coplanar ground state containing the minimal number of sublattices, inspired by the ground state of CeNiAsO. Our conclusions hold both in the absence and in the presence of spin-orbit coupling.
\end{abstract}

\maketitle

\section{Introduction}

Magnon spintronics aims to use the quanta of collective spin excitations as low-dissipation information carriers in future technological devices \cite{Chumak2015Jun, Chumak2019mgcomp}. Desired devices include spin-wave diodes \cite{Lan2015Diode}, spin-current rectifiers \cite{Chotorlishvili2022Rectification, Ezawa2026rectifier}, and magnon-based logic elements \cite{Fischer2017spinwavegate, Chumak2014allmg}. Much attention has been focused on ferromagnets, ferrimagnets, and antiferromagnets (AFMs) \cite{Bauer2012spincal, Uchida2021spincal,  Serga2010YIG, Cahaya2025spincomp, Xiao2010SSE, Rezende2016SSEAFM, Rezende2016SSEAFM2, Rezende2019AFM, Ciccarelli2025compFiMAFMrev}. The net magnetization of ferro- and ferrimagnets, and the spin degenerate magnon bands of AFMs may prove challenging \cite{Ciccarelli2025compFiMAFMrev}. 
Recently, the class of magnets with zero net magnetization have been expanded to include unconventional magnets with a nonrelativistic spin splitting. Altermagnets (AMs) have colinear ground states (GSs) like AFMs, but display an even-parity-wave spin splitting, both of electron and magnon bands \cite{Smejkal2022Feb, Smejkal2022Sep, Smejkal2022Dec, Brekke2023Aug, Maeland2024Feb, LeraandMaeland2025Feb, Leraand2026Mar, Petermann2025Dec}. Odd-parity-wave magnets (OPWMs) have noncolinear GSs and show odd-parity spin splitting \cite{Hellenes2023pwave, Brekke2024pwave, Yu2025pMmodel, Mitscherling2026Mar, Eikeland2026Jun, Sim2026Jul, Neumann2026pMmagnon, Kravchuk2026magmoment}.
AMs and OPWMs are primarily studied for their electronic properties \cite{Smejkal2022Dec, Hellenes2023pwave}, but the spin dynamics in insulating AMs and OPWMs receive attention for possible applications in magnon spintronics \cite{Cui2023AM, Aoyama2026AMFiMmag, Santos2026mgLieb, Neumann2026pMmagnon, Kravchuk2026magmoment}.

Compared to AFMs, AMs lack time-reversal symmetry combined with inversion or translation, and instead show time-reversal symmetry when combined with rotation. Thus, there can be an even-parity spin splitting of electron or magnon bands of $d,g,i,\dots$-wave type. 
Let the spin expectation value be $\boldsymbol{s}_{\boldsymbol{q}}$, which depends on momentum $\boldsymbol{q}$. In AMs, $\boldsymbol{s}_{\boldsymbol{q}} = \boldsymbol{s}_{-\boldsymbol{q}}$, due to inversion symmetry.
OPWMs on the other hand, lack inversion symmetry and instead show time-reversal $T$ combined with translation $\boldsymbol{\tau}$, which we refer to as $T\boldsymbol{\tau}$ symmetry for short. This symmetry classification allows for odd-parity spin splitting of electron or magnon bands ($p, f, h, \dots$ wave), if the magnetic GS is noncolinear. Thus, OPWMs obey $\boldsymbol{s}_{\boldsymbol{q}} = -\boldsymbol{s}_{-\boldsymbol{q}}$. OPWMs where $\boldsymbol{s}_{\boldsymbol{q}}$ is colinear are dubbed antialtermagnets \cite{Mitscherling2026Mar, Neumann2026pMmagnon, Jungwirth2024LiebRev, Lange2026aam}.

Candidate materials for OPWMs include CeNiAsO, Mn$_3$GaN, and Ce$_3$InN \cite{Hellenes2023pwavev2}. The coplanar (CeNiAsO) \cite{Wu2019CNAOexp} and noncoplanar (Mn$_3$GaN and Ce$_3$InN) \cite{Gabler2008Ce3InN, Shi2016Mn3GaN} magnetic GSs have been confirmed in experiments, and anisotropic resistivity has been observed in CeNiAsO \cite{Zhou2025CeNiAsO}. Recent angle-resolved photoemission spectroscopy (ARPES) evidence suggests quenched $p$-wave spin splitting of the electron bands in CeNiAsO close to the Fermi level \cite{ZhangShen2026JunCNAO, ZhangZhou2026JunCNAO}, while a spin-ARPES study observes the expected $p$-wave spin splitting below the Fermi level \cite{ZhangChen2026MayCNAO}. Additionally, iron-based superconductors contain OPWMs as part of their phase diagrams \cite{Dsouza2026Fe}. Furthermore, spin spirals are potential OPWMs provided they break inversion symmetry and display the $T\boldsymbol{\tau}$ symmetry. NiI$_2$ has incommensurate spin spirals and shows features related to OPWMs \cite{Song2025CominPwaveExp}. A commensurate spin spiral obeying the $T\boldsymbol{\tau}$ symmetry has been realized in Gd$_3$(Ru$_{1-\delta}$Rh$_\delta$)$_4$Al$_{12}$, i.e., by doping a known incommensurate spin spiral \cite{Yamada2025pMexp}.

In Ref.~\cite{Neumann2026pMmagnon}, they investigate noncoplanar OPWMs and predict $p$-wave and $f$-wave spin splittings of the magnons in two dimensional (2D) and 3D models. Following Refs.~\cite{Okuma2017magnonSpin, Neumann2026pMmagnon}, we refer to the magnon spin as the expectation value of the total spin in a single magnon state compared to vacuum. . 
In Ref.~\cite{Kravchuk2026magmoment}, they first consider a coplanar GS on a 2D triangular lattice which lacks the $T\boldsymbol{\tau}$ symmetry. Its magnons exhibit an $f$-wave splitting of the magnon magnetic moment and it belongs to a wider class of OPWMs \cite{Luo2025sym, Hayami2020fwave}. 
Antiferromagnetically stacking these layers yields a 3D OPWM in which $T\boldsymbol{\tau}$ corresponds to translation between adjacent layers \cite{Kravchuk2026magmoment}.
The magnon magnetic moment is defined by the response of the magnon bands to an external magnetic field \cite{Neumann2020magnonSpinOrbital}.

We show that all coplanar OPWMs with $T\boldsymbol{\tau}$ symmetry have magnetic moment nonzero only perpendicular to the coplanar GSs.
Since the magnon magnetic moment is directly proportional to the magnon spin  in colinear states \cite{Bauer2023Aug}, we conjecture that the magnetic moment is also closely related to the magnon spin in noncolinear cases.
Meanwhile, the magnon spin calculated within the approximation of linear spin wave theory (LSWT), yields zero spin also out-of-plane due to the coplanar nature of the GS. We view that as a limitation of the LSWT.
As a practical example, we consider a coplanar GS on a 2D rotated square lattice with four sublattices, inspired by the GS in CeNiAsO \cite{Hellenes2023pwave, Wu2019CNAOexp, Zhou2025CeNiAsO, Mitscherling2026Mar}. We find a $p$-wave splitting of the magnetic moment. This gives rise to spin currents under a temperature gradient, extending the result for the $f$-wave case \cite{Kravchuk2026magmoment}. Only even-order nonlinear response coefficients are nonzero. We point out that this represents a perfectly nonreciprocal spin current, relevant to spin-current diodes \cite{Lan2015Diode} and spin-current rectifiers \cite{Ezawa2026rectifier}. We also explain why the linear and other odd-order responses are zero. 

Furthermore, we consider the thermal Edelstein effect. A temperature gradient generates a net magnetization in the sample due to the magnetic-moment splitting in the magnon bands. 
We find that the thermal magnon Edelstein effect occurs to linear order in temperature gradient for $p$-wave splitting. 
In contrast, with $f$-wave splitting of the magnetic moment, the thermal Edelstein effect occurs to third order in temperature gradients \cite{Kravchuk2026magmoment}.
Thus, $p$-wave magnetic insulators are more efficient generators of magnetization under temperature gradients than higher-order-wave magnetic insulators. It is furthermore a nonrelativistic effect, unlike the usual case for the Edelstein effect \cite{Edelstein1990Jan, Bahari2026Apr, Chakraborty2025Aug}. OPWMs are originally described in the nonrelativistic limit where spin space and real space decouple. Thus, the effects of spin-orbit coupling (SOC) on OPWMs are interesting \cite{Hodt2025May}. 
We include terms originating from SOC in our Heisenberg Hamiltonian, and find that their impact is rather quantitative than qualitative.

In Sec.~\ref{sec:GS}, we present a general Heisenberg spin Hamiltonian and design it to have a specific noncolinear, coplanar GS. We derive the magnon spectrum via LSWT in Sec.~\ref{sec:magnon}. There, we also explain why there is zero magnon spin in all directions within LSWT, and derive the magnon magnetic moment. We present the nonequilibrium transport calculations in Sec.~\ref{sec:transport}.
We conclude in Sec.~\ref{sec:conclusion}, while appendices cover additional details.

\section{Coplanar ground state} \label{sec:GS}

\begin{figure}
    \centering
    \includegraphics[width=0.9\linewidth]{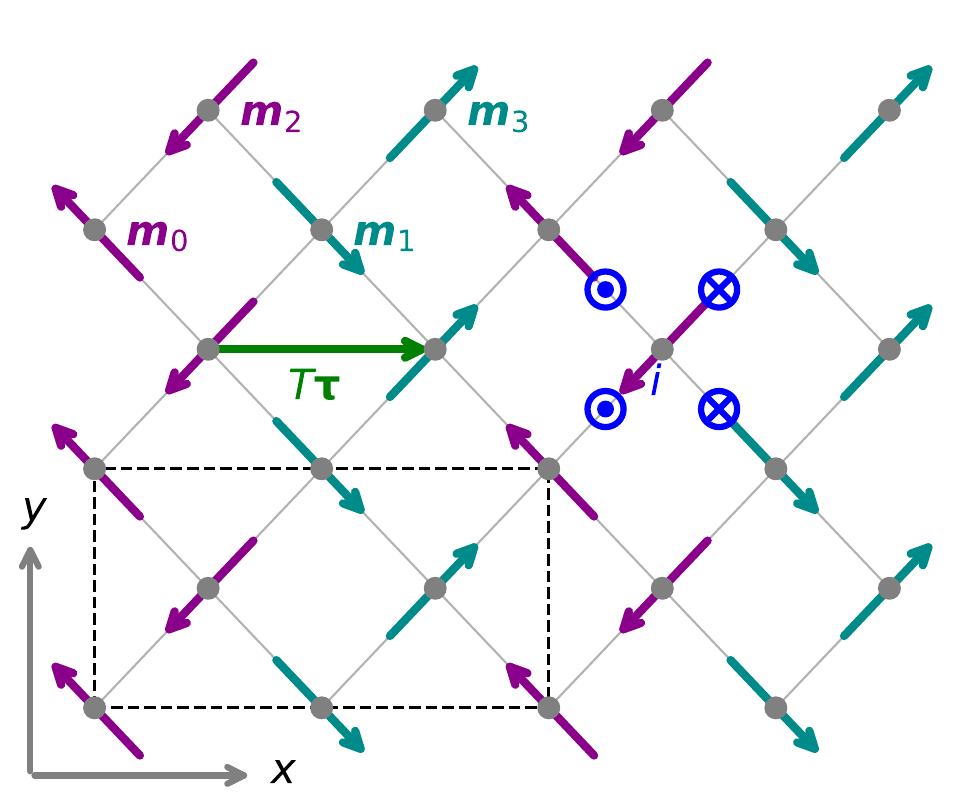}
    \caption{Schematic of the magnetic ground state on the rotated square lattice with spins represented by arrows. The magenta and cyan color on arrows emphasizes the $T\boldsymbol{\tau}$ symmetry illustrated by the green arrow. The dashed rectangle is the magnetic unit cell. The blue vectors denote the out-of-plane DMI vectors $\boldsymbol{D}_{i,j}$ from a site $i$ to its nearest neighbors $j$.}
    \label{fig:GS}
\end{figure}

We take the experimentally observed magnetic GS in CeNiAsO as our starting point \cite{Wu2019CNAOexp, Zhou2025CeNiAsO}. We ignore the nonmagnetic sites, and imagine a two-dimensional (2D) rotated square lattice with the classical magnetic GS shown in Fig.~\ref{fig:GS}. There are four sublattices with classical spin directions $\boldsymbol{m}_0 = (-1,1)/\sqrt{2}$, $\boldsymbol{m}_1 = (1,-1)/\sqrt{2}$, $\boldsymbol{m}_2 = (-1,-1)/\sqrt{2}$, $\boldsymbol{m}_3 = (1,1)/\sqrt{2}$. We refer to these as sublattice 0, 1, 2, and 3. This GS breaks inversion symmetry and has a $T\boldsymbol{\tau}$ symmetry, which makes it an OPWM. Specifically, we find it to be a $p$-wave magnet. While our GS is based on CeNiAsO, we believe our results are relevant to a wide range of coplanar $p$-wave magnets, be it true 2D magnets, thin films on a substrate, or near the surface of a 3D magnet. Our results are easily generalized to 3D by stacking the layers ferromagnetically. 
Note that iron-based superconductor OPWM candidates have a similar lattice structure to CeNiAsO \cite{Dsouza2026Fe, ZhangChen2026MayCNAO}. While CeNiAsO is metallic, we focus on the case of a magnetic insulator.

We set up a general Heisenberg model in terms of localized spins $\boldsymbol{S}_i$ with magnitude $S$,
\begin{align}
\label{eq:CNAOheisenberg}
    H &=\sum_{i,j} J_{i,j} \boldsymbol{S}_i \cdot \boldsymbol{S}_j + K_z \sum_i S_{iz}^2 -K\sum_i (\boldsymbol{S}_i \cdot \hat{n}_i)^2 \nonumber \\
    &+  \sum_{i,j} \frac{B_{i,j}}{S^2} (\boldsymbol{S}_i \cdot \boldsymbol{S}_j)^2 + \sum_{\langle i,j\rangle} \boldsymbol{D}_{i,j} \cdot (\boldsymbol{S}_i \times \boldsymbol{S}_j)  \nonumber \\
    &-  g\sum_{i} S\boldsymbol{H} \cdot \boldsymbol{S}_i.
\end{align}
We consider nearest-neighbor (NN) exchange $J_{i,j} = J_1$, next-nearest-neighbor (NNN) exchange $J_2$, and third-nearest-neighbor (TNN) exchange $J_3$. All TNN spins are antiparallel, and thus we assume there is an antiferromagnetic $J_3>0$. We take $J_3 S$ as our energy scale. Meanwhile NN spins are all perpendicular, which is a surprising magnetic state given that NN exchange often dominates. For the GS in Fig.~\ref{fig:GS} to be realized, it is likely that $|J_1|$ is small. In Ref.~\cite{Yu2025pMmodel}, they argue that $J_1=0$ by symmetry in certain OPWMs. Half the NNN are aligned ferromagnetically, the rest are aligned antiferromagnetically. Hence, this GS is unlikely if $J_2$ of either sign is the dominant energy scale in the system. If we consider the subspace of sublattice 0 and 1, or sublattice 2 and 3, we obtain stripe phase AFMs, which are stable if $J_2 < 2J_3$, at least when $J_2$ is antiferromagnetic \cite{Erlandsen2020stripe}.

We add an easy-plane anisotropy $K_z \geq 0$, which prefers the spins to be in the $xy$ plane. We further add a site-dependent easy-axis anisotropy $K\geq0$, where the  easy axes $\hat{n}_i$ are sublattice dependent, and point in the direction of the known magnetic GS of CeNiAsO.\footnote{The GS in CeNiAsO does not have perfectly perpendicular NN spins \cite{Mitscherling2026Mar}. For simplicity we consider the case of perpendicular spins. Rotated easy axes can explain the deviation from perpendicular spins.} A strong anisotropy of this kind is a possible stabilizing mechanism of this particular GS. Another term which can explain the perpendicular NN spins is a biquadratic term with $B_1>0$ \cite{Brown1971biquadratic, Neumann2026pMmagnon}. A positive NN biquadratic term is argued for in Ref.~\cite{Heinze2011nanoSk}, where a nanoscale skyrmion crystal is experimentally observed. The choice $B_1>0$ is also made in Refs.~\cite{Neumann2026pMmagnon, Pasrija2013biquadratic, Hirosawa2026biquad} to stabilize noncolinear magnetism. Note that in some systems the value of $B_1$ can be engineered \cite{Saleem2026biquadratic, Akagi2012biquad}. We also consider NNN $B_2$ and TNN $B_3$ biquadratic terms. Since NNN and TNN spins are colinear, we assume $B_2\leq 0$ and $B_3\leq 0$.

Dzyaloshinskii-Moriya interactions (DMI) originate from SOC and often result in noncolinear GSs \cite{DZYALOSHINSKY, Moriya,  Heinze2011nanoSk, dosSantosPRB, DiazPRL, DiazPRR, Maeland2022Jun, Maeland2022Aug, Maeland2023AprPRL, Maeland2023DecTSC}. OPWMs are defined by spin group symmetries which assume zero SOC to decouple spin space and real space. Even if SOC contributes to the stabilization, the symmetry classification of the GS can still be done using spin groups \cite{Jungwirth2024LiebRev}. We will consider DMI as a perturbation around the nonrelativistic limit, to explore the effect of SOC on OPWMs. Note that easy-axis and easy-plane anisotropies also originate from SOC, so similar arguments apply there \cite{Jungwirth2024LiebRev}. We consider only DMI for NN. We find that out-of-plane DMI vectors $\boldsymbol{D}_{i,j}$ can stabilize our choice of GS, while any in-plane component of the DMI vector results in noncoplanar GSs. We focus on the out-of-plane DMI vectors indicated in Fig.~\ref{fig:GS}. With NN vectors $\boldsymbol{a}_1 = (1,-1)/\sqrt{2}$ and $\boldsymbol{a}_2 = (1,1)/\sqrt{2}$, we have $\boldsymbol{D}_{\boldsymbol{a}_1} = \boldsymbol{D}_{\boldsymbol{a}_2} = -D\hat{z}$ and $\boldsymbol{D}_{-\boldsymbol{a}_1} = \boldsymbol{D}_{-\boldsymbol{a}_2} = D\hat{z}$. Out-of-plane DMI vectors are unusual, but appear in thin films \cite{Niu2024OOPDMI}. 

Finally, we include an external magnetic field $S\boldsymbol{H}$ with the Land\'e factor $g$ included in the term. Unless otherwise stated, we set the magnetic field to zero as it distorts the GS. The magnetic field is relevant to define and calculate the magnon magnetic moment. We set the reduced Planck's constant $\hbar = 1$, the Bohr magneton $ \mu_B = 1$, and the lattice constant $ a = 1$ throughout. Then, $S$ is dimensionless and $\boldsymbol{H}$ has unit energy.

We consider two methods of obtaining the magnetic GS, namely, simulated annealing and self-consistent iteration, as explained in Appendix~\ref{app:GS}. We set $\boldsymbol{S}_i = S\boldsymbol{m}_i$ and consider $\boldsymbol{m}_i$ to be classical unit vectors. Then, we minimize Eq.~\eqref{eq:CNAOheisenberg} with respect to their directions. 
Typically, we find that there are several degenerate GSs, where the state illustrated in Fig.~\ref{fig:GS} is one among them. That is sufficient for the proceeding analysis, as we can always consider a large domain where the state we consider is chosen. We can break the degeneracy by having nonzero easy-axis anisotropy and DMI, in which case we single out the state in Fig.~\ref{fig:GS} as the unique GS. If $J_1$ and $J_2$ are the dominant energy scales in the system we find states with lower energy than the one in Fig.~\ref{fig:GS}. Unless otherwise stated, we consider in the remainder of the paper choices of parameters where the state in Fig.~\ref{fig:GS} is the GS or one among a degenerate set of GSs.

The Mermin-Wagner theorem rules out long-range order in 2D at finite temperature if the system has a rotational symmetry \cite{Mermin1966Wagner}. 
Thus, our assumption of an ordered ground state is only valid at zero temperature or with nonzero easy-axis anisotropy, since this is the only term in the Hamiltonian which breaks a continuous rotation symmetry about the $z$-axis. Nevertheless, we find it instructive to also consider the limit of $K\to 0$, both to be able to eliminate SOC, and to show how the magnon gap develops upon introducing the easy-axis anisotropy. We also reiterate that our system can be generalized to 3D by stacking the layers ferromagnetically, where long-ranged order is possible without the easy-axis anisotropy.

If we consider an electron Hamiltonian akin to Eq.~(1) in Ref.~\cite{Brekke2024pwave}
describing itinerant electrons with NN hopping $t$, chemical potential $\mu$, and Kondo-like coupling $J_{sd}$ to the localized spins, we find a $p$-wave spin-split electron band structure. Only the $z$ component of the electron spin has a nonzero average $\langle s_z({\boldsymbol{k}}) \rangle$ with a $p$-wave momentum dependence. In this paper, we analyze a magnetic insulator, where the Fermi level lies within a large band gap of the electron sector. The physics is then dominated by fluctuations of the localized spins around the GS, quantized as magnons.

\section{Odd-parity magnon magnetic moment} \label{sec:magnon}

\subsection{Linear spin wave theory}
We analyze weak magnetic fluctuations around the GS spin configuration, captured by LSWT. Thus, we insert a Holstein-Primakoff (HP) transformation around the local spin directions \cite{HProtation_2009, RoldanMolina, RoldanMolinaHPdetails, Maeland2022Aug, Maeland2022Jun, Maeland2023AprPRL, Maeland2023DecTSC, Neumann2026pMmagnon}. We introduce a local frame $\hat{e}_3^i = \boldsymbol{m}_i =  (\sin\theta_i\cos\phi_i, \sin\theta_i\sin\phi_i, \cos\theta_i)$, $\hat{e}_1^i \perp \hat{e}_3^i$, $\hat{e}_1^i=  (\cos\theta_i\cos\phi_i, \cos\theta_i\sin\phi_i, -\sin\theta_i),$ and $\hat{e}_2^i = \hat{e}_3^i \cross \hat{e}_1^i = (-\sin\phi_i, \cos\phi_i,0)$. Then, the HP transformation is $\boldsymbol{S}_i \cdot \hat{e}_3^i = S - a_i^\dagger a_i$, $S_{i\pm} = \boldsymbol{S}_i \cdot \hat{e}_1^i \pm i \boldsymbol{S}_i \cdot \hat{e}_2^i$, $S_{i+} = \sqrt{2S-a_i^\dagger a_i}~a_i \approx \sqrt{2S}~a_i, S_{i-} = a_i^\dagger\sqrt{2S-a_i^\dagger a_i} \approx \sqrt{2S}~a_i^\dagger$. Here, $a_i^{(\dagger)}$ destroys (creates) a magnon on site $i$ located at $\boldsymbol{r}_i$. We rewrite Eq.~\eqref{eq:CNAOheisenberg} in terms of these local spin components and insert the HP. Then we keep terms up to second order in magnon operators. This assumption ignores magnon-magnon scattering which is weak at low temperature and especially weak for long-wavelength magnons \cite{Kittel1991Jan, DiazPRL, TopoMagnonRev}. The terms that contain no magnon operators correspond to the classical spin Hamiltonian, while those that are linear in magnon operators are zero in the GS \cite{HProtation_2009, Maeland2022Jun, Maeland2022Aug}. Thus, we focus on the terms that are quadratic in magnon operators. 

We introduce a Fourier transform (FT) to momentum space as
\begin{equation}
    \label{eq:FT}
    a_i = \frac{1}{\sqrt{N^{(r)}}} \sum_{\boldsymbol{q}} e^{i\boldsymbol{q}\cdot \boldsymbol{r}_i} a_{\boldsymbol{q}}^{(r)},
\end{equation}
where $N^{(r)}$ is the number of sites in sublattice $r$. We consider the case that all sublattices have the same number of lattice sites, i.e., $N^{(r)} = N_u$ is the number of unit cells.

With four sublattices, we introduce the operator vector $\boldsymbol{a}_{\boldsymbol{q}}^\dagger = (a_{\boldsymbol{q}}^{(0)\dagger}, a_{\boldsymbol{q}}^{(1)\dagger}, a_{\boldsymbol{q}}^{(2)\dagger}, a_{\boldsymbol{q}}^{(3)\dagger}, a_{-\boldsymbol{q}}^{(0)}, a_{-\boldsymbol{q}}^{(1)}, a_{-\boldsymbol{q}}^{(2)}, a_{-\boldsymbol{q}}^{(3)})$. 
Then the Hamiltonian can be written in matrix form as
\begin{equation}
    H = \frac{1}{2} \sum_{\boldsymbol{q}} \boldsymbol{a}_{\boldsymbol{q}}^\dagger M_{\boldsymbol{q}} \boldsymbol{a}_{\boldsymbol{q}},~M_{\boldsymbol{q}} = \begin{pmatrix}
        \eta_{\boldsymbol{q}} & \nu_{-\boldsymbol{q}}^* \\
        \nu_{\boldsymbol{q}} & \eta_{-\boldsymbol{q}}^*
    \end{pmatrix}.
\end{equation}
The matrix elements are
\begin{equation}
    \eta_{\boldsymbol{q}}^{r,s} = \eta_r \delta_{r,s} + S\Lambda_{\boldsymbol{q}+}^{r,s},~\nu_{\boldsymbol{q}}^{r,s} = \nu_r \delta_{r,s} + S(1+\delta_{r,s})\Lambda_{\boldsymbol{q}-}^{r,s},
\end{equation}
where,
\begin{align}
    \eta_r &= -2S\sum_s \sum_{\boldsymbol{\delta}_{(r,s)}} \bigg( J_{\boldsymbol{\delta}_{(r,s)}} \hat{e}_3^r \cdot \hat{e}_3^s + \boldsymbol{D}_{\boldsymbol{\delta}_{(r,s)}} \cdot (\hat{e}_3^r \times \hat{e}_3^s) \nonumber \\
    &- B_{\boldsymbol{\delta}_{(r,s)}} \big[ (\hat{e}_1^r \cdot \hat{e}_3^s)^2 +(\hat{e}_2^r \cdot \hat{e}_3^s)^2 -2 (\hat{e}_3^r \cdot \hat{e}_3^s)^2  \big] \bigg) \nonumber \\
    & +K_z S [(\hat{e}_1^r \cdot \hat{z})^2 + (\hat{e}_2^r \cdot \hat{z})^2 -2 (\hat{e}_3^r \cdot \hat{z})^2]+gS\boldsymbol{H} \cdot \hat{e}_3^r \nonumber\\
    &-KS \big[ (\hat{n}_r \cdot \hat{e}_1^r)^2 +(\hat{n}_r \cdot \hat{e}_2^r)^2 -2 (\hat{n}_r \cdot \hat{e}_3^r)^2 \big] ,
\end{align}
\begin{equation}
    \nu_r = K_z S (\hat{e}_-^r \cdot \hat{z})^2 - KS (\hat{n}_r \cdot \hat{e}_-^r)^2 +2S B_{\boldsymbol{\delta}_{(r,s)}} (\hat{e}_-^r \cdot \hat{e}_3^s)^2 ,
\end{equation}
\begin{align}
    \Lambda_{\boldsymbol{q}\pm}^{r,s} &= \sum_{\boldsymbol{\delta}_{(r,s)}} \bigg( J_{\boldsymbol{\delta}_{(r,s)}} e^{i\boldsymbol{q}\cdot \boldsymbol{\delta}_{(r,s)}} \hat{e}_\pm^r \cdot \hat{e}_-^s \nonumber \\
    &+ e^{i\boldsymbol{q}\cdot \boldsymbol{\delta}_{(r,s)}} \boldsymbol{D}_{\boldsymbol{\delta}_{(r,s)}} \cdot (\hat{e}_\pm^r \times \hat{e}_-^s) +2 B_{\boldsymbol{\delta}_{(r,s)}} e^{i\boldsymbol{q}\cdot \boldsymbol{\delta}_{(r,s)}} \nonumber \\
    &\times \big[(\hat{e}_\pm^r \cdot \hat{e}_-^s)(\hat{e}_3^r \cdot \hat{e}_3^s)+(\hat{e}_\pm^r \cdot \hat{e}_3^s) (\hat{e}_3^r \cdot \hat{e}_-^s)\big] \bigg).
\end{align}
Here, $\hat{e}_{\pm}^{r} = \hat{e}_{1}^{r} \pm i \hat{e}_{2}^{r}$,
\begin{equation}
    J_{\boldsymbol{\delta}_{(r,s)}} = \begin{cases}
        J_1 \text{ if } \exists~ i \in r, j \in s \text{ that are NN,} \\
        J_2 \text{ if } \exists~ i \in r, j \in s \text{ that are NNN,} \\
        J_3 \text{ if } \exists~ i \in r, j \in s \text{ that are TNN,} \\
        0 \text{ otherwise,}
    \end{cases}
\end{equation}
\begin{equation}
    \boldsymbol{D}_{\boldsymbol{\delta}_{(r,s)}} = \begin{cases}
        \boldsymbol{D}_{i,j} \text{ if } \exists~ i \in r, j \in s \text{ that are NN,} \\
        0 \text{ otherwise,}
    \end{cases}
\end{equation}
\begin{equation}
    B_{\boldsymbol{\delta}_{(r,s)}} = \begin{cases}
        B_1 \text{ if } \exists~ i \in r, j \in s \text{ that are NN,} \\
        B_2 \text{ if } \exists~ i \in r, j \in s \text{ that are NNN,} \\
        B_3 \text{ if } \exists~ i \in r, j \in s \text{ that are TNN,} \\
        0 \text{ otherwise.}
    \end{cases}
\end{equation}
We define $\boldsymbol{\delta}_{(r,s)}$ as the NN, NNN, or TNN vector connecting two inequivalent sites on the relevant sublattices. Note that $r=s$ is allowed in the above definitions. In the GS, every $i\in r$ has two NNNs $j\in s=r$ along $\boldsymbol{\delta}_{(r,s)} = (0, \pm \sqrt{2})$.

To preserve bosonic commutation relations requires a paraunitary transformation matrix $T_{\boldsymbol{q}}$ with the property $T_{\boldsymbol{q}}^{-1} = JT_{\boldsymbol{q}}^\dagger J$, where $J = \sigma_z \otimes I_{N_{\text{SL}}} = \operatorname{diag}(1,1,1,1,-1,-1,-1,-1)$ \cite{COLPA}. Here, $I_{N_{\text{SL}}}$ is the $N_{\text{SL}}\times N_{\text{SL}}$ unit matrix and $N_{\text{SL}} = 4$ is the number of sublattices. The transformation matrix can be written as 
\begin{equation}
\label{eq:ColpaT}
    T_{\boldsymbol{q}} = \begin{pmatrix} U_{\boldsymbol{q}} & V_{-\boldsymbol{q}}^* \\ V_{\boldsymbol{q}} & U_{-\boldsymbol{q}}^*    \end{pmatrix}.
\end{equation}
The diagonalization is set up as 
\begin{equation}
    \boldsymbol{a}_{\boldsymbol{q}}^\dagger M_{\boldsymbol{q}} \boldsymbol{a}_{\boldsymbol{q}} = (\boldsymbol{a}_{\boldsymbol{q}}^\dagger T_{\boldsymbol{q}}^\dagger)[(T_{\boldsymbol{q}}^\dagger)^{-1}M_{\boldsymbol{q}} T_{\boldsymbol{q}}^{-1}] (T_{\boldsymbol{q}} \boldsymbol{a}_{\boldsymbol{q}}) = \boldsymbol{b}_{\boldsymbol{q}}^\dagger D_{\boldsymbol{q}} \boldsymbol{b}_{\boldsymbol{q}},
\end{equation}
where $D_{\boldsymbol{q}} = \operatorname{diag}(\omega_{\boldsymbol{q}1}, \dots, \omega_{\boldsymbol{q}4}, \omega_{-\boldsymbol{q},1}, \dots, \omega_{-\boldsymbol{q},4})$ is diagonal and $\boldsymbol{b}_{\boldsymbol{q}}^\dagger = (b_{\boldsymbol{q}1}^\dagger, \dots, b_{\boldsymbol{q}4}^\dagger, b_{-\boldsymbol{q},1}, \dots, b_{-\boldsymbol{q},4})$. In the end,
\begin{equation}
    H = \sum_{\boldsymbol{q}}\sum_{m=1}^4 \omega_{\boldsymbol{q}m} b_{\boldsymbol{q}m}^\dagger b_{\boldsymbol{q}m},
\end{equation}
where we order the four magnon bands by energy in descending order at each $\boldsymbol{q}$. Since $M_{\boldsymbol{q}}$ is a dense $8\times8$ matrix, we perform the diagonalization numerically, as described in Refs.~\cite{Maeland2022Jun, Maeland2022Aug}.

\begin{figure*}
    \centering
    \includegraphics[width=0.99\linewidth]{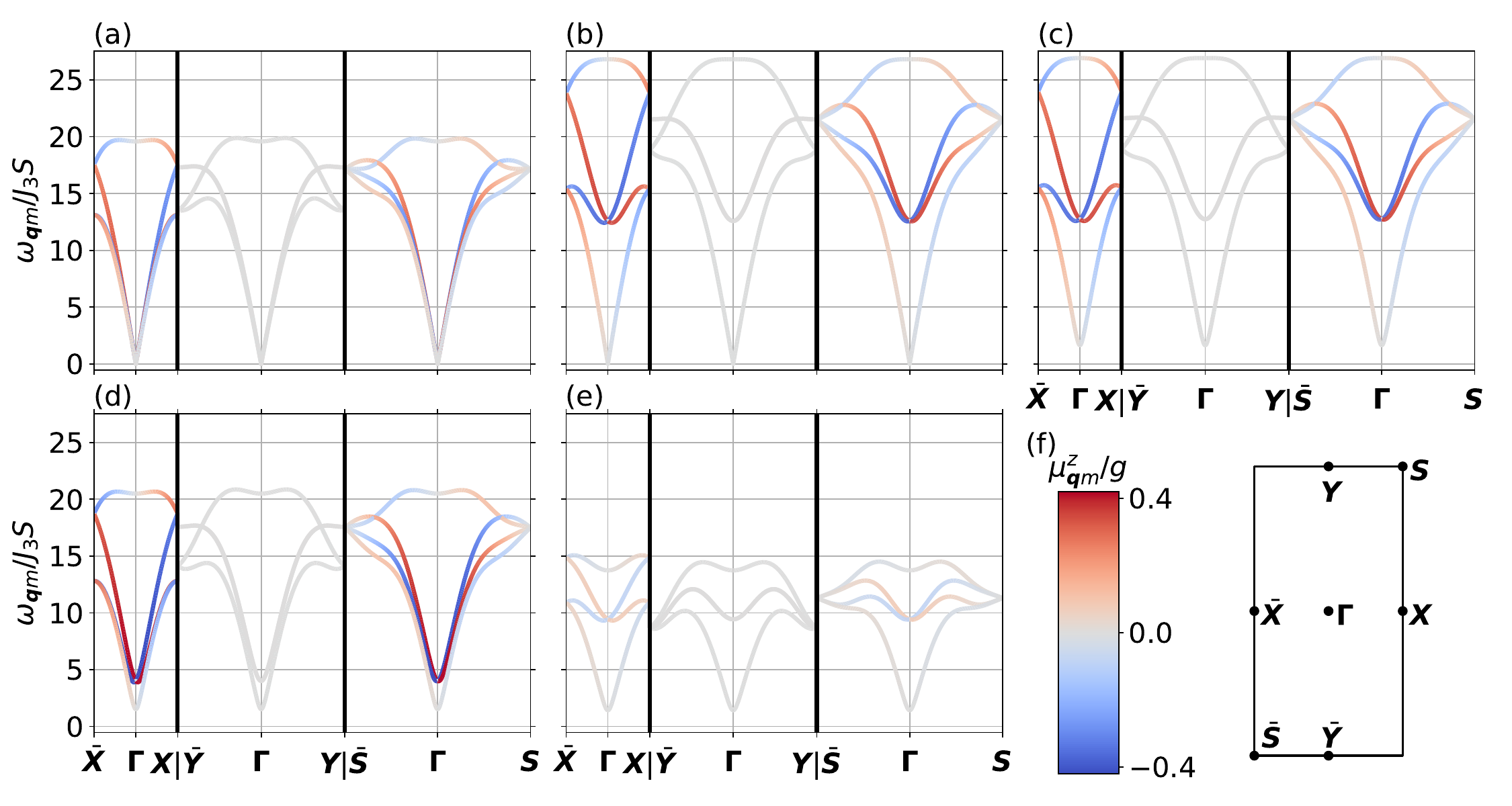}
    \caption{(a)-(e) Magnon spectrum $\omega_{\boldsymbol{q}m}$ along high-symmetry directions in the first Brillouin zone (1BZ) demonstrating a $p_x$-wave momentum dependence of the $z$ component of the magnon magnetic moment $\boldsymbol{\mu}_{\boldsymbol{q}m}$ in all four modes. (f) Colorbar for $\mu_{\boldsymbol{q}m}^z$ and an illustration of the 1BZ with high-symmetry points $\boldsymbol{\Gamma} = (0,0)$, $\boldsymbol{X} = (\pi/2\sqrt{2},0),$ $\boldsymbol{Y} = (0,\pi/\sqrt{2})$, and $\boldsymbol{S} = (\pi/2\sqrt{2},\pi/\sqrt{2})$ marked. We refer to the parameters in the panels as parameter set a-e, and list them in Table~\ref{tab:parameters}.
    Panel (a) includes no terms originating from SOC, panel (b) includes all terms except easy-axis anisotropy, and panel (c) demonstrates how an easy-axis anisotropy introduces a magnon gap. In panel (d), we remove DMI, while in panel (e) we include DMI but remove biquadratic terms and easy-plane anisotropy.    }
    \label{fig:spectrum}
\end{figure*}

\begin{figure}
    \centering
    \includegraphics[width=0.8\linewidth]{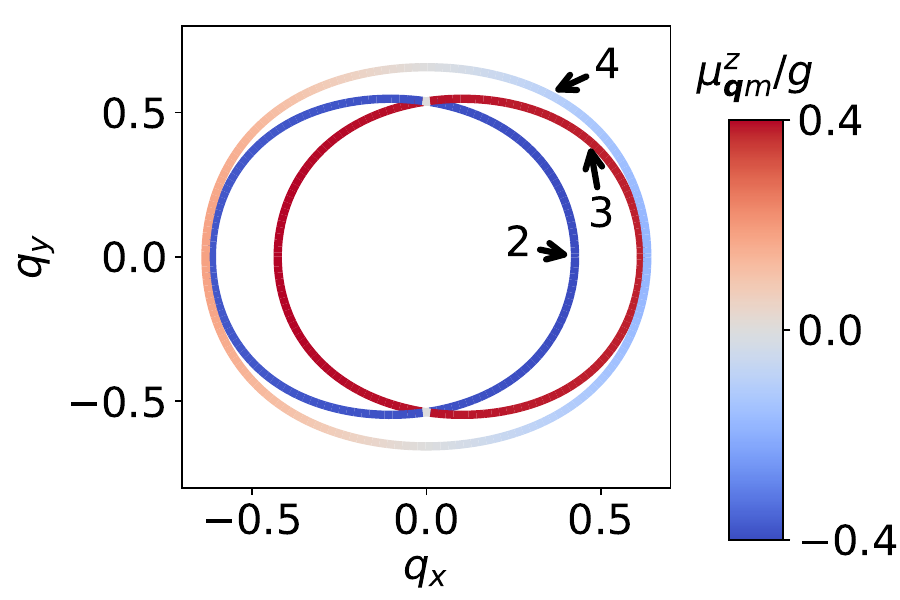}
    \caption{Constant energy contour at $\omega_{\boldsymbol{q}m} = 10J_3S$ of the magnon spectrum in Fig.~\ref{fig:spectrum}(d). Numbers with arrows indicate the band index $m$. Colors indicate $\mu_{\boldsymbol{q}m}^z$ and show the $p_x$-wave splitting of the magnon magnetic moment. Magnon band 4 shows a smooth $p_x$-wave dependence of $\mu_{\boldsymbol{q}4}^z$. Bands 2 and 3 have abrupt sign changes via degeneracy at $q_x = 0$, which we still interpret as $p_x$-wave magnetic moments.}
    \label{fig:contour}
\end{figure}

\subsection{Magnon spin} \label{sec:magnonspin}
The total spin is $\boldsymbol{S}_{\text{tot}} = \sum_i \boldsymbol{S}_i = \sum_r \sum_{i\in r} \boldsymbol{S}_i$. It is zero classically since we have zero net magnetization. The magnon spin is defined as the expectation value of $\boldsymbol{S}_{\text{tot}}$ in a one magnon state $\ket{\boldsymbol{q}m}$ compared to vacuum $\ket{0}$ \cite{Okuma2017magnonSpin, Neumann2020magnonSpinOrbital, Neumann2026pMmagnon},
\begin{equation}
\label{eq:sqm}
    \boldsymbol{s}_{\boldsymbol{q}m} = \bra{\boldsymbol{q}m} \boldsymbol{S}_{\text{tot}} \ket{\boldsymbol{q}m}-\bra{0} \boldsymbol{S}_{\text{tot}} \ket{0}.
\end{equation}

We now calculate the magnon spin within LSWT $\boldsymbol{s}_{\boldsymbol{q}m}^{\text{LSWT}}$. From $\boldsymbol{S}_i = \sum_\mu (\boldsymbol{S}_i \cdot \hat{e}_\mu^i) \hat{e}_\mu^i$, only the $\mu=3$ component gives terms to second order in magnon operators through $\boldsymbol{S}_i \cdot \hat{e}_3^i = S-a_i^\dagger a_i$.  
The linear terms give zero expectation value while the constant terms cancel between the one magnon state and vacuum.
We do an FT and transform to the band basis via
\begin{align}
    a_{\boldsymbol{q}}^{(r)} =& \sum_{m = 1}^{4} \bqty{(U_{\boldsymbol{q}}^\dagger)_{r,m} b_{\boldsymbol{q},m} -(V_{\boldsymbol{q}}^\dagger)_{r,m} b_{-\boldsymbol{q},m}^\dagger }, \\
    a_{-\boldsymbol{q}}^{(r)\dagger} =& \sum_{m = 1}^{4} \bqty{-(V_{-\boldsymbol{q}}^T)_{r,m} b_{\boldsymbol{q},m} +(U_{-\boldsymbol{q}}^T)_{r,m} b_{-\boldsymbol{q},m}^\dagger }.
\end{align}
The terms that give nonzero contributions to $\boldsymbol{s}_{\boldsymbol{q}m}^{\text{LSWT}}$ are $b_{\boldsymbol{q},m}^\dagger b_{\boldsymbol{q},m}$ and $b_{\boldsymbol{q},m}b_{\boldsymbol{q},m}^\dagger$.
We end up with
\begin{align}
    \boldsymbol{s}_{\boldsymbol{q}m}^{\text{LSWT}} &= - \sum_r \hat{e}_3^r [ (V_{-\boldsymbol{q}}^T)_{rm} (V_{-\boldsymbol{q}}^\dagger)_{rm} +  (U_{\boldsymbol{q}}^T)_{rm}  (U_{\boldsymbol{q}}^\dagger)_{rm}]. \label{eq:sqmLSWT}
\end{align}

Since the GS is coplanar in the $xy$ plane, none of the $\hat{e}_3^r$ unit vectors have a finite $z$ component. Hence, $s_{\boldsymbol{q}m,z}^{\text{LSWT}} = 0$. We further derive that $s_{\boldsymbol{q}m,x}^{\text{LSWT}} = s_{\boldsymbol{q}m,y}^{\text{LSWT}} = 0$. Thus, none of the components of the magnon spin have a nonzero average. 

We can explain the lack of in-plane magnon spin by symmetry arguments. We have a $[C_{2z}||E|\boldsymbol{\tau}]$ symmetry. Here, elements left of the double bar act only in spin space while elements right of double bar act only in real space. $C_{2z}$ means $180^\circ$ rotation around z axis, $E$ is identity operator, and $\boldsymbol{\tau}$ is a translation. 
The symmetry tells us that $(s_{\boldsymbol{q}m})_x = -(s_{\boldsymbol{q}m})_x$ and $(s_{\boldsymbol{q}m})_y = -(s_{\boldsymbol{q}m})_y$ so the in-plane components must be zero. 
For a coplanar state $T\boldsymbol{\tau}$ and $[C_{2z}||E|\boldsymbol{\tau}]$ symmetries will always appear together, as $T$ and $C_{2z}$ both flip the in-plane spins. Thus, we conclude that all coplanar OPWMs with the $T\boldsymbol{\tau}$ symmetry have zero in-plane magnon spin. 
There is no symmetry that forces $s_{\boldsymbol{q}m,z} = 0$.
Thus, $s_{\boldsymbol{q}m,z}^{\text{LSWT}} = 0$ may be a limitation of calculating the spin expectation value within the harmonic approximation of LSWT.
We move on to the magnon magnetic moment, which indeed has a nonzero $z$ component. 
In Appendix \ref{app:spin}, we show what perturbations lead to nonzero $\boldsymbol{s}_{\boldsymbol{q}m}^{\text{LSWT}}$.

\subsection{Magnon magnetic moment}

\begin{table}
    \centering
    \caption{Parameter values for the magnon spectra displayed in Figs.~\ref{fig:spectrum}(a)-\ref{fig:spectrum}(e).}
    \begin{ruledtabular}
    \begin{tabular}{cccccc}
         & a & b & c & d & e \\
         \hline
         $J_1/J_3$& 0 & -0.2 & -0.2 & -0.2 & -0.2 \\
         $J_2/J_3$& 0.2 & 0.2 & 0.2 & 0.2 & 0.2 \\
         $B_1/J_3$& 0.5 & 0.5 & 0.5 & 0.5 & 0 \\
         $B_2/J_3$& -0.2 & -0.2 & -0.2 & -0.2 & 0\\
         $B_3/J_3$& -0.1 & -0.1 & -0.1 & -0.1 & 0\\
         $K/J_3$& 0 & 0 & 0.05 & 0.05 & 0.05 \\
         $K_z/J_3$& 0 & 0.2 & 0.2 & 0.2 & 0 \\
         $D/J_3$& 0 & 0.5 & 0.5 & 0 & 0.5          
    \end{tabular}
    \end{ruledtabular}
    \label{tab:parameters}
\end{table}

The magnetic moment of a magnon is defined as
\begin{equation}
    \boldsymbol{\mu}_{\boldsymbol{q}m} = -\frac{1}{S}\left.\pdv{\omega_{\boldsymbol{q}m}}{\boldsymbol{H}}\right\vert_{\boldsymbol{H}=0}.
\end{equation}
Here, we must take into account that the magnetic GS changes when applying the magnetic field. We calculate the $x,y,$ and $z$ components by central difference, considering the magnon bands at small magnetic field in opposite directions. We then assign the value of the magnetic moment to the magnon spectrum at zero magnetic field. The symmetry arguments for magnon spin apply also to the magnon magnetic moment, and so $\mu_{\boldsymbol{q}m}^x = \mu_{\boldsymbol{q}m}^y = 0$, which we confirm by explicit calculation. The factor $1/S$ arises from our convention in Eq.~\eqref{eq:CNAOheisenberg}, where the magnetic field enters as $S\boldsymbol{H}$. That way, $\boldsymbol{\mu}_{\boldsymbol{q}m}$ is independent of $S$.

The $z$ component of the magnetic moment is nonzero and shows a $p_x$-wave momentum dependence in all magnon bands in Fig.~\ref{fig:spectrum}. Figure \ref{fig:spectrum}(a) illustrates the case with no SOC, $K=K_z=D=0$. In that case, we find that any nonzero $J_1$ destabilizes the sought after GS. We find three Goldstone modes (zero energy at zero momentum) due to the massively degenerate GS. There is a continuous rotation symmetry of the spins around the $x$, $y$, and $z$ axes. 
Figure \ref{fig:spectrum}(b) adds $K_z = 0.2J_3$ and $D=0.5J_3$ to stabilize the state, in which case $J_1 \neq 0$ is also allowed. A single Goldstone mode remains, due to the continuous rotation symmetry around the $z$ axis. The easy-plane anisotropy breaks the previous rotation symmetries around the $x$ and $y$ axes. Then, in Fig.~\ref{fig:spectrum}(c), we add easy-axis anisotropy and get a gapped magnon spectrum. 
The easy-axis anisotropy breaks all continuous rotation symmetries and thus is solely responsible for the magnon gap.

In Fig.~\ref{fig:spectrum}(d), we remove DMI. By comparing Figs.~\ref{fig:spectrum}(c) and \ref{fig:spectrum}(d), we see the role of DMI is to increase the energy of the three magnon modes with highest energy, especially close to $\boldsymbol{\Gamma}$. While we do not show it in the figure, the main role played by $B_2$ and $B_3$ is to increase the magnon energy at $\boldsymbol{Y}$. In Fig.~\ref{fig:spectrum}(e), we consider the case where DMI plays the main role in stabilizing the GS, i.e. no easy-plane anisotropy and no biquadratic interaction. Then, the absolute value of the magnetic moment is significantly smaller than the other cases. We conjecture that for this GS, the biquadratic term is most effective at generating magnon magnetic moment, while a stabilizing DMI leads to a decrease of the magnitude of the magnetic moment. If we consider nonzero biquadratic terms and a weak $D<0$, effectively flipping the direction of the DMI vectors, we find that a destabilizing DMI can increase the magnitude of the magnetic moment. 
Here, we ignore the fact that the GS changes for $D<0$, and expand around a metastable state.
Moreover, we see that parameter set e has no crossing of magnon bands apart from the degeneracies at high-symmetry points and the two degenerate modes along $\boldsymbol{\Bar{Y}\Gamma Y}$. In Fig.~\ref{fig:spectrum}, when magnon bands are degenerate in energy, we plot the average of their magnetic moments.

Figure \ref{fig:contour} shows a constant energy contour of the magnon spectrum to visualize the $p_x$-wave nature of the magnetic moment splitting. We can explain the $p_x$-wave splitting by symmetry. Our GS has a symmetry $[E || m_y]$ or, equivalently, $[E||C_{2x}]$, where $m_y$ is mirror through the $y=0$ plane, and $C_{2x}$ is twofold rotation about the $x$ axis. These symmetry operations flip $q_y$ but not $q_x$ and otherwise do nothing. These symmetries exist because the spins in the GS do not rotate between sites when moving in the $y$ direction. We know that $\mu_z(q_x,q_y) = -\mu_z(-q_x, -q_y)$ from $T\boldsymbol{\tau}$. The new symmetry tells us $\mu_z(q_x,q_y) = \mu_z(q_x, -q_y)$ which together must mean $\mu_z(q_x,q_y) = -\mu_z(-q_x, q_y)$ and so $\mu_z(0,q_y) = 0$. We find no symmetries that force other nodal lines and so, $\mu_{\boldsymbol{q}m}^z$ is $p_x$ wave. 
Given that we find the $p_x$-wave splitting for all parameter sets in Fig.~\ref{fig:spectrum}, we believe the splitting is tied to the symmetry of the magnetic GS, not any specific term in the Heisenberg Hamiltonian. It is of course essential that the state in Fig.~\ref{fig:GS} is indeed the GS of Eq.~\eqref{eq:CNAOheisenberg}.

\begin{figure}
    \centering
    \includegraphics[width=0.99\linewidth]{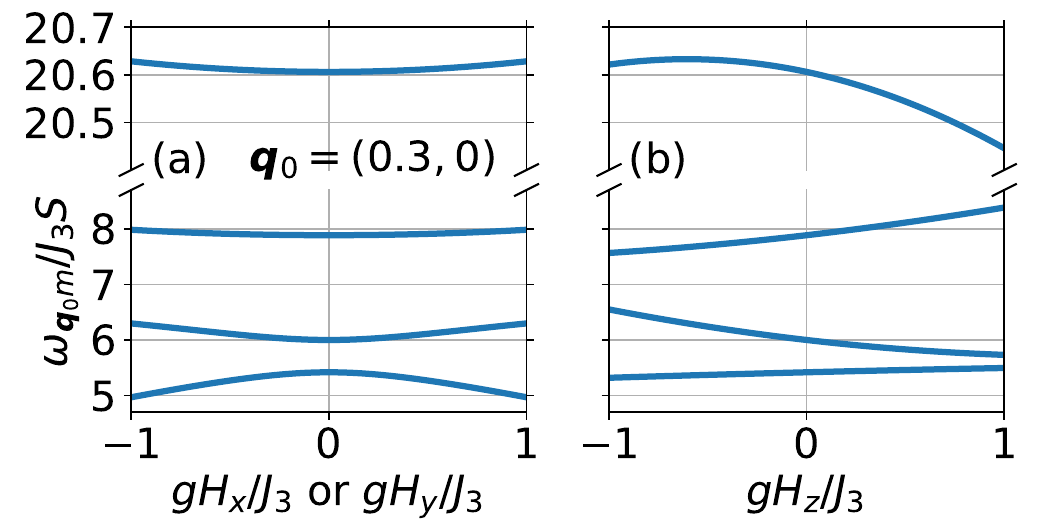}
    \caption{Magnon bands at $\boldsymbol{q}_0 = (0.3,0)$ as a function of magnetic field in (a) $x$ or $y$ direction and (b) $z$ direction. We consider parameter set d in Table~\ref{tab:parameters} and update the GS depending on the magnetic field.}
    \label{fig:wH}
\end{figure}

Figure~\ref{fig:wH} shows the four magnon bands at fixed momentum in response to a magnetic field. For an in-plane magnetic field, we observe that the bands respond symmetrically, and so there is no magnetic moment. For an out-of-plane magnetic field, the magnon bands respond asymmetrically, in line with their nonzero magnetic moment $\mu_{\boldsymbol{q}m}^z$. Note that in this figure, we keep track of the changing GS as a function of magnetic field. The spins all cant in the direction of the magnetic field compared to the GS in Fig.~\ref{fig:GS} at zero magnetic field.

\section{Nonreciprocal spin currents and linear thermal Edelstein effect} \label{sec:transport}

\begin{figure*}
    \centering
    \includegraphics[width=0.9\linewidth]{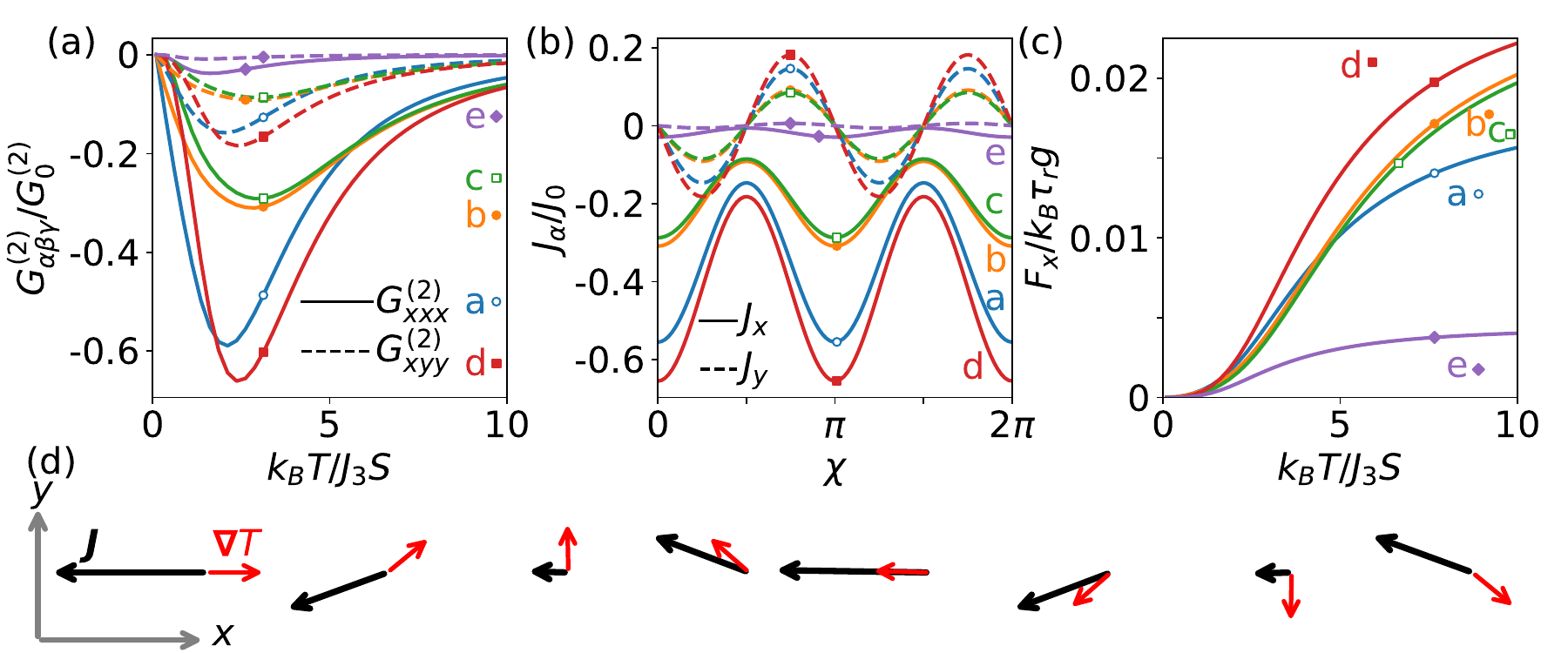}
    \caption{(a) The nonzero spin current transport coefficients $G_{xxx}^{(2)}$ (solid lines) and $G_{xyy}^{(2)}=G_{yxy}^{(2)}=G_{yyx}^{(2)}$ (dashed lines). (b) $x$ (solid lines) and $y$ (dashed lines) components of the spin current $\boldsymbol{J}$ as a function of the direction of the temperature gradient at $k_B T/J_3 S = 2.6$. In (a) and (b) we plot dimensionless values by dividing by the reference values $G_0^{(2)} = k_B^2 \tau_r^2 J_3 Sg$ and $J_0 = k_B^2 |\mathtt{\nabla}T|^2 \tau_r^2 J_3 Sg$. (c) The nonzero linear magnon Edelstein effect coefficient $F_x$. In panels (a)-(c) we show results for the five parameter sets a-e in Table~\ref{tab:parameters}, as indicated. (d) Schematic of the current $\boldsymbol{J}$ (black arrows, length showing variation in magnitude) depending on the direction of the applied temperature gradient $\boldsymbol{\nabla}T$ (red, thinner arrows).}
    \label{fig:GJF}
\end{figure*}

\subsection{Nonlinear spin current response}
We consider in-plane transport of the $z$ component of the magnon magnetic moment. In Ref.~\cite{Bauer2023Aug}, they show that the magnon magnetic moment in colinear states is directly proportional to the magnon spin. The magnetic field only couples to the spins via the Zeeman term allowing the use of the Hellmann-Feynman theorem to connect magnon magnetic moment and magnon spin. The proof for noncolinear states is likely more involved and we leave it outside the scope of the present paper. We conjecture that the magnetic moment we calculate is closely related to the spin magnetic moment and proportional to the true magnon spin via $\boldsymbol{\mu}_{\boldsymbol{q}m} \approx g \boldsymbol{s}_{\boldsymbol{q}m}$. Thus we refer to the transport of the magnetic moment as a spin current.

The spin current can be observed indirectly via an inverse spin Hall effect \cite{Uchida2021spincal}.
Furthermore, currents carried by magnons can couple to electric polarization, opening alternative measurement methods \cite{To2025orbitronic}.
As established in spin caloritronics \cite{Bauer2012spincal, Uchida2021spincal}, the spin current is driven by an in-plane temperature gradient, 
\begin{equation}
    \boldsymbol{\nabla} T = |\boldsymbol{\nabla} T|(\cos\chi, \sin\chi, 0),
\end{equation}
bringing the system out of equilibrium, and generating unequal magnon occupation in space. 

We calculate nonequilibrium transport phenomena within the semiclassical Boltzmann approach. 
At sufficiently low temperatures, weak magnon–magnon scattering ensures well-defined magnon quasiparticles, making the semiclassical Boltzmann approach an appropriate approximation to the more general Kubo formalism \cite{Nakata2021Boltzmann, Katsura2010Kubo, Neumann2026pMmagnon, Kravchuk2026magmoment}.
The magnon-driven magnetic moment current is \cite{Kravchuk2026magmoment}
\begin{equation}
    \boldsymbol{J} = \frac{1}{4N_u}\sum_{\boldsymbol{q}m} \mu_{\boldsymbol{q}m}^z \boldsymbol{v}_{\boldsymbol{q}m} \delta n_{\boldsymbol{q}m},
\end{equation}
where $4N_u$ is the area of the sample, $\boldsymbol{v}_{\boldsymbol{q}m} = \partial_{\boldsymbol{q}} \omega_{\boldsymbol{q}m}$ is the group velocity, and $ \delta n_{\boldsymbol{q}m}$ is the nonequilibrium part of the distribution function. We write the current as
\begin{equation}
    J_\alpha = G_{\alpha\beta} \partial_\beta T + G_{\alpha\beta\gamma}^{(2)} \partial_\beta T \partial_\gamma T + \dots
\end{equation}
with $\alpha,\beta \in \{x,y\}$.
We assume the temperature gradient is constant in space such that $\partial_\beta \partial_\gamma T = 0$.
$G_{\alpha\beta}$ represents the linear transport coefficient which is zero in our case. The next term is a nonlinear transport to second order with coefficient $G_{\alpha\beta\gamma}^{(2)}$ and the dots represent higher orders. The transport coefficients are
\begin{equation}
\label{eq:G}
    G_{\alpha\beta} = - \frac{\tau_r}{4N_u} \sum_{\boldsymbol{q}m}  \mu_{\boldsymbol{q}m}^z (v_{\boldsymbol{q}m})_\alpha (v_{\boldsymbol{q}m})_\beta \pdv{n_{\boldsymbol{q}m}^0 }{T},
\end{equation}
\begin{equation}
\label{eq:G2}
    G_{\alpha\beta\gamma}^{(2)} = \frac{\tau_r^2}{4N_u} \sum_{\boldsymbol{q}m} \mu_{\boldsymbol{q}m}^z (v_{\boldsymbol{q}m})_\alpha (v_{\boldsymbol{q}m})_\beta (v_{\boldsymbol{q}m})_\gamma \pdv[2]{n_{\boldsymbol{q}m}^0 }{T},
\end{equation}
where $\tau_r$ is the relaxation time and $n_{\boldsymbol{q}m}^0 = 1/(e^{\omega_{\boldsymbol{q}m}/k_B T}-1)$ is the equilibrium Bose-Einstein distribution at the average temperature $T$. We show the derivation in Appendix \ref{app:transport}.

We sort the magnon modes by energy in descending order at each $\boldsymbol{q}$. In that case, the magnon spectrum is inversion symmetric.\footnote{Since there is a sum over modes at each $\boldsymbol{q}$, the sorting is arbitrary.} Thus, the group velocity is odd in momentum. The same applies to the magnon magnetic moment. We then see that the summand in $G_{\alpha\beta}$ is overall odd in $\boldsymbol{q}$, so the sum is zero. Going to second order results in another factor of the group velocity, making the summand in $G_{\alpha\beta\gamma}^{(2)}$ overall even in $\boldsymbol{q}$. As we show in Appendix \ref{app:transport}, each higher order results in a new factor of the group velocity. With an odd-parity $\mu_{\boldsymbol{q}m}^z$ we conclude that transport to odd order is zero, while nonlinear transport to even order can be finite. That generates a perfectly nonreciprocal current, where the spin current direction and magnitude are unchanged if we flip the direction of the temperature gradient. Hence, OPWMs are good candidates for spin-current rectifiers. These ideas have previously been discussed for the electron-driven spin transport \cite{Ezawa2025nonreciprocal, Ezawa2026rectifier}. We predict that it also applies to the magnon-driven transport of magnetic moment.

From the definition in Eq.~\eqref{eq:G2}, we see that $G_{xxy}^{(2)}=G_{xyx}^{(2)} = G_{yxx}^{(2)}$ and $G_{xyy}^{(2)} = G_{yxy}^{(2)} = G_{yyx}^{(2)}$. Furthermore, since $\mu_{\boldsymbol{q}m}^z$ and $v_{\boldsymbol{q}m}^x$ are $p_x$ wave, and $v_{\boldsymbol{q}m}^y$ is $p_y$ wave, we find that $G_{xxy}^{(2)}=G_{xyx}^{(2)} = G_{yxx}^{(2)} = G_{yyy}^{(2)} = 0$. In these cases, while the summand is even in $\boldsymbol{q}$, it is odd in $q_x$ and $q_y$ individually, leading to zero sum.  

Figure~\ref{fig:GJF}(a) shows $G_{xxx}^{(2)}$ and $G_{xyy}^{(2)}=G_{yxy}^{(2)}=G_{yyx}^{(2)}$ for the same five parameter sets considered for the magnon spectrum in Fig.~\ref{fig:spectrum} and listed in Table~\ref{tab:parameters}. While the absolute values differ, all parameter sets show the same general behavior with an initial increase in the absolute value at low temperature, before converging to zero at large temperature. 
This shows that SOC is not necessary to generate the spin transport. 
Furthermore, we find that adding SOC in the form of anisotropies and DMI does not destroy the spin current. Parameter set e, where we remove biquadratic terms and consider only DMI as a stabilizing mechanism gives the smallest absolute value of the transport coefficients. This is due to the smaller magnitude of the magnetic moment in this case, as seen in Fig.~\ref{fig:spectrum}(e). We again stress that our calculations are most trustworthy at low temperature compared to the maximum magnon energy.

The transport coefficients are all negative, and $G_{xxx}^{(2)}$ has largest absolute value. Hence, the current is mostly in the negative $x$ direction, no matter the direction of the applied temperature gradient, as illustrated in Fig.~\ref{fig:GJF}(d). 
The spin current is
\begin{equation}
    \boldsymbol{J} = |\boldsymbol{\nabla}T|^2 (G_{xxx}^{(2)}\cos^2\chi + G_{xyy}^{(2)}\sin^2 \chi, 2G_{xyy}^{(2)}\cos\chi\sin\chi ,0).
\end{equation}
Note that $\chi \to \chi+\pi$ leaves the current invariant. This follows from the fact that the transport occurs to quadratic order in the temperature gradient. Figure \ref{fig:GJF}(b) shows $J_x$ and $J_y$ as functions of $\chi$. The $x$ component oscillates between $|\boldsymbol{\nabla}T|^2 G_{xxx}^{(2)}$ and $|\boldsymbol{\nabla}T|^2 G_{xyy}^{(2)}$, while the $y$ component oscillates between $\pm |\boldsymbol{\nabla}T|^2 G_{xyy}^{(2)}$. The specific cases where the spin current is colinear and perpendicular to the applied temperature gradient are called the spin Seebeck and the spin Nernst effect, respectively. In AMs, they appear as linear response and are reciprocal \cite{Cui2023AM}.

\subsection{Magnon Edelstein effect}
A temperature gradient may also generate a net magnetic moment in the sample. This phenomenon is called the thermal magnon Edelstein effect, in analogy with the electron-field-driven Edelstein effect in electronic systems \cite{Edelstein1990Jan, Bahari2026Apr, Li2020AFMEdelstein, Neumann2026pMmagnon, Kravchuk2026magmoment}. 
We obtain a magnetization in the $z$ direction given by \cite{Kravchuk2026magmoment}
\begin{equation}
    M_z = \frac{1}{4N_u} \sum_{\boldsymbol{q}m} \mu_{\boldsymbol{q}m}^z \delta n_{\boldsymbol{q}m}.
\end{equation}
We can write it as
\begin{equation}
    M_z = F_\alpha \partial_\alpha T + \dots,
\end{equation}
where the dots represent nonlinear responses and
\begin{equation}
\label{eq:F}
    F_\alpha = -\frac{\tau_r}{4N_u}\sum_{\boldsymbol{q}m} \mu_{\boldsymbol{q}m}^z (v_{\boldsymbol{q}m})_\alpha \partial_T n_{\boldsymbol{q}m}^0.  
\end{equation}

In contrast to Ref.~\cite{Kravchuk2026magmoment}, we find that the linear response $F_\alpha$ is nonzero. 
The difference arises because we have a $p$-wave splitting of the magnetic moment, while Ref.~\cite{Kravchuk2026magmoment} studies the $f$-wave case. The $f$ wave introduces more sign changes, leading to cancellations in the sum. 
Linear magnon Edelstein effect was found for $p$-wave spin-split magnons in Ref.~\cite{Neumann2026pMmagnon}, a result we extend to coplanar OPWMs.
We find that $F_y = 0$ since the summand is odd in $q_x$ and $q_y$ individually, while $F_x$ is nonzero since both $\mu_{\boldsymbol{q}m}^z$ and $v_{\boldsymbol{q}m}^x$ are $p_x$ wave leading to a summand that is even in both $q_x$ and $q_y$. By similar arguments we can conclude that all nonlinear responses to even orders are zero, while certain responses to odd order are nonzero. Thus, the effect is reciprocal, where the net magnetization flips sign if the temperature gradient changes sign.

We arrive at $M_z = F_x |\boldsymbol{\nabla}T| \cos \chi.$
Hence, we obtain the largest magnetization if the temperature gradient is applied along the $x$ axis, where the split in magnon magnetic moment is largest. Meanwhile, there is no net magnetization generated if the gradient is along the $y$ axis. Being a linear effect, the magnetization can become quite large. Furthermore, the thermal magnon Edelstein effect does not rely on SOC, as seen by parameter set a in Fig.~\ref{fig:GJF}(c). Both the electronic and magnonic Edelstein effects typically rely on SOC \cite{Edelstein1990Jan, Bahari2026Apr, Li2020AFMEdelstein}, but here it is  a nonrelativistic effect due to the splitting of magnetic moment in the magnon bands which occurs also at zero SOC, cf.~Fig~\ref{fig:spectrum}(a). Hence, this nonrelativistic thermal magnon Edelstein effect can be highly efficient, in analogy to the electronic version in metallic OPWMs \cite{Chakraborty2025Aug}.

\subsection{Measurability}

Note that our results are not material specific. Therefore, we use dimensionless units. Neverthelesss, we can relate our parameters to the choice done in Ref.~\cite{Kravchuk2026magmoment} to estimate the measurability of our spin currents and magnetization. Hence, we set $\tau_r \approx 1.3\times 10^{-10}$~s, $g\approx 2$, $a \approx 0.5~$nm, and $|\boldsymbol{\nabla}T| \approx 0.01~\text{K}/\text{nm}$. To compare to 3D calculations we imagine our 2D model represents a thin sheet of thickness $a$.
We estimate $J_3 \approx 1$~meV, $S = 5/2$ and consider the spin current at $k_B T = 2.6 J_3 S$, i.e., at $T \approx 75$~K. At that temperature, the maximum spin current $|\boldsymbol{J}|/J_0 \approx 0.65$. To get proper units we need to insert constants that we have set to one. 
Also, we calculate the magnetic moment current $\boldsymbol{J}$. To more directly match the literature we convert to a spin current $\boldsymbol{J}_S$ by assuming $\boldsymbol{\mu}_{\boldsymbol{q}m} = g\mu_B \boldsymbol{s}_{\boldsymbol{q}m}/\hbar$. 
The magnitude of the spin current density in units of angular momentum per time per area is $J_S \approx 0.65 (k_B |\boldsymbol{\nabla}T| \tau_r/\hbar)^2 J_3 S \approx 7.6\times 10^{-6}~\text{J}/\text{m}^2$. This is orders of magnitude greater than the current density $J_{\text{YIG}} \approx 10^{-11}~\text{J}/\text{m}^2$ estimated and measured in yttrium iron garnet (YIG) in linear response \cite{Rezende2014YIG}. The reason is that the theoretical estimate for YIG is done at higher temperature $T\approx 300$~K and weaker temperature gradient $|\boldsymbol{\nabla}T| \approx 300~\text{K}/\text{cm}$ \cite{Rezende2014YIG}.
The spin current in YIG was measured via an inverse spin Hall effect in Pt. We refer to Ref.~\cite{Rezende2014YIG} for a way to estimate the generated voltage across Pt from the spin current density.

The maximum value of the magnetization is $M_{\text{max}} = 0.022 k_B|\boldsymbol{\nabla}T|  \tau_r g$ for $k_B T \approx 10J_3 S$. 
Inserting constants previously set to one gives $M_{\text{max}} = 0.022  k_B|\boldsymbol{\nabla}T| \tau_r g \mu_B/\hbar a^2$, where $\mu_B$ is the Bohr magneton.  
Then, $M_{\text{max}} \approx 0.03 \mu_B/\text{nm}^3$, i.e., $\mu_0 M_{\text{max}} \approx 0.4~\text{mT}$, with $\mu_0$ the vacuum permeability. That is a substantial magnetic field. 
With the chosen temperature gradient and relaxation time, the factor relating linear response to third order response is $(k_B |\boldsymbol{\nabla}T|a \tau_r/\hbar)^{-2} \approx 137$. Thus, the linear response should be about two orders of magnitude greater than nonlinear response to third order.
By tuning the parameter values in Eq.~\eqref{eq:CNAOheisenberg} we conjecture that $\mu_0 M_{\text{max}}$ could be increased sufficiently to flip the magnetization of soft ferromagnetic thin films where the coercive field can be $0.5$~mT \cite{Fergen2002softFM}.

\section{Conclusion} \label{sec:conclusion}
We consider a noncolinear, coplanar magnetic ground state on a square lattice. The state has four sublattices, breaks inversion symmetry, and has an effective time-reversal symmetry when combined with translation. It is thus an OPWM. We consider the case of a magnetic insulator and focus on the magnon spectrum, quantizing the collective spin excitations around the classical ground state. We find that coplanar OPWMs have zero magnon spin expectation values in all directions within linear spin wave theory. Instead, they show a colinear, odd-parity momentum dependence of the magnon magnetic moment out of plane. 
Our minimal example is a $p$-wave magnet. We find that this gives rise to nonreciprocal spin currents where the spin current direction and magnitude does not change under reversal of the temperature gradient. Thus, OPWMs are potential spin-current rectifiers for use in magnon spintronics. Furthermore, we find an efficient linear magnon Edelstein effect. 
We include the effects of spin-orbit coupling, and find that they are neither necessary nor destructive to the magnetic moment splitting and particular transport phenomena.

\section*{Acknowledgments}
We thank Alexander Mook and Robin R.~Neumann  for useful discussions.
This work was supported by the Deutsche Forschungsgemeinschaft (DFG, German Research Foundation) through SFB 1170 (project ID 258499086) and the W{\"u}rzburg-Dresden Cluster of Excellence ctd.qmat (EXC 2147, project ID 390858490).

\appendix

\section{Obtaining the ground state} \label{app:GS}
To obtain the classical GS, we set $\boldsymbol{S}_i = S\boldsymbol{m}_i$, with $\boldsymbol{m}_i =  (\sin\theta_i\cos\phi_i, \sin\theta_i\sin\phi_i, \cos\theta_i)$. That yields a Hamiltonian $H(\{\boldsymbol{m}_i\})$ which we seek to minimize, giving the state with lowest free energy at zero temperature, i.e., the classical GS. The first challenge is to identify the periodicity of the GS. For that we first run a Monte Carlo related simulated annealing approach with various finite lattice sizes with periodic boundary conditions. We designed the Hamiltonian so that the GS in Fig.~\ref{fig:GS} is the most likely, and so we use its energy per site as a benchmark. If the simulated annealing finds no states with lower energy, we conclude the GS has the same periodicity as the one in Fig.~\ref{fig:GS}. We can then run simulations on that specific periodicity, and find that a self-consistent iteration yields the states with lowest energy \cite{Maeland2022Jun, Maeland2022Aug, Maeland2023AprPRL, Maeland2023DecTSC}. In most cases, the GS is exactly the one in Fig.~\ref{fig:GS}, up to rotationally symmetric GSs in certain cases as discussed in Sec.~\ref{sec:magnon}.

The simulated annealing approach is defined by the following steps \cite{simulatedannealing, DiazPRL, DiazPRR, Maeland2022Jun, Maeland2022Aug, Maeland2023AprPRL, Maeland2023DecTSC}:
\begin{enumerate}
    \item Start from a random state specified by the set $\{\boldsymbol{m}_i\}$. In practice, initialize the set $\{\theta_i, \phi_i\}$ to random angles.
    \item Choose a site $j$ at random and draw new random values for $\theta_j, \phi_j$. \label{item:update}
    \item Compute the change in energy, $\Delta H$, and $W = e^{-\beta\Delta H}$. Here, $\beta = 1/k_B T$ is an inverse temperature used for the simulation. Start from a high temperature, since the starting random state is a high-energy state.
    \item Find a random number, $r$, between $0$ and 1. If $W > r$, keep the change in $\theta_j, \phi_j$, giving a new state. If not, reject the new state.
    \item Reduce the temperature and repeat from step \ref{item:update} for a fixed number sweeps, usually a high number.
\end{enumerate}
A sweep is defined as testing random sites the same number of times as the number of sites. We note that in simulated annealing, the temperature is simply a tool used to obtain the global minimum of the function $H(\{\boldsymbol{m}_i\})$. Typically, we use the first tenth of the planned sweeps to thermalize the system, starting at a high temperature (e.g. $k_B T = 100J_3$), and ending at a low temperature (e.g. $k_B T = J_3/1000$). The idea is to avoid getting stuck in local minima, while also giving the simulation a chance to find the best possible state in the region where it freezes into by the initial thermalization. Usually, that will be close to the global minimum, but we run the simulation several times starting from different random starting distributions to be sure. Alternatively, the above procedure can be used by starting from a good guess for the GS, and keeping the temperature low. But in cases where we have a good guess for the GS, we prefer the self-consistent iteration, which typically finds the lowest energy states.

The self-consistent iteration is defined by the following steps \cite{dosSantosPRB, Maeland2022Jun, Maeland2022Aug, Maeland2023AprPRL, Maeland2023DecTSC}:
\begin{enumerate}
    \item Start from a random state specified by the set $\{\boldsymbol{m}_i\}$. In practice, initialize the set $\{\theta_i, \phi_i\}$ to random angles. Here we can also start from a good guess of the GS.
    \item Calculate the derivatives $T_i^\theta = \partial_{\theta_i} H$ and $T_i^\phi = \partial_{\phi_i} H$ for all sites. Then, set the angles to new values $\theta_i^{\text{new}} = \theta_i^{\text{old}}-\alpha T_i^\theta$ and $\phi_i^{\text{new}} = \phi_i^{\text{old}}-\alpha T_i^\phi$. Here, $\alpha$ is the mixing parameter. We find the best behavior using central difference to approximate the derivatives and $\alpha J_3 S \approx 10^{-6}$. \label{item:torque}
    \item Repeat step \ref{item:torque} until the derivatives are below a chosen threshold, indicating that self-consistency is reached, as the energy does not decrease by doing more steps.
\end{enumerate}
As an example, when we have nonzero magnetic field we start from the GS in Fig.~\ref{fig:GS}, and quickly reach self-consistency with a new state with lower energy and a net magnetization along the magnetic field. By choosing parameters where we believe the GS in Fig.~\ref{fig:GS} to be the true GS, self-consistency is reached at the first step of the iteration. That alone is not a proof that it is the GS, it just shows it is a local minimum. Additional checks with varying lattice sizes and starting from random distribution increase the confidence that we find the true GS.

\section{Generating magnon spin within LSWT} \label{app:spin}

\begin{figure}
    \centering
    \includegraphics[width=0.99\linewidth]{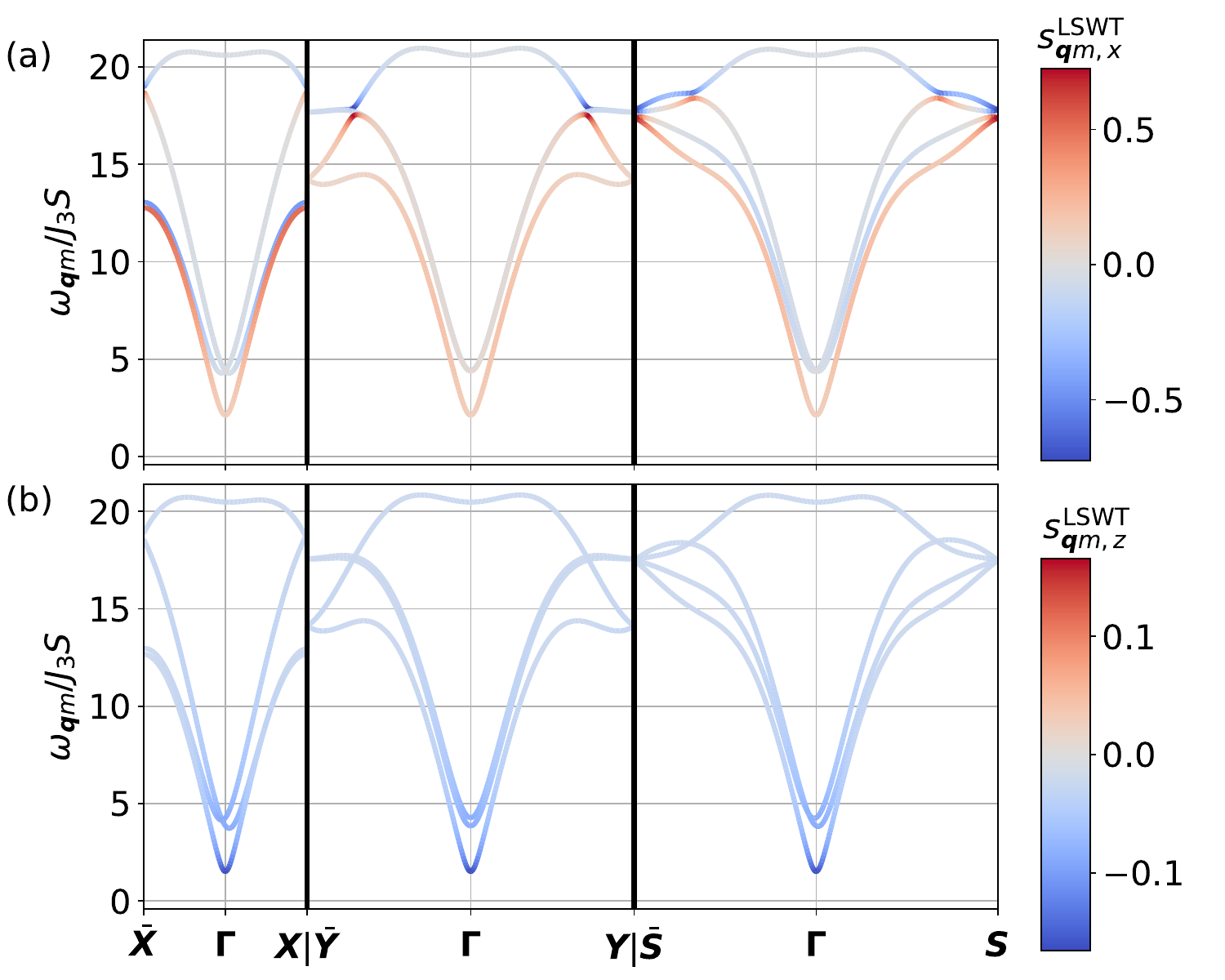}
    \caption{Magnon spin expectation values under symmetry-breaking perturbations. (a) Easy-axis anisotropy $K=0.2J_3$ only on sublattices corresponding to $\boldsymbol{m}_0$ and $\boldsymbol{m}_2$ in Fig.~\ref{fig:GS}. No magnetic field and same GS as Fig.~\ref{fig:GS}. (b) Magnetic field in positive $z$ direction $gH_z = 0.5J_3$ and easy-axis anisotropy $K=0.05J_3$ on all sites. Here, the GS is noncoplanar, with spins canted towards the magnetic field direction. Unspecified parameters correspond to parameter set d in Table~\ref{tab:parameters}.}
    \label{fig:wsxsy}
\end{figure}

Here, we ask what perturbations may affect our model to result in nonzero magnon spins $\boldsymbol{s}_{\boldsymbol{q}m}^{\text{LSWT}}$ within the harmonic approximation of Eq.~\eqref{eq:sqmLSWT}. As mentioned in Sec.~\ref{sec:magnonspin}, we find that the $[C_{2z}||E|\boldsymbol{\tau}]$ symmetry prevents in-plane magnon spin, while the coplanar GS rules out out-of-plane $\boldsymbol{s}_{\boldsymbol{q}m}^{\text{LSWT}}$. Thus we can either break $[C_{2z}||E|\boldsymbol{\tau}]$ symmetry, or make the spins noncoplanar.

In addition to a pure symmetry argument, let us provide additional arguments why there is zero magnon spin in-plane. The sum over $\hat{e}_3^r$ is zero because the GS has zero net magnetization. Still, the reason the spin is zero must also depend on the magnon eigenstates. We can rewrite the magnon spin as
\begin{align}
    \boldsymbol{s}_{\boldsymbol{q}m}^{\text{LSWT}}    &= - \sum_r \hat{e}_3^r \bigg[ \abs{(T_{\boldsymbol{q}}^{-1})_{r+N_{\text{SL}},m}}^2 +  \abs{(T_{\boldsymbol{q}}^{-1})_{rm}}^2 \bigg]. \label{eq:sqmT}
\end{align}
We can interpret $\abs{(T_{\boldsymbol{q}}^{-1})_{r+N_{\text{SL}},m}}^2$ and $\abs{(T_{\boldsymbol{q}}^{-1})_{rm}}^2$ as weights, describing how much sublattice $r$ contributes to magnon mode $m$. We find that these weights are independent of $r$ for each $m$, which is why the spin sums to zero. That means all sublattices see the same amount of spin fluctuation for all modes. By contrast, in AFMs and AMs, the corresponding weights depend on the sublattice index $r$ giving magnon bands with spin $\boldsymbol{s}_{\boldsymbol{q}\pm} = \pm \hat{N}$ along the N\'eel vector $\hat{N}$ despite zero net magnetization. These magnon spins correspond to those that can be inferred from electron-magnon coupling \cite{SunMaeland2023Aug, SunMaeland2024Mar, Maeland2024Feb}.

We can increase the spin fluctuation on only some sublattices by considering a site dependent strength of the easy-axis anisotropy, $-\sum_i K_i(\boldsymbol{S}_i \cdot \hat{n}_i)^2$. For simplicity, we keep it fixed on each sublattice. If we have easy-axis anisotropy only on sublattices 0 and 2, we expect more spin fluctuations on sublattices 1 and 3. Sublattices 1 and 3 have ground state spins corresponding to $\boldsymbol{m}_1$ and $\boldsymbol{m}_3$ in Fig.~\ref{fig:GS}, whose sum points in the $x$-direction. Indeed, Fig.~\ref{fig:wsxsy}(a) shows a nonzero spin expectation value in the $x$ direction, while the other directions still have zero magnon spin. We get an $s$-wave spin splitting of the magnon modes. Thus the system is not an OPWM anymore and the $s$-wave spin splitting is more reminiscent of ferrimagnets \cite{Aoyama2026AMFiMmag}. The reason is that the sublattice dependent easy-axis anisotropy breaks both $[C_{2z}||E|\boldsymbol{\tau}]$ and the defining $T\boldsymbol{\tau}$ symmetry.

If we apply a magnetic field the GS cants towards the field giving a net magnetization. Also here, $T\boldsymbol{\tau}$ symmetry is broken. Much like a ferromagnet, all the magnon bands then have a spin pointing in the opposite direction of the net magnetization \cite{Maeland2021Sep}. Figure~\ref{fig:wsxsy}(b) demonstrates the case for a magnetic field pointing in $z$ direction. We find nonzero $s_{\boldsymbol{q}m,z}^{\text{LSWT}} < 0$, while the other components are zero.

\section{Boltzmann transport calculation} \label{app:transport}
In the transport calculation, we rely on the nonequilibrium part of the distribution function $\delta n_{\boldsymbol{q}m} = n_{\boldsymbol{q}m}-n_{\boldsymbol{q}m}^0$. We model it using the Boltzmann equation in the relaxation time $\tau_r$ approximation as
\begin{equation}
\label{eq:dn}
    \delta n_{\boldsymbol{q}m} = -\tau_r \boldsymbol{v}_{\boldsymbol{q}m} \cdot \boldsymbol{\nabla} n_{\boldsymbol{q}m} = -\tau_r (v_{\boldsymbol{q}m})_\alpha \partial_\alpha n_{\boldsymbol{q}m}.
\end{equation}
We consider small changes from equilibrium, so that we can expand the distribution function as
\begin{equation}
    n_{\boldsymbol{q}m} = n_{\boldsymbol{q}m}^0 + \delta n_{\boldsymbol{q}m}^{(1)} + \delta n_{\boldsymbol{q}m}^{(2)} + \delta n_{\boldsymbol{q}m}^{(3)}  \dots,
\end{equation}
where each successive term is higher order in gradients. We insert this in Eq.~\eqref{eq:dn} and equate terms of the same order in gradients. That yields
\begin{align}
    \delta n_{\boldsymbol{q}m}^{(1)} &=  -\tau_r (v_{\boldsymbol{q}m})_\alpha \partial_\alpha n_{\boldsymbol{q}m}^0, \\
    \delta n_{\boldsymbol{q}m}^{(2)} &= -\tau_r (v_{\boldsymbol{q}m})_\alpha \partial_\alpha \delta n_{\boldsymbol{q}m}^{(1)} \nonumber\\
    &= \tau_r^2 (v_{\boldsymbol{q}m})_\alpha (v_{\boldsymbol{q}m})_\beta \partial_\alpha \partial_\beta n_{\boldsymbol{q}m}^0, \\
    \delta n_{\boldsymbol{q}m}^{(3)} &= -\tau_r (v_{\boldsymbol{q}m})_\alpha \partial_\alpha \delta n_{\boldsymbol{q}m}^{(2)} \nonumber\\
    &= -\tau_r^3 (v_{\boldsymbol{q}m})_\alpha (v_{\boldsymbol{q}m})_\beta (v_{\boldsymbol{q}m})_\gamma \partial_\alpha \partial_\beta \partial_\gamma n_{\boldsymbol{q}m}^0,
\end{align}
and so on. We see that higher order terms get new factors of $\boldsymbol{v}_{\boldsymbol{q}m}$ which is odd in $\boldsymbol{q}$. That is how in the main text we can draw conclusions about all orders of response from calculating the lowest order ones. We further rewrite $\partial_\alpha n_{\boldsymbol{q}m}^0 = \partial_\alpha T \partial_T n_{\boldsymbol{q}m}^0$ and $ \partial_\alpha \partial_\beta n_{\boldsymbol{q}m}^0 = \partial_\alpha T \partial_\beta T \partial_T^2 n_{\boldsymbol{q}m}^0$ to define the transport coefficients in Eqs.~\eqref{eq:G}, \eqref{eq:G2}, and \eqref{eq:F}. The derivatives of the distribution function are
\begin{equation}
    \pdv{n_{\boldsymbol{q}m}^0 }{T} = \frac{n_{\boldsymbol{q}m}^0(n_{\boldsymbol{q}m}^0+1)\omega_{\boldsymbol{q}m}}{k_B T^2}
\end{equation}
and
\begin{equation}
    \pdv[2]{n_{\boldsymbol{q}m}^0 }{T} = \frac{n_{\boldsymbol{q}m}^0(n_{\boldsymbol{q}m}^0+1)}{T^2} \bqty{\frac{\omega_{\boldsymbol{q}m}^2}{(k_B T)^2}(2n_{\boldsymbol{q}m}^0+1)-2\frac{\omega_{\boldsymbol{q}m}}{k_B T}}.
\end{equation}

\bibliography{main.bbl}

\begin{thebibliography}{92}%
\makeatletter
\providecommand \@ifxundefined [1]{%
 \@ifx{#1\undefined}
}%
\providecommand \@ifnum [1]{%
 \ifnum #1\expandafter \@firstoftwo
 \else \expandafter \@secondoftwo
 \fi
}%
\providecommand \@ifx [1]{%
 \ifx #1\expandafter \@firstoftwo
 \else \expandafter \@secondoftwo
 \fi
}%
\providecommand \natexlab [1]{#1}%
\providecommand \enquote  [1]{``#1''}%
\providecommand \bibnamefont  [1]{#1}%
\providecommand \bibfnamefont [1]{#1}%
\providecommand \citenamefont [1]{#1}%
\providecommand \href@noop [0]{\@secondoftwo}%
\providecommand \href [0]{\begingroup \@sanitize@url \@href}%
\providecommand \@href[1]{\@@startlink{#1}\@@href}%
\providecommand \@@href[1]{\endgroup#1\@@endlink}%
\providecommand \@sanitize@url [0]{\catcode `\\12\catcode `\$12\catcode `\&12\catcode `\#12\catcode `\^12\catcode `\_12\catcode `\%12\relax}%
\providecommand \@@startlink[1]{}%
\providecommand \@@endlink[0]{}%
\providecommand \url  [0]{\begingroup\@sanitize@url \@url }%
\providecommand \@url [1]{\endgroup\@href {#1}{\urlprefix }}%
\providecommand \urlprefix  [0]{URL }%
\providecommand \Eprint [0]{\href }%
\providecommand \doibase [0]{https://doi.org/}%
\providecommand \selectlanguage [0]{\@gobble}%
\providecommand \bibinfo  [0]{\@secondoftwo}%
\providecommand \bibfield  [0]{\@secondoftwo}%
\providecommand \translation [1]{[#1]}%
\providecommand \BibitemOpen [0]{}%
\providecommand \bibitemStop [0]{}%
\providecommand \bibitemNoStop [0]{.\EOS\space}%
\providecommand \EOS [0]{\spacefactor3000\relax}%
\providecommand \BibitemShut  [1]{\csname bibitem#1\endcsname}%
\let\auto@bib@innerbib\@empty
\bibitem [{\citenamefont {Chumak}\ \emph {et~al.}(2015)\citenamefont {Chumak}, \citenamefont {Vasyuchka}, \citenamefont {Serga},\ and\ \citenamefont {Hillebrands}}]{Chumak2015Jun}%
  \BibitemOpen
  \bibfield  {author} {\bibinfo {author} {\bibfnamefont {A.~V.}\ \bibnamefont {Chumak}}, \bibinfo {author} {\bibfnamefont {V.~I.}\ \bibnamefont {Vasyuchka}}, \bibinfo {author} {\bibfnamefont {A.~A.}\ \bibnamefont {Serga}},\ and\ \bibinfo {author} {\bibfnamefont {B.}~\bibnamefont {Hillebrands}},\ }\bibfield  {title} {\bibinfo {title} {{Magnon spintronics}},\ }\href {https://doi.org/10.1038/nphys3347} {\bibfield  {journal} {\bibinfo  {journal} {Nat. Phys.}\ }\textbf {\bibinfo {volume} {11}},\ \bibinfo {pages} {453} (\bibinfo {year} {2015})}\BibitemShut {NoStop}%
\bibitem [{\citenamefont {Chumak}(2019)}]{Chumak2019mgcomp}%
  \BibitemOpen
  \bibfield  {author} {\bibinfo {author} {\bibfnamefont {A.~V.}\ \bibnamefont {Chumak}},\ }\bibfield  {title} {\bibinfo {title} {{Magnon Spintronics : Fundamentals of Magnon-Based Computing}},\ }in\ \href {https://doi.org/10.1201/9780429423079-6} {\emph {\bibinfo {booktitle} {{Spintronics Handbook, Second Edition: Spin Transport and Magnetism}}}}\ (\bibinfo  {publisher} {CRC Press},\ \bibinfo {address} {Boca Raton, FL, USA},\ \bibinfo {year} {2019})\ pp.\ \bibinfo {pages} {247--302}\BibitemShut {NoStop}%
\bibitem [{\citenamefont {Lan}\ \emph {et~al.}(2015)\citenamefont {Lan}, \citenamefont {Yu}, \citenamefont {Wu},\ and\ \citenamefont {Xiao}}]{Lan2015Diode}%
  \BibitemOpen
  \bibfield  {author} {\bibinfo {author} {\bibfnamefont {J.}~\bibnamefont {Lan}}, \bibinfo {author} {\bibfnamefont {W.}~\bibnamefont {Yu}}, \bibinfo {author} {\bibfnamefont {R.}~\bibnamefont {Wu}},\ and\ \bibinfo {author} {\bibfnamefont {J.}~\bibnamefont {Xiao}},\ }\bibfield  {title} {\bibinfo {title} {{Spin-Wave Diode}},\ }\href {https://doi.org/10.1103/PhysRevX.5.041049} {\bibfield  {journal} {\bibinfo  {journal} {Phys. Rev. X}\ }\textbf {\bibinfo {volume} {5}},\ \bibinfo {pages} {041049} (\bibinfo {year} {2015})}\BibitemShut {NoStop}%
\bibitem [{\citenamefont {Chotorlishvili}\ \emph {et~al.}(2022)\citenamefont {Chotorlishvili}, \citenamefont {Wang}, \citenamefont {Dyrda{\l}}, \citenamefont {Guo}, \citenamefont {Dugaev}, \citenamefont {Barna{\ifmmode\acute{s}\else\'{s}\fi}},\ and\ \citenamefont {Berakdar}}]{Chotorlishvili2022Rectification}%
  \BibitemOpen
  \bibfield  {author} {\bibinfo {author} {\bibfnamefont {L.}~\bibnamefont {Chotorlishvili}}, \bibinfo {author} {\bibfnamefont {X.-g.}\ \bibnamefont {Wang}}, \bibinfo {author} {\bibfnamefont {A.}~\bibnamefont {Dyrda{\l}}}, \bibinfo {author} {\bibfnamefont {G.-h.}\ \bibnamefont {Guo}}, \bibinfo {author} {\bibfnamefont {V.~K.}\ \bibnamefont {Dugaev}}, \bibinfo {author} {\bibfnamefont {J.}~\bibnamefont {Barna{\ifmmode\acute{s}\else\'{s}\fi}}},\ and\ \bibinfo {author} {\bibfnamefont {J.}~\bibnamefont {Berakdar}},\ }\bibfield  {title} {\bibinfo {title} {{Rectification of the spin Seebeck current in noncollinear antiferromagnets}},\ }\href {https://doi.org/10.1103/PhysRevB.106.014417} {\bibfield  {journal} {\bibinfo  {journal} {Phys. Rev. B}\ }\textbf {\bibinfo {volume} {106}},\ \bibinfo {pages} {014417} (\bibinfo {year} {2022})}\BibitemShut {NoStop}%
\bibitem [{\citenamefont {Ezawa}(2026)}]{Ezawa2026rectifier}%
  \BibitemOpen
  \bibfield  {author} {\bibinfo {author} {\bibfnamefont {M.}~\bibnamefont {Ezawa}},\ }\bibfield  {title} {\bibinfo {title} {{Fourth- and sixth-order nonlinear spin current rectifier in three-dimensional $h$-wave and $j$-wave odd-parity magnets}},\ }\href {https://doi.org/10.1103/zxcv-xcyc} {\bibfield  {journal} {\bibinfo  {journal} {Phys. Rev. B}\ }\textbf {\bibinfo {volume} {114}},\ \bibinfo {pages} {115404} (\bibinfo {year} {2026})}\BibitemShut {NoStop}%
\bibitem [{\citenamefont {Fischer}\ \emph {et~al.}(2017)\citenamefont {Fischer}, \citenamefont {Kewenig}, \citenamefont {Bozhko}, \citenamefont {Serga}, \citenamefont {Syvorotka}, \citenamefont {Ciubotaru}, \citenamefont {Adelmann}, \citenamefont {Hillebrands},\ and\ \citenamefont {Chumak}}]{Fischer2017spinwavegate}%
  \BibitemOpen
  \bibfield  {author} {\bibinfo {author} {\bibfnamefont {T.}~\bibnamefont {Fischer}}, \bibinfo {author} {\bibfnamefont {M.}~\bibnamefont {Kewenig}}, \bibinfo {author} {\bibfnamefont {D.~A.}\ \bibnamefont {Bozhko}}, \bibinfo {author} {\bibfnamefont {A.~A.}\ \bibnamefont {Serga}}, \bibinfo {author} {\bibfnamefont {I.~I.}\ \bibnamefont {Syvorotka}}, \bibinfo {author} {\bibfnamefont {F.}~\bibnamefont {Ciubotaru}}, \bibinfo {author} {\bibfnamefont {C.}~\bibnamefont {Adelmann}}, \bibinfo {author} {\bibfnamefont {B.}~\bibnamefont {Hillebrands}},\ and\ \bibinfo {author} {\bibfnamefont {A.~V.}\ \bibnamefont {Chumak}},\ }\bibfield  {title} {\bibinfo {title} {{Experimental prototype of a spin-wave majority gate}},\ }\href {https://doi.org/10.1063/1.4979840} {\bibfield  {journal} {\bibinfo  {journal} {Appl. Phys. Lett.}\ }\textbf {\bibinfo {volume} {110}},\ \bibinfo {pages} {152401} (\bibinfo {year} {2017})}\BibitemShut {NoStop}%
\bibitem [{\citenamefont {Chumak}\ \emph {et~al.}(2014)\citenamefont {Chumak}, \citenamefont {Serga},\ and\ \citenamefont {Hillebrands}}]{Chumak2014allmg}%
  \BibitemOpen
  \bibfield  {author} {\bibinfo {author} {\bibfnamefont {A.~V.}\ \bibnamefont {Chumak}}, \bibinfo {author} {\bibfnamefont {A.~A.}\ \bibnamefont {Serga}},\ and\ \bibinfo {author} {\bibfnamefont {B.}~\bibnamefont {Hillebrands}},\ }\bibfield  {title} {\bibinfo {title} {{Magnon transistor for all-magnon data processing}},\ }\href {https://doi.org/10.1038/ncomms5700} {\bibfield  {journal} {\bibinfo  {journal} {Nat. Commun.}\ }\textbf {\bibinfo {volume} {5}},\ \bibinfo {pages} {4700} (\bibinfo {year} {2014})}\BibitemShut {NoStop}%
\bibitem [{\citenamefont {Bauer}\ \emph {et~al.}(2012)\citenamefont {Bauer}, \citenamefont {Saitoh},\ and\ \citenamefont {van Wees}}]{Bauer2012spincal}%
  \BibitemOpen
  \bibfield  {author} {\bibinfo {author} {\bibfnamefont {G.~E.~W.}\ \bibnamefont {Bauer}}, \bibinfo {author} {\bibfnamefont {E.}~\bibnamefont {Saitoh}},\ and\ \bibinfo {author} {\bibfnamefont {B.~J.}\ \bibnamefont {van Wees}},\ }\bibfield  {title} {\bibinfo {title} {{Spin caloritronics}},\ }\href {https://doi.org/10.1038/nmat3301} {\bibfield  {journal} {\bibinfo  {journal} {Nat. Mater.}\ }\textbf {\bibinfo {volume} {11}},\ \bibinfo {pages} {391} (\bibinfo {year} {2012})}\BibitemShut {NoStop}%
\bibitem [{\citenamefont {Uchida}(2021)}]{Uchida2021spincal}%
  \BibitemOpen
  \bibfield  {author} {\bibinfo {author} {\bibfnamefont {K.-i.}\ \bibnamefont {Uchida}},\ }\bibfield  {title} {\bibinfo {title} {{Transport phenomena in spin caloritronics}},\ }\href {https://doi.org/10.2183/pjab.97.004} {\bibfield  {journal} {\bibinfo  {journal} {Proc. Jpn. Acad., Ser. B}\ }\textbf {\bibinfo {volume} {97}},\ \bibinfo {pages} {69} (\bibinfo {year} {2021})}\BibitemShut {NoStop}%
\bibitem [{\citenamefont {Serga}\ \emph {et~al.}(2010)\citenamefont {Serga}, \citenamefont {Chumak},\ and\ \citenamefont {Hillebrands}}]{Serga2010YIG}%
  \BibitemOpen
  \bibfield  {author} {\bibinfo {author} {\bibfnamefont {A.~A.}\ \bibnamefont {Serga}}, \bibinfo {author} {\bibfnamefont {A.~V.}\ \bibnamefont {Chumak}},\ and\ \bibinfo {author} {\bibfnamefont {B.}~\bibnamefont {Hillebrands}},\ }\bibfield  {title} {\bibinfo {title} {{YIG magnonics}},\ }\href {https://doi.org/10.1088/0022-3727/43/26/264002} {\bibfield  {journal} {\bibinfo  {journal} {J. Phys. D: Appl. Phys.}\ }\textbf {\bibinfo {volume} {43}},\ \bibinfo {pages} {264002} (\bibinfo {year} {2010})}\BibitemShut {NoStop}%
\bibitem [{\citenamefont {Cahaya}(2025)}]{Cahaya2025spincomp}%
  \BibitemOpen
  \bibfield  {author} {\bibinfo {author} {\bibfnamefont {A.~B.}\ \bibnamefont {Cahaya}},\ }\bibfield  {title} {\bibinfo {title} {{Spin current compensation from competing magnon modes in ferrimagnets}},\ }\href {https://doi.org/10.1088/1361-6463/adfaa0} {\bibfield  {journal} {\bibinfo  {journal} {J. Phys. D: Appl. Phys.}\ }\textbf {\bibinfo {volume} {58}},\ \bibinfo {pages} {345002} (\bibinfo {year} {2025})}\BibitemShut {NoStop}%
\bibitem [{\citenamefont {Xiao}\ \emph {et~al.}(2010)\citenamefont {Xiao}, \citenamefont {Bauer}, \citenamefont {Uchida}, \citenamefont {Saitoh},\ and\ \citenamefont {Maekawa}}]{Xiao2010SSE}%
  \BibitemOpen
  \bibfield  {author} {\bibinfo {author} {\bibfnamefont {J.}~\bibnamefont {Xiao}}, \bibinfo {author} {\bibfnamefont {G.~E.~W.}\ \bibnamefont {Bauer}}, \bibinfo {author} {\bibfnamefont {K.-c.}\ \bibnamefont {Uchida}}, \bibinfo {author} {\bibfnamefont {E.}~\bibnamefont {Saitoh}},\ and\ \bibinfo {author} {\bibfnamefont {S.}~\bibnamefont {Maekawa}},\ }\bibfield  {title} {\bibinfo {title} {{Theory of magnon-driven spin Seebeck effect}},\ }\href {https://doi.org/10.1103/PhysRevB.81.214418} {\bibfield  {journal} {\bibinfo  {journal} {Phys. Rev. B}\ }\textbf {\bibinfo {volume} {81}},\ \bibinfo {pages} {214418} (\bibinfo {year} {2010})}\BibitemShut {NoStop}%
\bibitem [{\citenamefont {Rezende}\ \emph {et~al.}(2016{\natexlab{a}})\citenamefont {Rezende}, \citenamefont {Rodr{\ifmmode\acute{\imath}\else\'{\i}\fi}guez-Su{\ifmmode\acute{a}\else\'{a}\fi}rez},\ and\ \citenamefont {Azevedo}}]{Rezende2016SSEAFM}%
  \BibitemOpen
  \bibfield  {author} {\bibinfo {author} {\bibfnamefont {S.~M.}\ \bibnamefont {Rezende}}, \bibinfo {author} {\bibfnamefont {R.~L.}\ \bibnamefont {Rodr{\ifmmode\acute{\imath}\else\'{\i}\fi}guez-Su{\ifmmode\acute{a}\else\'{a}\fi}rez}},\ and\ \bibinfo {author} {\bibfnamefont {A.}~\bibnamefont {Azevedo}},\ }\bibfield  {title} {\bibinfo {title} {{Theory of the spin Seebeck effect in antiferromagnets}},\ }\href {https://doi.org/10.1103/PhysRevB.93.014425} {\bibfield  {journal} {\bibinfo  {journal} {Phys. Rev. B}\ }\textbf {\bibinfo {volume} {93}},\ \bibinfo {pages} {014425} (\bibinfo {year} {2016}{\natexlab{a}})}\BibitemShut {NoStop}%
\bibitem [{\citenamefont {Rezende}\ \emph {et~al.}(2016{\natexlab{b}})\citenamefont {Rezende}, \citenamefont {Rodr{\ifmmode\acute{\imath}\else\'{\i}\fi}guez-Su{\ifmmode\acute{a}\else\'{a}\fi}rez},\ and\ \citenamefont {Azevedo}}]{Rezende2016SSEAFM2}%
  \BibitemOpen
  \bibfield  {author} {\bibinfo {author} {\bibfnamefont {S.~M.}\ \bibnamefont {Rezende}}, \bibinfo {author} {\bibfnamefont {R.~L.}\ \bibnamefont {Rodr{\ifmmode\acute{\imath}\else\'{\i}\fi}guez-Su{\ifmmode\acute{a}\else\'{a}\fi}rez}},\ and\ \bibinfo {author} {\bibfnamefont {A.}~\bibnamefont {Azevedo}},\ }\bibfield  {title} {\bibinfo {title} {{Diffusive magnonic spin transport in antiferromagnetic insulators}},\ }\href {https://doi.org/10.1103/PhysRevB.93.054412} {\bibfield  {journal} {\bibinfo  {journal} {Phys. Rev. B}\ }\textbf {\bibinfo {volume} {93}},\ \bibinfo {pages} {054412} (\bibinfo {year} {2016}{\natexlab{b}})}\BibitemShut {NoStop}%
\bibitem [{\citenamefont {Rezende}\ \emph {et~al.}(2019)\citenamefont {Rezende}, \citenamefont {Azevedo},\ and\ \citenamefont {Rodr{\ifmmode\acute{\imath}\else\'{\i}\fi}guez-Su{\ifmmode\acute{a}\else\'{a}\fi}rez}}]{Rezende2019AFM}%
  \BibitemOpen
  \bibfield  {author} {\bibinfo {author} {\bibfnamefont {S.~M.}\ \bibnamefont {Rezende}}, \bibinfo {author} {\bibfnamefont {A.}~\bibnamefont {Azevedo}},\ and\ \bibinfo {author} {\bibfnamefont {R.~L.}\ \bibnamefont {Rodr{\ifmmode\acute{\imath}\else\'{\i}\fi}guez-Su{\ifmmode\acute{a}\else\'{a}\fi}rez}},\ }\bibfield  {title} {\bibinfo {title} {{Introduction to antiferromagnetic magnons}},\ }\href {https://doi.org/10.1063/1.5109132} {\bibfield  {journal} {\bibinfo  {journal} {J. Appl. Phys.}\ }\textbf {\bibinfo {volume} {126}},\ \bibinfo {pages} {151101} (\bibinfo {year} {2019})}\BibitemShut {NoStop}%
\bibitem [{\citenamefont {Ciccarelli}\ \emph {et~al.}(2025)\citenamefont {Ciccarelli}, \citenamefont {Nava~Antonio},\ and\ \citenamefont {Barker}}]{Ciccarelli2025compFiMAFMrev}%
  \BibitemOpen
  \bibfield  {author} {\bibinfo {author} {\bibfnamefont {C.}~\bibnamefont {Ciccarelli}}, \bibinfo {author} {\bibfnamefont {G.}~\bibnamefont {Nava~Antonio}},\ and\ \bibinfo {author} {\bibfnamefont {J.}~\bibnamefont {Barker}},\ }\bibfield  {title} {\bibinfo {title} {{Spin emission from antiferromagnets and compensated ferrimagnets}},\ }\href {https://doi.org/10.1063/5.0273489} {\bibfield  {journal} {\bibinfo  {journal} {Appl. Phys. Rev.}\ }\textbf {\bibinfo {volume} {12}},\ \bibinfo {pages} {041306} (\bibinfo {year} {2025})}\BibitemShut {NoStop}%
\bibitem [{\citenamefont {{\ifmmode\check{S}\else\v{S}\fi}mejkal}\ \emph {et~al.}(2022{\natexlab{a}})\citenamefont {{\ifmmode\check{S}\else\v{S}\fi}mejkal}, \citenamefont {Hellenes}, \citenamefont {Gonz{\ifmmode\acute{a}\else\'{a}\fi}lez-Hern{\ifmmode\acute{a}\else\'{a}\fi}ndez}, \citenamefont {Sinova},\ and\ \citenamefont {Jungwirth}}]{Smejkal2022Feb}%
  \BibitemOpen
  \bibfield  {author} {\bibinfo {author} {\bibfnamefont {L.}~\bibnamefont {{\ifmmode\check{S}\else\v{S}\fi}mejkal}}, \bibinfo {author} {\bibfnamefont {A.~B.}\ \bibnamefont {Hellenes}}, \bibinfo {author} {\bibfnamefont {R.}~\bibnamefont {Gonz{\ifmmode\acute{a}\else\'{a}\fi}lez-Hern{\ifmmode\acute{a}\else\'{a}\fi}ndez}}, \bibinfo {author} {\bibfnamefont {J.}~\bibnamefont {Sinova}},\ and\ \bibinfo {author} {\bibfnamefont {T.}~\bibnamefont {Jungwirth}},\ }\bibfield  {title} {\bibinfo {title} {{Giant and Tunneling Magnetoresistance in Unconventional Collinear Antiferromagnets with Nonrelativistic Spin-Momentum Coupling}},\ }\href {https://doi.org/10.1103/PhysRevX.12.011028} {\bibfield  {journal} {\bibinfo  {journal} {Phys. Rev. X}\ }\textbf {\bibinfo {volume} {12}},\ \bibinfo {pages} {011028} (\bibinfo {year} {2022}{\natexlab{a}})}\BibitemShut {NoStop}%
\bibitem [{\citenamefont {{\ifmmode\check{S}\else\v{S}\fi}mejkal}\ \emph {et~al.}(2022{\natexlab{b}})\citenamefont {{\ifmmode\check{S}\else\v{S}\fi}mejkal}, \citenamefont {Sinova},\ and\ \citenamefont {Jungwirth}}]{Smejkal2022Sep}%
  \BibitemOpen
  \bibfield  {author} {\bibinfo {author} {\bibfnamefont {L.}~\bibnamefont {{\ifmmode\check{S}\else\v{S}\fi}mejkal}}, \bibinfo {author} {\bibfnamefont {J.}~\bibnamefont {Sinova}},\ and\ \bibinfo {author} {\bibfnamefont {T.}~\bibnamefont {Jungwirth}},\ }\bibfield  {title} {\bibinfo {title} {{Beyond Conventional Ferromagnetism and Antiferromagnetism: A Phase with Nonrelativistic Spin and Crystal Rotation Symmetry}},\ }\href {https://doi.org/10.1103/PhysRevX.12.031042} {\bibfield  {journal} {\bibinfo  {journal} {Phys. Rev. X}\ }\textbf {\bibinfo {volume} {12}},\ \bibinfo {pages} {031042} (\bibinfo {year} {2022}{\natexlab{b}})}\BibitemShut {NoStop}%
\bibitem [{\citenamefont {{\ifmmode\check{S}\else\v{S}\fi}mejkal}\ \emph {et~al.}(2022{\natexlab{c}})\citenamefont {{\ifmmode\check{S}\else\v{S}\fi}mejkal}, \citenamefont {Sinova},\ and\ \citenamefont {Jungwirth}}]{Smejkal2022Dec}%
  \BibitemOpen
  \bibfield  {author} {\bibinfo {author} {\bibfnamefont {L.}~\bibnamefont {{\ifmmode\check{S}\else\v{S}\fi}mejkal}}, \bibinfo {author} {\bibfnamefont {J.}~\bibnamefont {Sinova}},\ and\ \bibinfo {author} {\bibfnamefont {T.}~\bibnamefont {Jungwirth}},\ }\bibfield  {title} {\bibinfo {title} {{Emerging Research Landscape of Altermagnetism}},\ }\href {https://doi.org/10.1103/PhysRevX.12.040501} {\bibfield  {journal} {\bibinfo  {journal} {Phys. Rev. X}\ }\textbf {\bibinfo {volume} {12}},\ \bibinfo {pages} {040501} (\bibinfo {year} {2022}{\natexlab{c}})}\BibitemShut {NoStop}%
\bibitem [{\citenamefont {Brekke}\ \emph {et~al.}(2023)\citenamefont {Brekke}, \citenamefont {Brataas},\ and\ \citenamefont {Sudb{\o}}}]{Brekke2023Aug}%
  \BibitemOpen
  \bibfield  {author} {\bibinfo {author} {\bibfnamefont {B.}~\bibnamefont {Brekke}}, \bibinfo {author} {\bibfnamefont {A.}~\bibnamefont {Brataas}},\ and\ \bibinfo {author} {\bibfnamefont {A.}~\bibnamefont {Sudb{\o}}},\ }\bibfield  {title} {\bibinfo {title} {{Two-dimensional altermagnets: Superconductivity in a minimal microscopic model}},\ }\href {https://doi.org/10.1103/PhysRevB.108.224421} {\bibfield  {journal} {\bibinfo  {journal} {Phys. Rev. B}\ }\textbf {\bibinfo {volume} {108}},\ \bibinfo {pages} {224421} (\bibinfo {year} {2023})}\BibitemShut {NoStop}%
\bibitem [{\citenamefont {M{\ae}land}\ \emph {et~al.}(2024)\citenamefont {M{\ae}land}, \citenamefont {Brekke},\ and\ \citenamefont {Sudb{\o}}}]{Maeland2024Feb}%
  \BibitemOpen
  \bibfield  {author} {\bibinfo {author} {\bibfnamefont {K.}~\bibnamefont {M{\ae}land}}, \bibinfo {author} {\bibfnamefont {B.}~\bibnamefont {Brekke}},\ and\ \bibinfo {author} {\bibfnamefont {A.}~\bibnamefont {Sudb{\o}}},\ }\bibfield  {title} {\bibinfo {title} {{Many-body effects on superconductivity mediated by double-magnon processes in altermagnets}},\ }\href {https://doi.org/10.1103/PhysRevB.109.134515} {\bibfield  {journal} {\bibinfo  {journal} {Phys. Rev. B}\ }\textbf {\bibinfo {volume} {109}},\ \bibinfo {pages} {134515} (\bibinfo {year} {2024})}\BibitemShut {NoStop}%
\bibitem [{\citenamefont {Leraand}\ \emph {et~al.}(2025)\citenamefont {Leraand}, \citenamefont {M{\ae}land},\ and\ \citenamefont {Sudb{\o}}}]{LeraandMaeland2025Feb}%
  \BibitemOpen
  \bibfield  {author} {\bibinfo {author} {\bibfnamefont {K.}~\bibnamefont {Leraand}}, \bibinfo {author} {\bibfnamefont {K.}~\bibnamefont {M{\ae}land}},\ and\ \bibinfo {author} {\bibfnamefont {A.}~\bibnamefont {Sudb{\o}}},\ }\bibfield  {title} {\bibinfo {title} {{Phonon-mediated spin-polarized superconductivity in altermagnets}},\ }\href {https://doi.org/10.1103/g4dl-1ff2} {\bibfield  {journal} {\bibinfo  {journal} {Phys. Rev. B}\ }\textbf {\bibinfo {volume} {112}},\ \bibinfo {pages} {104510} (\bibinfo {year} {2025})}\BibitemShut {NoStop}%
\bibitem [{\citenamefont {Leraand}\ \emph {et~al.}(2026)\citenamefont {Leraand}, \citenamefont {M{\ae}land},\ and\ \citenamefont {Sudb{\o}}}]{Leraand2026Mar}%
  \BibitemOpen
  \bibfield  {author} {\bibinfo {author} {\bibfnamefont {K.}~\bibnamefont {Leraand}}, \bibinfo {author} {\bibfnamefont {K.}~\bibnamefont {M{\ae}land}},\ and\ \bibinfo {author} {\bibfnamefont {A.}~\bibnamefont {Sudb{\o}}},\ }\bibfield  {title} {\bibinfo {title} {{Spin-dependent quasiparticle lifetimes in altermagnets}},\ }\href {https://doi.org/10.1103/r3vm-m1m3} {\bibfield  {journal} {\bibinfo  {journal} {Phys. Rev. B}\ }\textbf {\bibinfo {volume} {113}},\ \bibinfo {pages} {115148} (\bibinfo {year} {2026})}\BibitemShut {NoStop}%
\bibitem [{\citenamefont {Petermann}\ \emph {et~al.}(2025)\citenamefont {Petermann}, \citenamefont {M{\ae}land},\ and\ \citenamefont {Trauzettel}}]{Petermann2025Dec}%
  \BibitemOpen
  \bibfield  {author} {\bibinfo {author} {\bibfnamefont {E.}~\bibnamefont {Petermann}}, \bibinfo {author} {\bibfnamefont {K.}~\bibnamefont {M{\ae}land}},\ and\ \bibinfo {author} {\bibfnamefont {B.}~\bibnamefont {Trauzettel}},\ }\bibfield  {title} {\bibinfo {title} {{Spin-resolved quasiparticle interference patterns on altermagnets via non-spin-resolved scanning tunneling microscopy}},\ }\href {https://doi.org/10.1103/sg3g-crcz} {\bibfield  {journal} {\bibinfo  {journal} {Phys. Rev. B}\ }\textbf {\bibinfo {volume} {112}},\ \bibinfo {pages} {214450} (\bibinfo {year} {2025})}\BibitemShut {NoStop}%
\bibitem [{\citenamefont {Hellenes}\ \emph {et~al.}(2024{\natexlab{a}})\citenamefont {Hellenes}, \citenamefont {Jungwirth}, \citenamefont {Jaeschke-Ubiergo}, \citenamefont {Chakraborty}, \citenamefont {Sinova},\ and\ \citenamefont {{\ifmmode\check{S}\else\v{S}\fi}mejkal}}]{Hellenes2023pwave}%
  \BibitemOpen
  \bibfield  {author} {\bibinfo {author} {\bibfnamefont {A.~B.}\ \bibnamefont {Hellenes}}, \bibinfo {author} {\bibfnamefont {T.}~\bibnamefont {Jungwirth}}, \bibinfo {author} {\bibfnamefont {R.}~\bibnamefont {Jaeschke-Ubiergo}}, \bibinfo {author} {\bibfnamefont {A.}~\bibnamefont {Chakraborty}}, \bibinfo {author} {\bibfnamefont {J.}~\bibnamefont {Sinova}},\ and\ \bibinfo {author} {\bibfnamefont {L.}~\bibnamefont {{\ifmmode\check{S}\else\v{S}\fi}mejkal}},\ }\bibfield  {title} {\bibinfo {title} {{P-wave magnets}},\ }\href {https://arxiv.org/abs/2309.01607v3} {\bibfield  {journal} {\bibinfo  {journal} {arXiv:2309.01607v3}\ } (\bibinfo {year} {2024}{\natexlab{a}})}\BibitemShut {NoStop}%
\bibitem [{\citenamefont {Brekke}\ \emph {et~al.}(2024)\citenamefont {Brekke}, \citenamefont {Sukhachov}, \citenamefont {Giil}, \citenamefont {Brataas},\ and\ \citenamefont {Linder}}]{Brekke2024pwave}%
  \BibitemOpen
  \bibfield  {author} {\bibinfo {author} {\bibfnamefont {B.}~\bibnamefont {Brekke}}, \bibinfo {author} {\bibfnamefont {P.}~\bibnamefont {Sukhachov}}, \bibinfo {author} {\bibfnamefont {H.~G.}\ \bibnamefont {Giil}}, \bibinfo {author} {\bibfnamefont {A.}~\bibnamefont {Brataas}},\ and\ \bibinfo {author} {\bibfnamefont {J.}~\bibnamefont {Linder}},\ }\bibfield  {title} {\bibinfo {title} {{Minimal Models and Transport Properties of Unconventional $p$-Wave Magnets}},\ }\href {https://doi.org/10.1103/PhysRevLett.133.236703} {\bibfield  {journal} {\bibinfo  {journal} {Phys. Rev. Lett.}\ }\textbf {\bibinfo {volume} {133}},\ \bibinfo {pages} {236703} (\bibinfo {year} {2024})}\BibitemShut {NoStop}%
\bibitem [{\citenamefont {Yu}\ \emph {et~al.}(2025)\citenamefont {Yu}, \citenamefont {Lyngby}, \citenamefont {Shishidou}, \citenamefont {Roig}, \citenamefont {Kreisel}, \citenamefont {Weinert}, \citenamefont {Andersen},\ and\ \citenamefont {Agterberg}}]{Yu2025pMmodel}%
  \BibitemOpen
  \bibfield  {author} {\bibinfo {author} {\bibfnamefont {Y.}~\bibnamefont {Yu}}, \bibinfo {author} {\bibfnamefont {M.~B.}\ \bibnamefont {Lyngby}}, \bibinfo {author} {\bibfnamefont {T.}~\bibnamefont {Shishidou}}, \bibinfo {author} {\bibfnamefont {M.}~\bibnamefont {Roig}}, \bibinfo {author} {\bibfnamefont {A.}~\bibnamefont {Kreisel}}, \bibinfo {author} {\bibfnamefont {M.}~\bibnamefont {Weinert}}, \bibinfo {author} {\bibfnamefont {B.~M.}\ \bibnamefont {Andersen}},\ and\ \bibinfo {author} {\bibfnamefont {D.~F.}\ \bibnamefont {Agterberg}},\ }\bibfield  {title} {\bibinfo {title} {{Odd-Parity Magnetism Driven by Antiferromagnetic Exchange}},\ }\href {https://doi.org/10.1103/zk69-k6b2} {\bibfield  {journal} {\bibinfo  {journal} {Phys. Rev. Lett.}\ }\textbf {\bibinfo {volume} {135}},\ \bibinfo {pages} {046701} (\bibinfo {year} {2025})}\BibitemShut {NoStop}%
\bibitem [{\citenamefont {Mitscherling}\ \emph {et~al.}(2026)\citenamefont {Mitscherling}, \citenamefont {Priessnitz}, \citenamefont {Geschner},\ and\ \citenamefont {{\ifmmode\check{S}\else\v{S}\fi}mejkal}}]{Mitscherling2026Mar}%
  \BibitemOpen
  \bibfield  {author} {\bibinfo {author} {\bibfnamefont {J.}~\bibnamefont {Mitscherling}}, \bibinfo {author} {\bibfnamefont {J.}~\bibnamefont {Priessnitz}}, \bibinfo {author} {\bibfnamefont {C.~K.}\ \bibnamefont {Geschner}},\ and\ \bibinfo {author} {\bibfnamefont {L.}~\bibnamefont {{\ifmmode\check{S}\else\v{S}\fi}mejkal}},\ }\bibfield  {title} {\bibinfo {title} {{Microscopic origin of $p$-wave magnetism}},\ }\href {https://doi.org/10.48550/arXiv.2603.09736} {\bibfield  {journal} {\bibinfo  {journal} {arXiv:2603.09736}\ } (\bibinfo {year} {2026})}\BibitemShut {NoStop}%
\bibitem [{\citenamefont {Eikeland}\ \emph {et~al.}(2026)\citenamefont {Eikeland}, \citenamefont {Lundemo},\ and\ \citenamefont {Sudb{\o}}}]{Eikeland2026Jun}%
  \BibitemOpen
  \bibfield  {author} {\bibinfo {author} {\bibfnamefont {K.~R.}\ \bibnamefont {Eikeland}}, \bibinfo {author} {\bibfnamefont {S.~D.}\ \bibnamefont {Lundemo}},\ and\ \bibinfo {author} {\bibfnamefont {A.}~\bibnamefont {Sudb{\o}}},\ }\bibfield  {title} {\bibinfo {title} {{The fate of odd-parity magnetism in one dimension}},\ }\href {https://doi.org/10.48550/arXiv.2606.26222} {\bibfield  {journal} {\bibinfo  {journal} {arXiv:2606.26222}\ } (\bibinfo {year} {2026})}\BibitemShut {NoStop}%
\bibitem [{\citenamefont {Sim}\ and\ \citenamefont {Rachel}(2026)}]{Sim2026Jul}%
  \BibitemOpen
  \bibfield  {author} {\bibinfo {author} {\bibfnamefont {G.}~\bibnamefont {Sim}}\ and\ \bibinfo {author} {\bibfnamefont {S.}~\bibnamefont {Rachel}},\ }\bibfield  {title} {\bibinfo {title} {{Quantum Spin Models of Commensurate $p$-Wave Magnets}},\ }\href {https://doi.org/10.1103/7cpd-mv33} {\bibfield  {journal} {\bibinfo  {journal} {Phys. Rev. Lett.}\ }\textbf {\bibinfo {volume} {137}},\ \bibinfo {pages} {026503} (\bibinfo {year} {2026})}\BibitemShut {NoStop}%
\bibitem [{\citenamefont {Neumann}\ \emph {et~al.}(2026)\citenamefont {Neumann}, \citenamefont {Jaeschke-Ubiergo}, \citenamefont {Zarzuela}, \citenamefont {{\ifmmode\check{S}\else\v{S}\fi}mejkal}, \citenamefont {Sinova},\ and\ \citenamefont {Mook}}]{Neumann2026pMmagnon}%
  \BibitemOpen
  \bibfield  {author} {\bibinfo {author} {\bibfnamefont {R.~R.}\ \bibnamefont {Neumann}}, \bibinfo {author} {\bibfnamefont {R.}~\bibnamefont {Jaeschke-Ubiergo}}, \bibinfo {author} {\bibfnamefont {R.}~\bibnamefont {Zarzuela}}, \bibinfo {author} {\bibfnamefont {L.}~\bibnamefont {{\ifmmode\check{S}\else\v{S}\fi}mejkal}}, \bibinfo {author} {\bibfnamefont {J.}~\bibnamefont {Sinova}},\ and\ \bibinfo {author} {\bibfnamefont {A.}~\bibnamefont {Mook}},\ }\bibfield  {title} {\bibinfo {title} {{Odd-Parity-Wave Magnons and Nonrelativistic Thermal Edelstein Effect}},\ }\href {https://doi.org/10.48550/arXiv.2603.05415} {\bibfield  {journal} {\bibinfo  {journal} {arXiv:2603.05415}\ } (\bibinfo {year} {2026})}\BibitemShut {NoStop}%
\bibitem [{\citenamefont {Kravchuk}\ \emph {et~al.}(2026)\citenamefont {Kravchuk}, \citenamefont {Yershov}, \citenamefont {Pradenas}, \citenamefont {Neumann}, \citenamefont {Jaeschke-Ubiergo}, \citenamefont {Zarzuela}, \citenamefont {Sinova}, \citenamefont {Brink},\ and\ \citenamefont {Mook}}]{Kravchuk2026magmoment}%
  \BibitemOpen
  \bibfield  {author} {\bibinfo {author} {\bibfnamefont {V.~P.}\ \bibnamefont {Kravchuk}}, \bibinfo {author} {\bibfnamefont {K.~V.}\ \bibnamefont {Yershov}}, \bibinfo {author} {\bibfnamefont {B.}~\bibnamefont {Pradenas}}, \bibinfo {author} {\bibfnamefont {R.~R.}\ \bibnamefont {Neumann}}, \bibinfo {author} {\bibfnamefont {R.}~\bibnamefont {Jaeschke-Ubiergo}}, \bibinfo {author} {\bibfnamefont {R.}~\bibnamefont {Zarzuela}}, \bibinfo {author} {\bibfnamefont {J.}~\bibnamefont {Sinova}}, \bibinfo {author} {\bibfnamefont {J.~v.~d.}\ \bibnamefont {Brink}},\ and\ \bibinfo {author} {\bibfnamefont {A.}~\bibnamefont {Mook}},\ }\bibfield  {title} {\bibinfo {title} {{Nonlinear Magnon Magnetic Moment Transport in Triangular-Lattice f-Wave Antialtermagnets}},\ }\href {https://doi.org/10.48550/arXiv.2605.22614} {\bibfield  {journal} {\bibinfo  {journal} {arXiv:2605.22614}\ } (\bibinfo {year} {2026})}\BibitemShut {NoStop}%
\bibitem [{\citenamefont {Cui}\ \emph {et~al.}(2023)\citenamefont {Cui}, \citenamefont {Zeng}, \citenamefont {Cui}, \citenamefont {Yu},\ and\ \citenamefont {Yang}}]{Cui2023AM}%
  \BibitemOpen
  \bibfield  {author} {\bibinfo {author} {\bibfnamefont {Q.}~\bibnamefont {Cui}}, \bibinfo {author} {\bibfnamefont {B.}~\bibnamefont {Zeng}}, \bibinfo {author} {\bibfnamefont {P.}~\bibnamefont {Cui}}, \bibinfo {author} {\bibfnamefont {T.}~\bibnamefont {Yu}},\ and\ \bibinfo {author} {\bibfnamefont {H.}~\bibnamefont {Yang}},\ }\bibfield  {title} {\bibinfo {title} {{Efficient spin Seebeck and spin Nernst effects of magnons in altermagnets}},\ }\href {https://doi.org/10.1103/PhysRevB.108.L180401} {\bibfield  {journal} {\bibinfo  {journal} {Phys. Rev. B}\ }\textbf {\bibinfo {volume} {108}},\ \bibinfo {pages} {L180401} (\bibinfo {year} {2023})}\BibitemShut {NoStop}%
\bibitem [{\citenamefont {Aoyama}\ and\ \citenamefont {Kawamura}(2026)}]{Aoyama2026AMFiMmag}%
  \BibitemOpen
  \bibfield  {author} {\bibinfo {author} {\bibfnamefont {K.}~\bibnamefont {Aoyama}}\ and\ \bibinfo {author} {\bibfnamefont {H.}~\bibnamefont {Kawamura}},\ }\bibfield  {title} {\bibinfo {title} {{Magnetic field effects on spin-split band and magnon transport in altermagnets and emergent compensated ferrimagnets}},\ }\href {https://doi.org/10.48550/arXiv.2606.03451} {\bibfield  {journal} {\bibinfo  {journal} {arXiv:2606.03451}\ } (\bibinfo {year} {2026})}\BibitemShut {NoStop}%
\bibitem [{\citenamefont {Santos}\ and\ \citenamefont {Costa}(2026)}]{Santos2026mgLieb}%
  \BibitemOpen
  \bibfield  {author} {\bibinfo {author} {\bibfnamefont {D.~L.~R.}\ \bibnamefont {Santos}}\ and\ \bibinfo {author} {\bibfnamefont {A.~T.}\ \bibnamefont {Costa}},\ }\bibfield  {title} {\bibinfo {title} {{Magnons in Metallic Altermagnetic $\text{KV}_2\text{Se}_2\text{O}$}},\ }\href {https://doi.org/10.48550/arXiv.2609.01853} {\bibfield  {journal} {\bibinfo  {journal} {arXiv:2609.01853}\ } (\bibinfo {year} {2026})}\BibitemShut {NoStop}%
\bibitem [{\citenamefont {Jungwirth}\ \emph {et~al.}(2025)\citenamefont {Jungwirth}, \citenamefont {Fernandes}, \citenamefont {Fradkin}, \citenamefont {MacDonald}, \citenamefont {Sinova},\ and\ \citenamefont {{\ifmmode\check{S}\else\v{S}\fi}mejkal}}]{Jungwirth2024LiebRev}%
  \BibitemOpen
  \bibfield  {author} {\bibinfo {author} {\bibfnamefont {T.}~\bibnamefont {Jungwirth}}, \bibinfo {author} {\bibfnamefont {R.~M.}\ \bibnamefont {Fernandes}}, \bibinfo {author} {\bibfnamefont {E.}~\bibnamefont {Fradkin}}, \bibinfo {author} {\bibfnamefont {A.~H.}\ \bibnamefont {MacDonald}}, \bibinfo {author} {\bibfnamefont {J.}~\bibnamefont {Sinova}},\ and\ \bibinfo {author} {\bibfnamefont {L.}~\bibnamefont {{\ifmmode\check{S}\else\v{S}\fi}mejkal}},\ }\bibfield  {title} {\bibinfo {title} {{Altermagnetism: An unconventional spin-ordered phase of matter}},\ }\href {https://doi.org/10.1016/j.newton.2025.100162} {\bibfield  {journal} {\bibinfo  {journal} {Newton}\ }\textbf {\bibinfo {volume} {1}},\ \bibinfo {pages} {100162} (\bibinfo {year} {2025})}\BibitemShut {NoStop}%
\bibitem [{\citenamefont {Lange}\ \emph {et~al.}(2026)\citenamefont {Lange}, \citenamefont {Jaeschke-Ubiergo}, \citenamefont {Mook},\ and\ \citenamefont {Sinova}}]{Lange2026aam}%
  \BibitemOpen
  \bibfield  {author} {\bibinfo {author} {\bibfnamefont {C.}~\bibnamefont {Lange}}, \bibinfo {author} {\bibfnamefont {R.}~\bibnamefont {Jaeschke-Ubiergo}}, \bibinfo {author} {\bibfnamefont {A.}~\bibnamefont {Mook}},\ and\ \bibinfo {author} {\bibfnamefont {J.}~\bibnamefont {Sinova}},\ }\bibfield  {title} {\bibinfo {title} {{Anti-spin Laue groups: classification of anti-altermagnets and their representative minimal models}},\ }\href {https://doi.org/10.48550/arXiv.2608.19056} {\bibfield  {journal} {\bibinfo  {journal} {arXiv:2608.19056}\ } (\bibinfo {year} {2026})}\BibitemShut {NoStop}%
\bibitem [{\citenamefont {Hellenes}\ \emph {et~al.}(2024{\natexlab{b}})\citenamefont {Hellenes}, \citenamefont {Jungwirth}, \citenamefont {Sinova},\ and\ \citenamefont {{\ifmmode\check{S}\else\v{S}\fi}mejkal}}]{Hellenes2023pwavev2}%
  \BibitemOpen
  \bibfield  {author} {\bibinfo {author} {\bibfnamefont {A.~B.}\ \bibnamefont {Hellenes}}, \bibinfo {author} {\bibfnamefont {T.}~\bibnamefont {Jungwirth}}, \bibinfo {author} {\bibfnamefont {J.}~\bibnamefont {Sinova}},\ and\ \bibinfo {author} {\bibfnamefont {L.}~\bibnamefont {{\ifmmode\check{S}\else\v{S}\fi}mejkal}},\ }\bibfield  {title} {\bibinfo {title} {{Unconventional p-wave magnets}},\ }\href {https://arxiv.org/abs/2309.01607v2} {\bibfield  {journal} {\bibinfo  {journal} {arXiv:2309.01607v2}\ } (\bibinfo {year} {2024}{\natexlab{b}})}\BibitemShut {NoStop}%
\bibitem [{\citenamefont {Wu}\ \emph {et~al.}(2019)\citenamefont {Wu}, \citenamefont {Phelan}, \citenamefont {Liu}, \citenamefont {Morey}, \citenamefont {Tutmaher}, \citenamefont {Neuefeind}, \citenamefont {Huq}, \citenamefont {Stone}, \citenamefont {Feygenson}, \citenamefont {Tam}, \citenamefont {Frandsen}, \citenamefont {Trump}, \citenamefont {Wan}, \citenamefont {Dunsiger}, \citenamefont {McQueen}, \citenamefont {Uemura},\ and\ \citenamefont {Broholm}}]{Wu2019CNAOexp}%
  \BibitemOpen
  \bibfield  {author} {\bibinfo {author} {\bibfnamefont {S.}~\bibnamefont {Wu}}, \bibinfo {author} {\bibfnamefont {W.~A.}\ \bibnamefont {Phelan}}, \bibinfo {author} {\bibfnamefont {L.}~\bibnamefont {Liu}}, \bibinfo {author} {\bibfnamefont {J.~R.}\ \bibnamefont {Morey}}, \bibinfo {author} {\bibfnamefont {J.~A.}\ \bibnamefont {Tutmaher}}, \bibinfo {author} {\bibfnamefont {J.~C.}\ \bibnamefont {Neuefeind}}, \bibinfo {author} {\bibfnamefont {A.}~\bibnamefont {Huq}}, \bibinfo {author} {\bibfnamefont {M.~B.}\ \bibnamefont {Stone}}, \bibinfo {author} {\bibfnamefont {M.}~\bibnamefont {Feygenson}}, \bibinfo {author} {\bibfnamefont {D.~W.}\ \bibnamefont {Tam}}, \bibinfo {author} {\bibfnamefont {B.~A.}\ \bibnamefont {Frandsen}}, \bibinfo {author} {\bibfnamefont {B.}~\bibnamefont {Trump}}, \bibinfo {author} {\bibfnamefont {C.}~\bibnamefont {Wan}}, \bibinfo {author} {\bibfnamefont {S.~R.}\ \bibnamefont {Dunsiger}}, \bibinfo {author} {\bibfnamefont {T.~M.}\ \bibnamefont {McQueen}}, \bibinfo {author} {\bibfnamefont {Y.~J.}\
  \bibnamefont {Uemura}},\ and\ \bibinfo {author} {\bibfnamefont {C.~L.}\ \bibnamefont {Broholm}},\ }\bibfield  {title} {\bibinfo {title} {{Incommensurate Magnetism Near Quantum Criticality in CeNiAsO}},\ }\href {https://doi.org/10.1103/PhysRevLett.122.197203} {\bibfield  {journal} {\bibinfo  {journal} {Phys. Rev. Lett.}\ }\textbf {\bibinfo {volume} {122}},\ \bibinfo {pages} {197203} (\bibinfo {year} {2019})}\BibitemShut {NoStop}%
\bibitem [{\citenamefont {G{\ifmmode\ddot{a}\else\"{a}\fi}bler}\ \emph {et~al.}(2008)\citenamefont {G{\ifmmode\ddot{a}\else\"{a}\fi}bler}, \citenamefont {Schnelle}, \citenamefont {Senyshyn},\ and\ \citenamefont {Niewa}}]{Gabler2008Ce3InN}%
  \BibitemOpen
  \bibfield  {author} {\bibinfo {author} {\bibfnamefont {F.}~\bibnamefont {G{\ifmmode\ddot{a}\else\"{a}\fi}bler}}, \bibinfo {author} {\bibfnamefont {W.}~\bibnamefont {Schnelle}}, \bibinfo {author} {\bibfnamefont {A.}~\bibnamefont {Senyshyn}},\ and\ \bibinfo {author} {\bibfnamefont {R.}~\bibnamefont {Niewa}},\ }\bibfield  {title} {\bibinfo {title} {{Magnetic structure of the inverse perovskite (Ce$_3$N)In}},\ }\href {https://doi.org/10.1016/j.solidstatesciences.2008.03.010} {\bibfield  {journal} {\bibinfo  {journal} {Solid State Sci.}\ }\textbf {\bibinfo {volume} {10}},\ \bibinfo {pages} {1910} (\bibinfo {year} {2008})}\BibitemShut {NoStop}%
\bibitem [{\citenamefont {Shi}\ \emph {et~al.}(2016)\citenamefont {Shi}, \citenamefont {Sun}, \citenamefont {Yan}, \citenamefont {Deng}, \citenamefont {Wang}, \citenamefont {Wu}, \citenamefont {Hu}, \citenamefont {Lu}, \citenamefont {Malik}, \citenamefont {Huang},\ and\ \citenamefont {Wang}}]{Shi2016Mn3GaN}%
  \BibitemOpen
  \bibfield  {author} {\bibinfo {author} {\bibfnamefont {K.}~\bibnamefont {Shi}}, \bibinfo {author} {\bibfnamefont {Y.}~\bibnamefont {Sun}}, \bibinfo {author} {\bibfnamefont {J.}~\bibnamefont {Yan}}, \bibinfo {author} {\bibfnamefont {S.}~\bibnamefont {Deng}}, \bibinfo {author} {\bibfnamefont {L.}~\bibnamefont {Wang}}, \bibinfo {author} {\bibfnamefont {H.}~\bibnamefont {Wu}}, \bibinfo {author} {\bibfnamefont {P.}~\bibnamefont {Hu}}, \bibinfo {author} {\bibfnamefont {H.}~\bibnamefont {Lu}}, \bibinfo {author} {\bibfnamefont {M.~I.}\ \bibnamefont {Malik}}, \bibinfo {author} {\bibfnamefont {Q.}~\bibnamefont {Huang}},\ and\ \bibinfo {author} {\bibfnamefont {C.}~\bibnamefont {Wang}},\ }\bibfield  {title} {\bibinfo {title} {{Baromagnetic Effect in Antiperovskite Mn$_3$Ga$_{0.95}$N$_{0.94}$ by Neutron Powder Diffraction Analysis}},\ }\href {https://doi.org/10.1002/adma.201600310} {\bibfield  {journal} {\bibinfo  {journal} {Adv. Mater.}\ }\textbf {\bibinfo {volume} {28}},\ \bibinfo {pages} {3761} (\bibinfo {year}
  {2016})}\BibitemShut {NoStop}%
\bibitem [{\citenamefont {Zhou}\ \emph {et~al.}(2025)\citenamefont {Zhou}, \citenamefont {Wang}, \citenamefont {Ma}, \citenamefont {Li}, \citenamefont {Shao}, \citenamefont {Liu},\ and\ \citenamefont {Li}}]{Zhou2025CeNiAsO}%
  \BibitemOpen
  \bibfield  {author} {\bibinfo {author} {\bibfnamefont {H.}~\bibnamefont {Zhou}}, \bibinfo {author} {\bibfnamefont {M.}~\bibnamefont {Wang}}, \bibinfo {author} {\bibfnamefont {X.}~\bibnamefont {Ma}}, \bibinfo {author} {\bibfnamefont {G.}~\bibnamefont {Li}}, \bibinfo {author} {\bibfnamefont {D.-F.}\ \bibnamefont {Shao}}, \bibinfo {author} {\bibfnamefont {B.}~\bibnamefont {Liu}},\ and\ \bibinfo {author} {\bibfnamefont {S.}~\bibnamefont {Li}},\ }\bibfield  {title} {\bibinfo {title} {{Anisotropic resistivity of a $p$-wave magnet candidate CeNiAsO}},\ }\href {https://doi.org/10.48550/arXiv.2509.07351} {\bibfield  {journal} {\bibinfo  {journal} {arXiv:2509.07351}\ } (\bibinfo {year} {2025})}\BibitemShut {NoStop}%
\bibitem [{\citenamefont {Zhang}\ \emph {et~al.}(2026{\natexlab{a}})\citenamefont {Zhang}, \citenamefont {Jiang}, \citenamefont {Shen}, \citenamefont {Yuan}, \citenamefont {Yoo}, \citenamefont {Ma}, \citenamefont {Ye}, \citenamefont {Liu}, \citenamefont {Liu}, \citenamefont {Kim}, \citenamefont {Guo}, \citenamefont {Wang},\ and\ \citenamefont {Shen}}]{ZhangShen2026JunCNAO}%
  \BibitemOpen
  \bibfield  {author} {\bibinfo {author} {\bibfnamefont {X.}~\bibnamefont {Zhang}}, \bibinfo {author} {\bibfnamefont {Z.}~\bibnamefont {Jiang}}, \bibinfo {author} {\bibfnamefont {S.}~\bibnamefont {Shen}}, \bibinfo {author} {\bibfnamefont {J.}~\bibnamefont {Yuan}}, \bibinfo {author} {\bibfnamefont {J.}~\bibnamefont {Yoo}}, \bibinfo {author} {\bibfnamefont {X.}~\bibnamefont {Ma}}, \bibinfo {author} {\bibfnamefont {M.}~\bibnamefont {Ye}}, \bibinfo {author} {\bibfnamefont {J.}~\bibnamefont {Liu}}, \bibinfo {author} {\bibfnamefont {Z.}~\bibnamefont {Liu}}, \bibinfo {author} {\bibfnamefont {C.}~\bibnamefont {Kim}}, \bibinfo {author} {\bibfnamefont {Y.}~\bibnamefont {Guo}}, \bibinfo {author} {\bibfnamefont {Y.}~\bibnamefont {Wang}},\ and\ \bibinfo {author} {\bibfnamefont {D.}~\bibnamefont {Shen}},\ }\bibfield  {title} {\bibinfo {title} {{Quenching of Nonrelativistic $p$-Wave Spin Splitting by Reduced $c\text{-}f$ Coupling in $\text{CeNiAsO}$}},\ }\href {https://doi.org/10.48550/arXiv.2606.02420} {\bibfield
  {journal} {\bibinfo  {journal} {arXiv:2606.02420}\ } (\bibinfo {year} {2026}{\natexlab{a}})}\BibitemShut {NoStop}%
\bibitem [{\citenamefont {Zhang}\ \emph {et~al.}(2026{\natexlab{b}})\citenamefont {Zhang}, \citenamefont {Sun}, \citenamefont {Zhou}, \citenamefont {Shi}, \citenamefont {Wu}, \citenamefont {Gu}, \citenamefont {Mao}, \citenamefont {Dong}, \citenamefont {Xu}, \citenamefont {Li}, \citenamefont {Cao}, \citenamefont {Miao}, \citenamefont {Liang}, \citenamefont {Cai}, \citenamefont {Zhu}, \citenamefont {Xu}, \citenamefont {Chen}, \citenamefont {Deng}, \citenamefont {Liu}, \citenamefont {Ma}, \citenamefont {Liu}, \citenamefont {Ye}, \citenamefont {Zhang}, \citenamefont {Wang}, \citenamefont {Zhang}, \citenamefont {Yang}, \citenamefont {Peng}, \citenamefont {Xu}, \citenamefont {Liu}, \citenamefont {Li}, \citenamefont {Mao}, \citenamefont {Li}, \citenamefont {Weng}, \citenamefont {Zhao},\ and\ \citenamefont {Zhou}}]{ZhangZhou2026JunCNAO}%
  \BibitemOpen
  \bibfield  {author} {\bibinfo {author} {\bibfnamefont {J.}~\bibnamefont {Zhang}}, \bibinfo {author} {\bibfnamefont {Y.}~\bibnamefont {Sun}}, \bibinfo {author} {\bibfnamefont {H.}~\bibnamefont {Zhou}}, \bibinfo {author} {\bibfnamefont {J.}~\bibnamefont {Shi}}, \bibinfo {author} {\bibfnamefont {D.}~\bibnamefont {Wu}}, \bibinfo {author} {\bibfnamefont {H.}~\bibnamefont {Gu}}, \bibinfo {author} {\bibfnamefont {W.}~\bibnamefont {Mao}}, \bibinfo {author} {\bibfnamefont {H.}~\bibnamefont {Dong}}, \bibinfo {author} {\bibfnamefont {Y.}~\bibnamefont {Xu}}, \bibinfo {author} {\bibfnamefont {Y.}~\bibnamefont {Li}}, \bibinfo {author} {\bibfnamefont {Z.}~\bibnamefont {Cao}}, \bibinfo {author} {\bibfnamefont {T.}~\bibnamefont {Miao}}, \bibinfo {author} {\bibfnamefont {B.}~\bibnamefont {Liang}}, \bibinfo {author} {\bibfnamefont {N.}~\bibnamefont {Cai}}, \bibinfo {author} {\bibfnamefont {W.}~\bibnamefont {Zhu}}, \bibinfo {author} {\bibfnamefont {M.}~\bibnamefont {Xu}}, \bibinfo {author} {\bibfnamefont {J.}~\bibnamefont
  {Chen}}, \bibinfo {author} {\bibfnamefont {C.}~\bibnamefont {Deng}}, \bibinfo {author} {\bibfnamefont {B.}~\bibnamefont {Liu}}, \bibinfo {author} {\bibfnamefont {X.}~\bibnamefont {Ma}}, \bibinfo {author} {\bibfnamefont {Z.}~\bibnamefont {Liu}}, \bibinfo {author} {\bibfnamefont {M.}~\bibnamefont {Ye}}, \bibinfo {author} {\bibfnamefont {S.}~\bibnamefont {Zhang}}, \bibinfo {author} {\bibfnamefont {Z.}~\bibnamefont {Wang}}, \bibinfo {author} {\bibfnamefont {F.}~\bibnamefont {Zhang}}, \bibinfo {author} {\bibfnamefont {F.}~\bibnamefont {Yang}}, \bibinfo {author} {\bibfnamefont {Q.}~\bibnamefont {Peng}}, \bibinfo {author} {\bibfnamefont {Z.}~\bibnamefont {Xu}}, \bibinfo {author} {\bibfnamefont {G.}~\bibnamefont {Liu}}, \bibinfo {author} {\bibfnamefont {X.}~\bibnamefont {Li}}, \bibinfo {author} {\bibfnamefont {H.}~\bibnamefont {Mao}}, \bibinfo {author} {\bibfnamefont {S.}~\bibnamefont {Li}}, \bibinfo {author} {\bibfnamefont {H.}~\bibnamefont {Weng}}, \bibinfo {author} {\bibfnamefont {L.}~\bibnamefont {Zhao}},\ and\
  \bibinfo {author} {\bibfnamefont {X.~J.}\ \bibnamefont {Zhou}},\ }\bibfield  {title} {\bibinfo {title} {{Suppression of p-Wave Altermagnetism by Localized 4f Electrons in CeNiAsO}},\ }\href {https://doi.org/10.48550/arXiv.2606.02422} {\bibfield  {journal} {\bibinfo  {journal} {arXiv:2606.02422}\ } (\bibinfo {year} {2026}{\natexlab{b}})}\BibitemShut {NoStop}%
\bibitem [{\citenamefont {Zhang}\ \emph {et~al.}(2026{\natexlab{c}})\citenamefont {Zhang}, \citenamefont {Li}, \citenamefont {Cheng}, \citenamefont {Fan}, \citenamefont {Yin}, \citenamefont {Gao}, \citenamefont {Liu}, \citenamefont {Cui}, \citenamefont {Yin}, \citenamefont {Zhao}, \citenamefont {Lin}, \citenamefont {Liu}, \citenamefont {Ye}, \citenamefont {Huang}, \citenamefont {Qiao}, \citenamefont {Xie}, \citenamefont {Miao}, \citenamefont {Wu}, \citenamefont {Liu}, \citenamefont {Cao},\ and\ \citenamefont {Chen}}]{ZhangChen2026MayCNAO}%
  \BibitemOpen
  \bibfield  {author} {\bibinfo {author} {\bibfnamefont {F.}~\bibnamefont {Zhang}}, \bibinfo {author} {\bibfnamefont {H.}~\bibnamefont {Li}}, \bibinfo {author} {\bibfnamefont {X.}~\bibnamefont {Cheng}}, \bibinfo {author} {\bibfnamefont {Y.}~\bibnamefont {Fan}}, \bibinfo {author} {\bibfnamefont {Y.}~\bibnamefont {Yin}}, \bibinfo {author} {\bibfnamefont {Y.}~\bibnamefont {Gao}}, \bibinfo {author} {\bibfnamefont {Z.}~\bibnamefont {Liu}}, \bibinfo {author} {\bibfnamefont {S.}~\bibnamefont {Cui}}, \bibinfo {author} {\bibfnamefont {Z.}~\bibnamefont {Yin}}, \bibinfo {author} {\bibfnamefont {Y.}~\bibnamefont {Zhao}}, \bibinfo {author} {\bibfnamefont {J.}~\bibnamefont {Lin}}, \bibinfo {author} {\bibfnamefont {Z.}~\bibnamefont {Liu}}, \bibinfo {author} {\bibfnamefont {M.}~\bibnamefont {Ye}}, \bibinfo {author} {\bibfnamefont {Y.}~\bibnamefont {Huang}}, \bibinfo {author} {\bibfnamefont {S.}~\bibnamefont {Qiao}}, \bibinfo {author} {\bibfnamefont {W.}~\bibnamefont {Xie}}, \bibinfo {author} {\bibfnamefont {P.}~\bibnamefont
  {Miao}}, \bibinfo {author} {\bibfnamefont {H.}~\bibnamefont {Wu}}, \bibinfo {author} {\bibfnamefont {J.}~\bibnamefont {Liu}}, \bibinfo {author} {\bibfnamefont {G.}~\bibnamefont {Cao}},\ and\ \bibinfo {author} {\bibfnamefont {C.}~\bibnamefont {Chen}},\ }\bibfield  {title} {\bibinfo {title} {{Odd spin symmetry and anisotropy switching in p-wave magnet CeNiAsO}},\ }\href {https://doi.org/10.48550/arXiv.2605.28701} {\bibfield  {journal} {\bibinfo  {journal} {arXiv:2605.28701}\ } (\bibinfo {year} {2026}{\natexlab{c}})}\BibitemShut {NoStop}%
\bibitem [{\citenamefont {Dsouza}\ \emph {et~al.}(2026)\citenamefont {Dsouza}, \citenamefont {Kreisel}, \citenamefont {Andersen}, \citenamefont {Agterberg},\ and\ \citenamefont {Christensen}}]{Dsouza2026Fe}%
  \BibitemOpen
  \bibfield  {author} {\bibinfo {author} {\bibfnamefont {R.}~\bibnamefont {Dsouza}}, \bibinfo {author} {\bibfnamefont {A.}~\bibnamefont {Kreisel}}, \bibinfo {author} {\bibfnamefont {B.~M.}\ \bibnamefont {Andersen}}, \bibinfo {author} {\bibfnamefont {D.~F.}\ \bibnamefont {Agterberg}},\ and\ \bibinfo {author} {\bibfnamefont {M.~H.}\ \bibnamefont {Christensen}},\ }\bibfield  {title} {\bibinfo {title} {{Odd-parity magnetism in Fe-based superconductors with coplanar magnetic order}},\ }\href {https://doi.org/10.1103/8hq8-38qs} {\bibfield  {journal} {\bibinfo  {journal} {Phys. Rev. B}\ }\textbf {\bibinfo {volume} {113}},\ \bibinfo {pages} {144509} (\bibinfo {year} {2026})}\BibitemShut {NoStop}%
\bibitem [{\citenamefont {Song}\ \emph {et~al.}(2025)\citenamefont {Song}, \citenamefont {Stavri{\ifmmode\acute{c}\else\'{c}\fi}}, \citenamefont {Barone}, \citenamefont {Droghetti}, \citenamefont {Antonenko}, \citenamefont {Venderbos}, \citenamefont {Occhialini}, \citenamefont {Ilyas}, \citenamefont {Erge{\ifmmode\mbox{\c{c}}\else\c{c}\fi}en}, \citenamefont {Gedik}, \citenamefont {Cheong}, \citenamefont {Fernandes}, \citenamefont {Picozzi},\ and\ \citenamefont {Comin}}]{Song2025CominPwaveExp}%
  \BibitemOpen
  \bibfield  {author} {\bibinfo {author} {\bibfnamefont {Q.}~\bibnamefont {Song}}, \bibinfo {author} {\bibfnamefont {S.}~\bibnamefont {Stavri{\ifmmode\acute{c}\else\'{c}\fi}}}, \bibinfo {author} {\bibfnamefont {P.}~\bibnamefont {Barone}}, \bibinfo {author} {\bibfnamefont {A.}~\bibnamefont {Droghetti}}, \bibinfo {author} {\bibfnamefont {D.~S.}\ \bibnamefont {Antonenko}}, \bibinfo {author} {\bibfnamefont {J.~W.~F.}\ \bibnamefont {Venderbos}}, \bibinfo {author} {\bibfnamefont {C.~A.}\ \bibnamefont {Occhialini}}, \bibinfo {author} {\bibfnamefont {B.}~\bibnamefont {Ilyas}}, \bibinfo {author} {\bibfnamefont {E.}~\bibnamefont {Erge{\ifmmode\mbox{\c{c}}\else\c{c}\fi}en}}, \bibinfo {author} {\bibfnamefont {N.}~\bibnamefont {Gedik}}, \bibinfo {author} {\bibfnamefont {S.-W.}\ \bibnamefont {Cheong}}, \bibinfo {author} {\bibfnamefont {R.~M.}\ \bibnamefont {Fernandes}}, \bibinfo {author} {\bibfnamefont {S.}~\bibnamefont {Picozzi}},\ and\ \bibinfo {author} {\bibfnamefont {R.}~\bibnamefont {Comin}},\ }\bibfield  {title}
  {\bibinfo {title} {{Electrical switching of a p-wave magnet}},\ }\href {https://doi.org/10.1038/s41586-025-09034-7} {\bibfield  {journal} {\bibinfo  {journal} {Nature}\ }\textbf {\bibinfo {volume} {642}},\ \bibinfo {pages} {64} (\bibinfo {year} {2025})}\BibitemShut {NoStop}%
\bibitem [{\citenamefont {Yamada}\ \emph {et~al.}(2025)\citenamefont {Yamada}, \citenamefont {Birch}, \citenamefont {Baral}, \citenamefont {Okumura}, \citenamefont {Nakano}, \citenamefont {Gao}, \citenamefont {Ezawa}, \citenamefont {Nomoto}, \citenamefont {Masell}, \citenamefont {Ishihara}, \citenamefont {Kolincio}, \citenamefont {Belopolski}, \citenamefont {Sagayama}, \citenamefont {Nakao}, \citenamefont {Ohishi}, \citenamefont {Ohhara}, \citenamefont {Kiyanagi}, \citenamefont {Nakajima}, \citenamefont {Tokura}, \citenamefont {Arima}, \citenamefont {Motome}, \citenamefont {Hirschmann},\ and\ \citenamefont {Hirschberger}}]{Yamada2025pMexp}%
  \BibitemOpen
  \bibfield  {author} {\bibinfo {author} {\bibfnamefont {R.}~\bibnamefont {Yamada}}, \bibinfo {author} {\bibfnamefont {M.~T.}\ \bibnamefont {Birch}}, \bibinfo {author} {\bibfnamefont {P.~R.}\ \bibnamefont {Baral}}, \bibinfo {author} {\bibfnamefont {S.}~\bibnamefont {Okumura}}, \bibinfo {author} {\bibfnamefont {R.}~\bibnamefont {Nakano}}, \bibinfo {author} {\bibfnamefont {S.}~\bibnamefont {Gao}}, \bibinfo {author} {\bibfnamefont {M.}~\bibnamefont {Ezawa}}, \bibinfo {author} {\bibfnamefont {T.}~\bibnamefont {Nomoto}}, \bibinfo {author} {\bibfnamefont {J.}~\bibnamefont {Masell}}, \bibinfo {author} {\bibfnamefont {Y.}~\bibnamefont {Ishihara}}, \bibinfo {author} {\bibfnamefont {K.~K.}\ \bibnamefont {Kolincio}}, \bibinfo {author} {\bibfnamefont {I.}~\bibnamefont {Belopolski}}, \bibinfo {author} {\bibfnamefont {H.}~\bibnamefont {Sagayama}}, \bibinfo {author} {\bibfnamefont {H.}~\bibnamefont {Nakao}}, \bibinfo {author} {\bibfnamefont {K.}~\bibnamefont {Ohishi}}, \bibinfo {author} {\bibfnamefont {T.}~\bibnamefont
  {Ohhara}}, \bibinfo {author} {\bibfnamefont {R.}~\bibnamefont {Kiyanagi}}, \bibinfo {author} {\bibfnamefont {T.}~\bibnamefont {Nakajima}}, \bibinfo {author} {\bibfnamefont {Y.}~\bibnamefont {Tokura}}, \bibinfo {author} {\bibfnamefont {T.-h.}\ \bibnamefont {Arima}}, \bibinfo {author} {\bibfnamefont {Y.}~\bibnamefont {Motome}}, \bibinfo {author} {\bibfnamefont {M.~M.}\ \bibnamefont {Hirschmann}},\ and\ \bibinfo {author} {\bibfnamefont {M.}~\bibnamefont {Hirschberger}},\ }\bibfield  {title} {\bibinfo {title} {{A metallic p-wave magnet with commensurate spin helix}},\ }\href {https://doi.org/10.1038/s41586-025-09633-4} {\bibfield  {journal} {\bibinfo  {journal} {Nature}\ }\textbf {\bibinfo {volume} {646}},\ \bibinfo {pages} {837} (\bibinfo {year} {2025})}\BibitemShut {NoStop}%
\bibitem [{\citenamefont {Okuma}(2017)}]{Okuma2017magnonSpin}%
  \BibitemOpen
  \bibfield  {author} {\bibinfo {author} {\bibfnamefont {N.}~\bibnamefont {Okuma}},\ }\bibfield  {title} {\bibinfo {title} {{Magnon Spin-Momentum Locking: Various Spin Vortices and Dirac magnons in Noncollinear Antiferromagnets}},\ }\href {https://doi.org/10.1103/PhysRevLett.119.107205} {\bibfield  {journal} {\bibinfo  {journal} {Phys. Rev. Lett.}\ }\textbf {\bibinfo {volume} {119}},\ \bibinfo {pages} {107205} (\bibinfo {year} {2017})}\BibitemShut {NoStop}%
\bibitem [{\citenamefont {Luo}\ \emph {et~al.}(2025)\citenamefont {Luo}, \citenamefont {Hu}, \citenamefont {Hu},\ and\ \citenamefont {Law}}]{Luo2025sym}%
  \BibitemOpen
  \bibfield  {author} {\bibinfo {author} {\bibfnamefont {X.-J.}\ \bibnamefont {Luo}}, \bibinfo {author} {\bibfnamefont {J.-X.}\ \bibnamefont {Hu}}, \bibinfo {author} {\bibfnamefont {M.-L.}\ \bibnamefont {Hu}},\ and\ \bibinfo {author} {\bibfnamefont {K.~T.}\ \bibnamefont {Law}},\ }\bibfield  {title} {\bibinfo {title} {{Spin Group Symmetry Criteria for Odd-parity Magnets}},\ }\href {https://doi.org/10.48550/arXiv.2510.05512} {\bibfield  {journal} {\bibinfo  {journal} {arXiv:2510.05512}\ } (\bibinfo {year} {2025})}\BibitemShut {NoStop}%
\bibitem [{\citenamefont {Hayami}\ \emph {et~al.}(2020)\citenamefont {Hayami}, \citenamefont {Yanagi},\ and\ \citenamefont {Kusunose}}]{Hayami2020fwave}%
  \BibitemOpen
  \bibfield  {author} {\bibinfo {author} {\bibfnamefont {S.}~\bibnamefont {Hayami}}, \bibinfo {author} {\bibfnamefont {Y.}~\bibnamefont {Yanagi}},\ and\ \bibinfo {author} {\bibfnamefont {H.}~\bibnamefont {Kusunose}},\ }\bibfield  {title} {\bibinfo {title} {{Spontaneous antisymmetric spin splitting in noncollinear antiferromagnets without spin-orbit coupling}},\ }\href {https://doi.org/10.1103/PhysRevB.101.220403} {\bibfield  {journal} {\bibinfo  {journal} {Phys. Rev. B}\ }\textbf {\bibinfo {volume} {101}},\ \bibinfo {pages} {220403} (\bibinfo {year} {2020})}\BibitemShut {NoStop}%
\bibitem [{\citenamefont {Neumann}\ \emph {et~al.}(2020)\citenamefont {Neumann}, \citenamefont {Mook}, \citenamefont {Henk},\ and\ \citenamefont {Mertig}}]{Neumann2020magnonSpinOrbital}%
  \BibitemOpen
  \bibfield  {author} {\bibinfo {author} {\bibfnamefont {R.~R.}\ \bibnamefont {Neumann}}, \bibinfo {author} {\bibfnamefont {A.}~\bibnamefont {Mook}}, \bibinfo {author} {\bibfnamefont {J.}~\bibnamefont {Henk}},\ and\ \bibinfo {author} {\bibfnamefont {I.}~\bibnamefont {Mertig}},\ }\bibfield  {title} {\bibinfo {title} {{Orbital Magnetic Moment of Magnons}},\ }\href {https://doi.org/10.1103/PhysRevLett.125.117209} {\bibfield  {journal} {\bibinfo  {journal} {Phys. Rev. Lett.}\ }\textbf {\bibinfo {volume} {125}},\ \bibinfo {pages} {117209} (\bibinfo {year} {2020})}\BibitemShut {NoStop}%
\bibitem [{\citenamefont {Bauer}\ \emph {et~al.}(2023)\citenamefont {Bauer}, \citenamefont {Tang}, \citenamefont {Elyasi}, \citenamefont {Blanter},\ and\ \citenamefont {van Wees}}]{Bauer2023Aug}%
  \BibitemOpen
  \bibfield  {author} {\bibinfo {author} {\bibfnamefont {G.~E.~W.}\ \bibnamefont {Bauer}}, \bibinfo {author} {\bibfnamefont {P.}~\bibnamefont {Tang}}, \bibinfo {author} {\bibfnamefont {M.}~\bibnamefont {Elyasi}}, \bibinfo {author} {\bibfnamefont {Y.~M.}\ \bibnamefont {Blanter}},\ and\ \bibinfo {author} {\bibfnamefont {B.~J.}\ \bibnamefont {van Wees}},\ }\bibfield  {title} {\bibinfo {title} {{Soft magnons in anisotropic ferromagnets}},\ }\href {https://doi.org/10.1103/PhysRevB.108.064431} {\bibfield  {journal} {\bibinfo  {journal} {Phys. Rev. B}\ }\textbf {\bibinfo {volume} {108}},\ \bibinfo {pages} {064431} (\bibinfo {year} {2023})}\BibitemShut {NoStop}%
\bibitem [{\citenamefont {Edelstein}(1990)}]{Edelstein1990Jan}%
  \BibitemOpen
  \bibfield  {author} {\bibinfo {author} {\bibfnamefont {V.~M.}\ \bibnamefont {Edelstein}},\ }\bibfield  {title} {\bibinfo {title} {{Spin polarization of conduction electrons induced by electric current in two-dimensional asymmetric electron systems}},\ }\href {https://doi.org/10.1016/0038-1098(90)90963-C} {\bibfield  {journal} {\bibinfo  {journal} {Solid State Commun.}\ }\textbf {\bibinfo {volume} {73}},\ \bibinfo {pages} {233} (\bibinfo {year} {1990})}\BibitemShut {NoStop}%
\bibitem [{\citenamefont {Bahari}\ \emph {et~al.}(2026)\citenamefont {Bahari}, \citenamefont {M{\ae}land}, \citenamefont {Timm},\ and\ \citenamefont {Trauzettel}}]{Bahari2026Apr}%
  \BibitemOpen
  \bibfield  {author} {\bibinfo {author} {\bibfnamefont {M.}~\bibnamefont {Bahari}}, \bibinfo {author} {\bibfnamefont {K.}~\bibnamefont {M{\ae}land}}, \bibinfo {author} {\bibfnamefont {C.}~\bibnamefont {Timm}},\ and\ \bibinfo {author} {\bibfnamefont {B.}~\bibnamefont {Trauzettel}},\ }\bibfield  {title} {\bibinfo {title} {{Multipolar spin-orbit coupling in noncentrosymmetric crystals with time-reversal symmetry: Beyond spin $\frac{1}{2}$}},\ }\href {https://doi.org/10.1103/2jln-lctf} {\bibfield  {journal} {\bibinfo  {journal} {Phys. Rev. B}\ }\textbf {\bibinfo {volume} {113}},\ \bibinfo {pages} {165137} (\bibinfo {year} {2026})}\BibitemShut {NoStop}%
\bibitem [{\citenamefont {Chakraborty}\ \emph {et~al.}(2025)\citenamefont {Chakraborty}, \citenamefont {Birk~Hellenes}, \citenamefont {Jaeschke-Ubiergo}, \citenamefont {Jungwirth}, \citenamefont {{\ifmmode\check{S}\else\v{S}\fi}mejkal},\ and\ \citenamefont {Sinova}}]{Chakraborty2025Aug}%
  \BibitemOpen
  \bibfield  {author} {\bibinfo {author} {\bibfnamefont {A.}~\bibnamefont {Chakraborty}}, \bibinfo {author} {\bibfnamefont {A.}~\bibnamefont {Birk~Hellenes}}, \bibinfo {author} {\bibfnamefont {R.}~\bibnamefont {Jaeschke-Ubiergo}}, \bibinfo {author} {\bibfnamefont {T.}~\bibnamefont {Jungwirth}}, \bibinfo {author} {\bibfnamefont {L.}~\bibnamefont {{\ifmmode\check{S}\else\v{S}\fi}mejkal}},\ and\ \bibinfo {author} {\bibfnamefont {J.}~\bibnamefont {Sinova}},\ }\bibfield  {title} {\bibinfo {title} {{Highly efficient non-relativistic Edelstein effect in nodal p-wave magnets}},\ }\href {https://doi.org/10.1038/s41467-025-62516-0} {\bibfield  {journal} {\bibinfo  {journal} {Nat. Commun.}\ }\textbf {\bibinfo {volume} {16}},\ \bibinfo {pages} {7270} (\bibinfo {year} {2025})}\BibitemShut {NoStop}%
\bibitem [{\citenamefont {Hodt}\ \emph {et~al.}(2025)\citenamefont {Hodt}, \citenamefont {Bentmann},\ and\ \citenamefont {Linder}}]{Hodt2025May}%
  \BibitemOpen
  \bibfield  {author} {\bibinfo {author} {\bibfnamefont {E.~W.}\ \bibnamefont {Hodt}}, \bibinfo {author} {\bibfnamefont {H.}~\bibnamefont {Bentmann}},\ and\ \bibinfo {author} {\bibfnamefont {J.}~\bibnamefont {Linder}},\ }\bibfield  {title} {\bibinfo {title} {{Fate of $p$-wave spin polarization in helimagnets with Rashba spin-orbit coupling}},\ }\href {https://doi.org/10.1103/PhysRevB.111.205416} {\bibfield  {journal} {\bibinfo  {journal} {Phys. Rev. B}\ }\textbf {\bibinfo {volume} {111}},\ \bibinfo {pages} {205416} (\bibinfo {year} {2025})}\BibitemShut {NoStop}%
\bibitem [{\citenamefont {Erlandsen}\ and\ \citenamefont {Sudb{\o}}(2020)}]{Erlandsen2020stripe}%
  \BibitemOpen
  \bibfield  {author} {\bibinfo {author} {\bibfnamefont {E.}~\bibnamefont {Erlandsen}}\ and\ \bibinfo {author} {\bibfnamefont {A.}~\bibnamefont {Sudb{\o}}},\ }\bibfield  {title} {\bibinfo {title} {{Schwinger boson study of superconductivity mediated by antiferromagnetic spin fluctuations}},\ }\href {https://doi.org/10.1103/PhysRevB.102.214502} {\bibfield  {journal} {\bibinfo  {journal} {Phys. Rev. B}\ }\textbf {\bibinfo {volume} {102}},\ \bibinfo {pages} {214502} (\bibinfo {year} {2020})}\BibitemShut {NoStop}%
\bibitem [{\citenamefont {Brown}(1971)}]{Brown1971biquadratic}%
  \BibitemOpen
  \bibfield  {author} {\bibinfo {author} {\bibfnamefont {H.~A.}\ \bibnamefont {Brown}},\ }\bibfield  {title} {\bibinfo {title} {{Heisenberg Ferromagnet with Biquadratic Exchange}},\ }\href {https://doi.org/10.1103/PhysRevB.4.115} {\bibfield  {journal} {\bibinfo  {journal} {Phys. Rev. B}\ }\textbf {\bibinfo {volume} {4}},\ \bibinfo {pages} {115} (\bibinfo {year} {1971})}\BibitemShut {NoStop}%
\bibitem [{\citenamefont {Heinze}\ \emph {et~al.}(2011)\citenamefont {Heinze}, \citenamefont {von Bergmann}, \citenamefont {Menzel}, \citenamefont {Brede}, \citenamefont {Kubetzka}, \citenamefont {Wiesendanger}, \citenamefont {Bihlmayer},\ and\ \citenamefont {Bl{\ifmmode\ddot{u}\else\"{u}\fi}gel}}]{Heinze2011nanoSk}%
  \BibitemOpen
  \bibfield  {author} {\bibinfo {author} {\bibfnamefont {S.}~\bibnamefont {Heinze}}, \bibinfo {author} {\bibfnamefont {K.}~\bibnamefont {von Bergmann}}, \bibinfo {author} {\bibfnamefont {M.}~\bibnamefont {Menzel}}, \bibinfo {author} {\bibfnamefont {J.}~\bibnamefont {Brede}}, \bibinfo {author} {\bibfnamefont {A.}~\bibnamefont {Kubetzka}}, \bibinfo {author} {\bibfnamefont {R.}~\bibnamefont {Wiesendanger}}, \bibinfo {author} {\bibfnamefont {G.}~\bibnamefont {Bihlmayer}},\ and\ \bibinfo {author} {\bibfnamefont {S.}~\bibnamefont {Bl{\ifmmode\ddot{u}\else\"{u}\fi}gel}},\ }\bibfield  {title} {\bibinfo {title} {{Spontaneous atomic-scale magnetic skyrmion lattice in two dimensions}},\ }\href {https://doi.org/10.1038/nphys2045} {\bibfield  {journal} {\bibinfo  {journal} {Nat. Phys.}\ }\textbf {\bibinfo {volume} {7}},\ \bibinfo {pages} {713} (\bibinfo {year} {2011})}\BibitemShut {NoStop}%
\bibitem [{\citenamefont {Pasrija}\ and\ \citenamefont {Kumar}(2013)}]{Pasrija2013biquadratic}%
  \BibitemOpen
  \bibfield  {author} {\bibinfo {author} {\bibfnamefont {K.}~\bibnamefont {Pasrija}}\ and\ \bibinfo {author} {\bibfnamefont {S.}~\bibnamefont {Kumar}},\ }\bibfield  {title} {\bibinfo {title} {{High-temperature noncollinear magnetism in a classical bilinear-biquadratic Heisenberg model}},\ }\href {https://doi.org/10.1103/PhysRevB.88.144418} {\bibfield  {journal} {\bibinfo  {journal} {Phys. Rev. B}\ }\textbf {\bibinfo {volume} {88}},\ \bibinfo {pages} {144418} (\bibinfo {year} {2013})}\BibitemShut {NoStop}%
\bibitem [{\citenamefont {Hirosawa}\ \emph {et~al.}(2026)\citenamefont {Hirosawa}, \citenamefont {Mook},\ and\ \citenamefont {Azhar}}]{Hirosawa2026biquad}%
  \BibitemOpen
  \bibfield  {author} {\bibinfo {author} {\bibfnamefont {T.}~\bibnamefont {Hirosawa}}, \bibinfo {author} {\bibfnamefont {A.}~\bibnamefont {Mook}},\ and\ \bibinfo {author} {\bibfnamefont {M.}~\bibnamefont {Azhar}},\ }\bibfield  {title} {\bibinfo {title} {{${\mathbb{Z}}_{2}$ vortex crystals in tetrahedral antiferromagnets: Fractional charges and topological magnons}},\ }\href {https://doi.org/10.1103/2467-g47w} {\bibfield  {journal} {\bibinfo  {journal} {Phys. Rev. B}\ }\textbf {\bibinfo {volume} {113}},\ \bibinfo {pages} {L020413} (\bibinfo {year} {2026})}\BibitemShut {NoStop}%
\bibitem [{\citenamefont {Saleem}\ \emph {et~al.}(2026)\citenamefont {Saleem}, \citenamefont {Pasek}, \citenamefont {Korkusinski}, \citenamefont {Cygorek},\ and\ \citenamefont {Potasz}}]{Saleem2026biquadratic}%
  \BibitemOpen
  \bibfield  {author} {\bibinfo {author} {\bibfnamefont {Y.}~\bibnamefont {Saleem}}, \bibinfo {author} {\bibfnamefont {W.}~\bibnamefont {Pasek}}, \bibinfo {author} {\bibfnamefont {M.}~\bibnamefont {Korkusinski}}, \bibinfo {author} {\bibfnamefont {M.}~\bibnamefont {Cygorek}},\ and\ \bibinfo {author} {\bibfnamefont {P.}~\bibnamefont {Potasz}},\ }\bibfield  {title} {\bibinfo {title} {{Engineering biquadratic interactions in spin-1 chains by spin-$\frac{1}{2}$ spacers}},\ }\href {https://doi.org/10.1103/h7z2-rft1} {\bibfield  {journal} {\bibinfo  {journal} {Phys. Rev. B}\ }\textbf {\bibinfo {volume} {113}},\ \bibinfo {pages} {235143} (\bibinfo {year} {2026})}\BibitemShut {NoStop}%
\bibitem [{\citenamefont {Akagi}\ \emph {et~al.}(2012)\citenamefont {Akagi}, \citenamefont {Udagawa},\ and\ \citenamefont {Motome}}]{Akagi2012biquad}%
  \BibitemOpen
  \bibfield  {author} {\bibinfo {author} {\bibfnamefont {Y.}~\bibnamefont {Akagi}}, \bibinfo {author} {\bibfnamefont {M.}~\bibnamefont {Udagawa}},\ and\ \bibinfo {author} {\bibfnamefont {Y.}~\bibnamefont {Motome}},\ }\bibfield  {title} {\bibinfo {title} {{Hidden Multiple-Spin Interactions as an Origin of Spin Scalar Chiral Order in Frustrated Kondo Lattice Models}},\ }\href {https://doi.org/10.1103/PhysRevLett.108.096401} {\bibfield  {journal} {\bibinfo  {journal} {Phys. Rev. Lett.}\ }\textbf {\bibinfo {volume} {108}},\ \bibinfo {pages} {096401} (\bibinfo {year} {2012})}\BibitemShut {NoStop}%
\bibitem [{\citenamefont {Dzyaloshinsky}(1958)}]{DZYALOSHINSKY}%
  \BibitemOpen
  \bibfield  {author} {\bibinfo {author} {\bibfnamefont {I.}~\bibnamefont {Dzyaloshinsky}},\ }\bibfield  {title} {\bibinfo {title} {A thermodynamic theory of ``weak" ferromagnetism of antiferromagnetics},\ }\href {https://doi.org/https://doi.org/10.1016/0022-3697(58)90076-3} {\bibfield  {journal} {\bibinfo  {journal} {J. Phys. Chem. Solids}\ }\textbf {\bibinfo {volume} {4}},\ \bibinfo {pages} {241} (\bibinfo {year} {1958})}\BibitemShut {NoStop}%
\bibitem [{\citenamefont {Moriya}(1960)}]{Moriya}%
  \BibitemOpen
  \bibfield  {author} {\bibinfo {author} {\bibfnamefont {T.}~\bibnamefont {Moriya}},\ }\bibfield  {title} {\bibinfo {title} {Anisotropic superexchange interaction and weak ferromagnetism},\ }\href {https://doi.org/10.1103/PhysRev.120.91} {\bibfield  {journal} {\bibinfo  {journal} {Phys. Rev.}\ }\textbf {\bibinfo {volume} {120}},\ \bibinfo {pages} {91} (\bibinfo {year} {1960})}\BibitemShut {NoStop}%
\bibitem [{\citenamefont {dos Santos}\ \emph {et~al.}(2018)\citenamefont {dos Santos}, \citenamefont {dos Santos~Dias}, \citenamefont {Guimar\~aes}, \citenamefont {Bouaziz},\ and\ \citenamefont {Lounis}}]{dosSantosPRB}%
  \BibitemOpen
  \bibfield  {author} {\bibinfo {author} {\bibfnamefont {F.~J.}\ \bibnamefont {dos Santos}}, \bibinfo {author} {\bibfnamefont {M.}~\bibnamefont {dos Santos~Dias}}, \bibinfo {author} {\bibfnamefont {F.~S.~M.}\ \bibnamefont {Guimar\~aes}}, \bibinfo {author} {\bibfnamefont {J.}~\bibnamefont {Bouaziz}},\ and\ \bibinfo {author} {\bibfnamefont {S.}~\bibnamefont {Lounis}},\ }\bibfield  {title} {\bibinfo {title} {Spin-resolved inelastic electron scattering by spin waves in noncollinear magnets},\ }\href {https://doi.org/10.1103/PhysRevB.97.024431} {\bibfield  {journal} {\bibinfo  {journal} {Phys. Rev. B}\ }\textbf {\bibinfo {volume} {97}},\ \bibinfo {pages} {024431} (\bibinfo {year} {2018})}\BibitemShut {NoStop}%
\bibitem [{\citenamefont {D\'{\i}az}\ \emph {et~al.}(2019)\citenamefont {D\'{\i}az}, \citenamefont {Klinovaja},\ and\ \citenamefont {Loss}}]{DiazPRL}%
  \BibitemOpen
  \bibfield  {author} {\bibinfo {author} {\bibfnamefont {S.~A.}\ \bibnamefont {D\'{\i}az}}, \bibinfo {author} {\bibfnamefont {J.}~\bibnamefont {Klinovaja}},\ and\ \bibinfo {author} {\bibfnamefont {D.}~\bibnamefont {Loss}},\ }\bibfield  {title} {\bibinfo {title} {{Topological Magnons and Edge States in Antiferromagnetic Skyrmion Crystals}},\ }\href {https://doi.org/10.1103/PhysRevLett.122.187203} {\bibfield  {journal} {\bibinfo  {journal} {Phys. Rev. Lett.}\ }\textbf {\bibinfo {volume} {122}},\ \bibinfo {pages} {187203} (\bibinfo {year} {2019})}\BibitemShut {NoStop}%
\bibitem [{\citenamefont {D\'{\i}az}\ \emph {et~al.}(2020)\citenamefont {D\'{\i}az}, \citenamefont {Hirosawa}, \citenamefont {Klinovaja},\ and\ \citenamefont {Loss}}]{DiazPRR}%
  \BibitemOpen
  \bibfield  {author} {\bibinfo {author} {\bibfnamefont {S.~A.}\ \bibnamefont {D\'{\i}az}}, \bibinfo {author} {\bibfnamefont {T.}~\bibnamefont {Hirosawa}}, \bibinfo {author} {\bibfnamefont {J.}~\bibnamefont {Klinovaja}},\ and\ \bibinfo {author} {\bibfnamefont {D.}~\bibnamefont {Loss}},\ }\bibfield  {title} {\bibinfo {title} {Chiral magnonic edge states in ferromagnetic skyrmion crystals controlled by magnetic fields},\ }\href {https://doi.org/10.1103/PhysRevResearch.2.013231} {\bibfield  {journal} {\bibinfo  {journal} {Phys. Rev. Res.}\ }\textbf {\bibinfo {volume} {2}},\ \bibinfo {pages} {013231} (\bibinfo {year} {2020})}\BibitemShut {NoStop}%
\bibitem [{\citenamefont {M{\ae}land}\ and\ \citenamefont {Sudb{\o}}(2022{\natexlab{a}})}]{Maeland2022Jun}%
  \BibitemOpen
  \bibfield  {author} {\bibinfo {author} {\bibfnamefont {K.}~\bibnamefont {M{\ae}land}}\ and\ \bibinfo {author} {\bibfnamefont {A.}~\bibnamefont {Sudb{\o}}},\ }\bibfield  {title} {\bibinfo {title} {{Quantum fluctuations in the order parameter of quantum skyrmion crystals}},\ }\href {https://doi.org/10.1103/PhysRevB.105.224416} {\bibfield  {journal} {\bibinfo  {journal} {Phys. Rev. B}\ }\textbf {\bibinfo {volume} {105}},\ \bibinfo {pages} {224416} (\bibinfo {year} {2022}{\natexlab{a}})}\BibitemShut {NoStop}%
\bibitem [{\citenamefont {M{\ae}land}\ and\ \citenamefont {Sudb{\o}}(2022{\natexlab{b}})}]{Maeland2022Aug}%
  \BibitemOpen
  \bibfield  {author} {\bibinfo {author} {\bibfnamefont {K.}~\bibnamefont {M{\ae}land}}\ and\ \bibinfo {author} {\bibfnamefont {A.}~\bibnamefont {Sudb{\o}}},\ }\bibfield  {title} {\bibinfo {title} {{Quantum topological phase transitions in skyrmion crystals}},\ }\href {https://doi.org/10.1103/PhysRevResearch.4.L032025} {\bibfield  {journal} {\bibinfo  {journal} {Phys. Rev. Res.}\ }\textbf {\bibinfo {volume} {4}},\ \bibinfo {pages} {L032025} (\bibinfo {year} {2022}{\natexlab{b}})}\BibitemShut {NoStop}%
\bibitem [{\citenamefont {M{\ae}land}\ and\ \citenamefont {Sudb{\o}}(2023)}]{Maeland2023AprPRL}%
  \BibitemOpen
  \bibfield  {author} {\bibinfo {author} {\bibfnamefont {K.}~\bibnamefont {M{\ae}land}}\ and\ \bibinfo {author} {\bibfnamefont {A.}~\bibnamefont {Sudb{\o}}},\ }\bibfield  {title} {\bibinfo {title} {{Topological Superconductivity Mediated by Skyrmionic Magnons}},\ }\href {https://doi.org/10.1103/PhysRevLett.130.156002} {\bibfield  {journal} {\bibinfo  {journal} {Phys. Rev. Lett.}\ }\textbf {\bibinfo {volume} {130}},\ \bibinfo {pages} {156002} (\bibinfo {year} {2023})}\BibitemShut {NoStop}%
\bibitem [{\citenamefont {M{\ae}land}\ \emph {et~al.}(2023)\citenamefont {M{\ae}land}, \citenamefont {Abnar}, \citenamefont {Benestad},\ and\ \citenamefont {Sudb{\o}}}]{Maeland2023DecTSC}%
  \BibitemOpen
  \bibfield  {author} {\bibinfo {author} {\bibfnamefont {K.}~\bibnamefont {M{\ae}land}}, \bibinfo {author} {\bibfnamefont {S.}~\bibnamefont {Abnar}}, \bibinfo {author} {\bibfnamefont {J.}~\bibnamefont {Benestad}},\ and\ \bibinfo {author} {\bibfnamefont {A.}~\bibnamefont {Sudb{\o}}},\ }\bibfield  {title} {\bibinfo {title} {{Topological superconductivity mediated by magnons of helical magnetic states}},\ }\href {https://doi.org/10.1103/PhysRevB.108.224515} {\bibfield  {journal} {\bibinfo  {journal} {Phys. Rev. B}\ }\textbf {\bibinfo {volume} {108}},\ \bibinfo {pages} {224515} (\bibinfo {year} {2023})}\BibitemShut {NoStop}%
\bibitem [{\citenamefont {Niu}\ \emph {et~al.}(2024)\citenamefont {Niu}, \citenamefont {Kwon}, \citenamefont {Ma}, \citenamefont {Cheng}, \citenamefont {Ophus}, \citenamefont {Miao}, \citenamefont {Sun}, \citenamefont {Wu}, \citenamefont {Liu}, \citenamefont {Parkin}, \citenamefont {Won}, \citenamefont {Schmid}, \citenamefont {Ding},\ and\ \citenamefont {Chen}}]{Niu2024OOPDMI}%
  \BibitemOpen
  \bibfield  {author} {\bibinfo {author} {\bibfnamefont {H.}~\bibnamefont {Niu}}, \bibinfo {author} {\bibfnamefont {H.~Y.}\ \bibnamefont {Kwon}}, \bibinfo {author} {\bibfnamefont {T.}~\bibnamefont {Ma}}, \bibinfo {author} {\bibfnamefont {Z.}~\bibnamefont {Cheng}}, \bibinfo {author} {\bibfnamefont {C.}~\bibnamefont {Ophus}}, \bibinfo {author} {\bibfnamefont {B.}~\bibnamefont {Miao}}, \bibinfo {author} {\bibfnamefont {L.}~\bibnamefont {Sun}}, \bibinfo {author} {\bibfnamefont {Y.}~\bibnamefont {Wu}}, \bibinfo {author} {\bibfnamefont {K.}~\bibnamefont {Liu}}, \bibinfo {author} {\bibfnamefont {S.~S.~P.}\ \bibnamefont {Parkin}}, \bibinfo {author} {\bibfnamefont {C.}~\bibnamefont {Won}}, \bibinfo {author} {\bibfnamefont {A.~K.}\ \bibnamefont {Schmid}}, \bibinfo {author} {\bibfnamefont {H.}~\bibnamefont {Ding}},\ and\ \bibinfo {author} {\bibfnamefont {G.}~\bibnamefont {Chen}},\ }\bibfield  {title} {\bibinfo {title} {{Reducing crystal symmetry to generate out-of-plane Dzyaloshinskii{\textendash}Moriya interaction}},\
  }\href {https://doi.org/10.1038/s41467-024-54521-6} {\bibfield  {journal} {\bibinfo  {journal} {Nat. Commun.}\ }\textbf {\bibinfo {volume} {15}},\ \bibinfo {pages} {10199} (\bibinfo {year} {2024})}\BibitemShut {NoStop}%
\bibitem [{\citenamefont {Mermin}\ and\ \citenamefont {Wagner}(1966)}]{Mermin1966Wagner}%
  \BibitemOpen
  \bibfield  {author} {\bibinfo {author} {\bibfnamefont {N.~D.}\ \bibnamefont {Mermin}}\ and\ \bibinfo {author} {\bibfnamefont {H.}~\bibnamefont {Wagner}},\ }\bibfield  {title} {\bibinfo {title} {{Absence of Ferromagnetism or Antiferromagnetism in One- or Two-Dimensional Isotropic Heisenberg Models}},\ }\href {https://doi.org/10.1103/PhysRevLett.17.1133} {\bibfield  {journal} {\bibinfo  {journal} {Phys. Rev. Lett.}\ }\textbf {\bibinfo {volume} {17}},\ \bibinfo {pages} {1133} (\bibinfo {year} {1966})}\BibitemShut {NoStop}%
\bibitem [{\citenamefont {Haraldsen}\ and\ \citenamefont {Fishman}(2009)}]{HProtation_2009}%
  \BibitemOpen
  \bibfield  {author} {\bibinfo {author} {\bibfnamefont {J.~T.}\ \bibnamefont {Haraldsen}}\ and\ \bibinfo {author} {\bibfnamefont {R.~S.}\ \bibnamefont {Fishman}},\ }\bibfield  {title} {\bibinfo {title} {{Spin rotation technique for non-collinear magnetic systems: application to the generalizedVillain model}},\ }\href {https://doi.org/10.1088/0953-8984/21/21/216001} {\bibfield  {journal} {\bibinfo  {journal} {J. Phys.: Condens. Matter}\ }\textbf {\bibinfo {volume} {21}},\ \bibinfo {pages} {216001} (\bibinfo {year} {2009})}\BibitemShut {NoStop}%
\bibitem [{\citenamefont {Rold\'an-Molina}\ \emph {et~al.}(2015)\citenamefont {Rold\'an-Molina}, \citenamefont {Santander}, \citenamefont {Nunez},\ and\ \citenamefont {Fern\'andez-Rossier}}]{RoldanMolina}%
  \BibitemOpen
  \bibfield  {author} {\bibinfo {author} {\bibfnamefont {A.}~\bibnamefont {Rold\'an-Molina}}, \bibinfo {author} {\bibfnamefont {M.~J.}\ \bibnamefont {Santander}}, \bibinfo {author} {\bibfnamefont {A.~S.}\ \bibnamefont {Nunez}},\ and\ \bibinfo {author} {\bibfnamefont {J.}~\bibnamefont {Fern\'andez-Rossier}},\ }\bibfield  {title} {\bibinfo {title} {Quantum fluctuations stabilize skyrmion textures},\ }\href {https://doi.org/10.1103/PhysRevB.92.245436} {\bibfield  {journal} {\bibinfo  {journal} {Phys. Rev. B}\ }\textbf {\bibinfo {volume} {92}},\ \bibinfo {pages} {245436} (\bibinfo {year} {2015})}\BibitemShut {NoStop}%
\bibitem [{\citenamefont {Rold\'an-Molina}\ \emph {et~al.}(2014)\citenamefont {Rold\'an-Molina}, \citenamefont {Santander}, \citenamefont {N\'u\~nez},\ and\ \citenamefont {Fern\'andez-Rossier}}]{RoldanMolinaHPdetails}%
  \BibitemOpen
  \bibfield  {author} {\bibinfo {author} {\bibfnamefont {A.}~\bibnamefont {Rold\'an-Molina}}, \bibinfo {author} {\bibfnamefont {M.~J.}\ \bibnamefont {Santander}}, \bibinfo {author} {\bibfnamefont {A.~S.}\ \bibnamefont {N\'u\~nez}},\ and\ \bibinfo {author} {\bibfnamefont {J.}~\bibnamefont {Fern\'andez-Rossier}},\ }\bibfield  {title} {\bibinfo {title} {Quantum theory of spin waves in finite chiral spin chains},\ }\href {https://doi.org/10.1103/PhysRevB.89.054403} {\bibfield  {journal} {\bibinfo  {journal} {Phys. Rev. B}\ }\textbf {\bibinfo {volume} {89}},\ \bibinfo {pages} {054403} (\bibinfo {year} {2014})}\BibitemShut {NoStop}%
\bibitem [{\citenamefont {Kittel}(1991)}]{Kittel1991Jan}%
  \BibitemOpen
  \bibfield  {author} {\bibinfo {author} {\bibfnamefont {C.}~\bibnamefont {Kittel}},\ }\href@noop {} {\emph {\bibinfo {title} {{Quantum Theory of Solids}}}}\ (\bibinfo  {publisher} {Wiley},\ \bibinfo {address} {Hoboken, NJ},\ \bibinfo {year} {1991})\BibitemShut {NoStop}%
\bibitem [{\citenamefont {McClarty}(2022)}]{TopoMagnonRev}%
  \BibitemOpen
  \bibfield  {author} {\bibinfo {author} {\bibfnamefont {P.~A.}\ \bibnamefont {McClarty}},\ }\bibfield  {title} {\bibinfo {title} {{Topological Magnons: A Review}},\ }\href {https://doi.org/10.1146/annurev-conmatphys-031620-104715} {\bibfield  {journal} {\bibinfo  {journal} {Annu. Rev. Condens. Matter Phys.}\ }\textbf {\bibinfo {volume} {13}},\ \bibinfo {pages} {171} (\bibinfo {year} {2022})}\BibitemShut {NoStop}%
\bibitem [{\citenamefont {Colpa}(1978)}]{COLPA}%
  \BibitemOpen
  \bibfield  {author} {\bibinfo {author} {\bibfnamefont {J.~H.~P.}\ \bibnamefont {Colpa}},\ }\bibfield  {title} {\bibinfo {title} {{Diagonalization of the quadratic boson Hamiltonian}},\ }\href {https://doi.org/10.1016/0378-4371(78)90160-7} {\bibfield  {journal} {\bibinfo  {journal} {Physica}\ }\textbf {\bibinfo {volume} {93A}},\ \bibinfo {pages} {327} (\bibinfo {year} {1978})}\BibitemShut {NoStop}%
\bibitem [{\citenamefont {To}\ \emph {et~al.}(2025)\citenamefont {To}, \citenamefont {Garcia-Gaitan}, \citenamefont {Ren}, \citenamefont {Zide}, \citenamefont {Jungfleisch}, \citenamefont {Xiao}, \citenamefont {Nikoli{\ifmmode\acute{c}\else\'{c}\fi}}, \citenamefont {Bryant},\ and\ \citenamefont {Doty}}]{To2025orbitronic}%
  \BibitemOpen
  \bibfield  {author} {\bibinfo {author} {\bibfnamefont {D.~Q.}\ \bibnamefont {To}}, \bibinfo {author} {\bibfnamefont {F.}~\bibnamefont {Garcia-Gaitan}}, \bibinfo {author} {\bibfnamefont {Y.}~\bibnamefont {Ren}}, \bibinfo {author} {\bibfnamefont {J.~M.~O.}\ \bibnamefont {Zide}}, \bibinfo {author} {\bibfnamefont {M.~B.}\ \bibnamefont {Jungfleisch}}, \bibinfo {author} {\bibfnamefont {J.~Q.}\ \bibnamefont {Xiao}}, \bibinfo {author} {\bibfnamefont {B.~K.}\ \bibnamefont {Nikoli{\ifmmode\acute{c}\else\'{c}\fi}}}, \bibinfo {author} {\bibfnamefont {G.~W.}\ \bibnamefont {Bryant}},\ and\ \bibinfo {author} {\bibfnamefont {M.~F.}\ \bibnamefont {Doty}},\ }\bibfield  {title} {\bibinfo {title} {{Magnon-induced electric polarization and magnon Nernst effects}},\ }\href {https://doi.org/10.1073/pnas.2507255122} {\bibfield  {journal} {\bibinfo  {journal} {Proc. Natl. Acad. Sci. U.S.A.}\ }\textbf {\bibinfo {volume} {122}},\ \bibinfo {pages} {e2507255122} (\bibinfo {year} {2025})}\BibitemShut {NoStop}%
\bibitem [{\citenamefont {Nakata}\ and\ \citenamefont {Ohnuma}(2021)}]{Nakata2021Boltzmann}%
  \BibitemOpen
  \bibfield  {author} {\bibinfo {author} {\bibfnamefont {K.}~\bibnamefont {Nakata}}\ and\ \bibinfo {author} {\bibfnamefont {Y.}~\bibnamefont {Ohnuma}},\ }\bibfield  {title} {\bibinfo {title} {{Magnonic thermal transport using the quantum Boltzmann equation}},\ }\href {https://doi.org/10.1103/PhysRevB.104.064408} {\bibfield  {journal} {\bibinfo  {journal} {Phys. Rev. B}\ }\textbf {\bibinfo {volume} {104}},\ \bibinfo {pages} {064408} (\bibinfo {year} {2021})}\BibitemShut {NoStop}%
\bibitem [{\citenamefont {Katsura}\ \emph {et~al.}(2010)\citenamefont {Katsura}, \citenamefont {Nagaosa},\ and\ \citenamefont {Lee}}]{Katsura2010Kubo}%
  \BibitemOpen
  \bibfield  {author} {\bibinfo {author} {\bibfnamefont {H.}~\bibnamefont {Katsura}}, \bibinfo {author} {\bibfnamefont {N.}~\bibnamefont {Nagaosa}},\ and\ \bibinfo {author} {\bibfnamefont {P.~A.}\ \bibnamefont {Lee}},\ }\bibfield  {title} {\bibinfo {title} {{Theory of the Thermal Hall Effect in Quantum Magnets}},\ }\href {https://doi.org/10.1103/PhysRevLett.104.066403} {\bibfield  {journal} {\bibinfo  {journal} {Phys. Rev. Lett.}\ }\textbf {\bibinfo {volume} {104}},\ \bibinfo {pages} {066403} (\bibinfo {year} {2010})}\BibitemShut {NoStop}%
\bibitem [{\citenamefont {Ezawa}(2025)}]{Ezawa2025nonreciprocal}%
  \BibitemOpen
  \bibfield  {author} {\bibinfo {author} {\bibfnamefont {M.}~\bibnamefont {Ezawa}},\ }\bibfield  {title} {\bibinfo {title} {{Third-order and fifth-order nonlinear spin-current generation in $g$-wave and $i$-wave altermagnets and perfectly nonreciprocal spin current in $f$-wave magnets}},\ }\href {https://doi.org/10.1103/PhysRevB.111.125420} {\bibfield  {journal} {\bibinfo  {journal} {Phys. Rev. B}\ }\textbf {\bibinfo {volume} {111}},\ \bibinfo {pages} {125420} (\bibinfo {year} {2025})}\BibitemShut {NoStop}%
\bibitem [{\citenamefont {Li}\ \emph {et~al.}(2020)\citenamefont {Li}, \citenamefont {Mook}, \citenamefont {Raeliarijaona},\ and\ \citenamefont {Kovalev}}]{Li2020AFMEdelstein}%
  \BibitemOpen
  \bibfield  {author} {\bibinfo {author} {\bibfnamefont {B.}~\bibnamefont {Li}}, \bibinfo {author} {\bibfnamefont {A.}~\bibnamefont {Mook}}, \bibinfo {author} {\bibfnamefont {A.}~\bibnamefont {Raeliarijaona}},\ and\ \bibinfo {author} {\bibfnamefont {A.~A.}\ \bibnamefont {Kovalev}},\ }\bibfield  {title} {\bibinfo {title} {{Magnonic analog of the Edelstein effect in antiferromagnetic insulators}},\ }\href {https://doi.org/10.1103/PhysRevB.101.024427} {\bibfield  {journal} {\bibinfo  {journal} {Phys. Rev. B}\ }\textbf {\bibinfo {volume} {101}},\ \bibinfo {pages} {024427} (\bibinfo {year} {2020})}\BibitemShut {NoStop}%
\bibitem [{\citenamefont {Rezende}\ \emph {et~al.}(2014)\citenamefont {Rezende}, \citenamefont {Rodr{\ifmmode\acute{\imath}\else\'{\i}\fi}guez-Su{\ifmmode\acute{a}\else\'{a}\fi}rez}, \citenamefont {Cunha}, \citenamefont {Rodrigues}, \citenamefont {Machado}, \citenamefont {Guerra}, \citenamefont {Ortiz},\ and\ \citenamefont {Azevedo}}]{Rezende2014YIG}%
  \BibitemOpen
  \bibfield  {author} {\bibinfo {author} {\bibfnamefont {S.~M.}\ \bibnamefont {Rezende}}, \bibinfo {author} {\bibfnamefont {R.~L.}\ \bibnamefont {Rodr{\ifmmode\acute{\imath}\else\'{\i}\fi}guez-Su{\ifmmode\acute{a}\else\'{a}\fi}rez}}, \bibinfo {author} {\bibfnamefont {R.~O.}\ \bibnamefont {Cunha}}, \bibinfo {author} {\bibfnamefont {A.~R.}\ \bibnamefont {Rodrigues}}, \bibinfo {author} {\bibfnamefont {F.~L.~A.}\ \bibnamefont {Machado}}, \bibinfo {author} {\bibfnamefont {G.~A.~F.}\ \bibnamefont {Guerra}}, \bibinfo {author} {\bibfnamefont {J.~C.~L.}\ \bibnamefont {Ortiz}},\ and\ \bibinfo {author} {\bibfnamefont {A.}~\bibnamefont {Azevedo}},\ }\bibfield  {title} {\bibinfo {title} {{Magnon spin-current theory for the longitudinal spin-Seebeck effect}},\ }\href {https://doi.org/10.1103/PhysRevB.89.014416} {\bibfield  {journal} {\bibinfo  {journal} {Phys. Rev. B}\ }\textbf {\bibinfo {volume} {89}},\ \bibinfo {pages} {014416} (\bibinfo {year} {2014})}\BibitemShut {NoStop}%
\bibitem [{\citenamefont {Fergen}\ \emph {et~al.}(2002)\citenamefont {Fergen}, \citenamefont {Seemann}, \citenamefont {Weth},\ and\ \citenamefont {Sch{\ifmmode\ddot{u}\else\"{u}\fi}ppen}}]{Fergen2002softFM}%
  \BibitemOpen
  \bibfield  {author} {\bibinfo {author} {\bibfnamefont {I.}~\bibnamefont {Fergen}}, \bibinfo {author} {\bibfnamefont {K.}~\bibnamefont {Seemann}}, \bibinfo {author} {\bibfnamefont {A.~v.~d.}\ \bibnamefont {Weth}},\ and\ \bibinfo {author} {\bibfnamefont {A.}~\bibnamefont {Sch{\ifmmode\ddot{u}\else\"{u}\fi}ppen}},\ }\bibfield  {title} {\bibinfo {title} {{Soft ferromagnetic thin films for high frequency applications}},\ }\href {https://doi.org/10.1016/S0304-8853(01)01185-4} {\bibfield  {journal} {\bibinfo  {journal} {J. Magn. Magn. Mater.}\ }\textbf {\bibinfo {volume} {242-245}},\ \bibinfo {pages} {146} (\bibinfo {year} {2002})}\BibitemShut {NoStop}%
\bibitem [{\citenamefont {Kirkpatrick}\ \emph {et~al.}(1983)\citenamefont {Kirkpatrick}, \citenamefont {Gelatt},\ and\ \citenamefont {Vecchi}}]{simulatedannealing}%
  \BibitemOpen
  \bibfield  {author} {\bibinfo {author} {\bibfnamefont {S.}~\bibnamefont {Kirkpatrick}}, \bibinfo {author} {\bibfnamefont {C.~D.}\ \bibnamefont {Gelatt}},\ and\ \bibinfo {author} {\bibfnamefont {M.~P.}\ \bibnamefont {Vecchi}},\ }\bibfield  {title} {\bibinfo {title} {{Optimization by Simulated Annealing}},\ }\href {https://doi.org/10.1126/science.220.4598.671} {\bibfield  {journal} {\bibinfo  {journal} {Science}\ }\textbf {\bibinfo {volume} {220}},\ \bibinfo {pages} {671} (\bibinfo {year} {1983})}\BibitemShut {NoStop}%
\bibitem [{\citenamefont {Sun}\ \emph {et~al.}(2023)\citenamefont {Sun}, \citenamefont {M{\ae}land},\ and\ \citenamefont {Sudb{\o}}}]{SunMaeland2023Aug}%
  \BibitemOpen
  \bibfield  {author} {\bibinfo {author} {\bibfnamefont {C.}~\bibnamefont {Sun}}, \bibinfo {author} {\bibfnamefont {K.}~\bibnamefont {M{\ae}land}},\ and\ \bibinfo {author} {\bibfnamefont {A.}~\bibnamefont {Sudb{\o}}},\ }\bibfield  {title} {\bibinfo {title} {{Stability of superconducting gap symmetries arising from antiferromagnetic magnons}},\ }\href {https://doi.org/10.1103/PhysRevB.108.054520} {\bibfield  {journal} {\bibinfo  {journal} {Phys. Rev. B}\ }\textbf {\bibinfo {volume} {108}},\ \bibinfo {pages} {054520} (\bibinfo {year} {2023})}\BibitemShut {NoStop}%
\bibitem [{\citenamefont {Sun}\ \emph {et~al.}(2024)\citenamefont {Sun}, \citenamefont {M{\ae}land}, \citenamefont {Thingstad},\ and\ \citenamefont {Sudb{\o}}}]{SunMaeland2024Mar}%
  \BibitemOpen
  \bibfield  {author} {\bibinfo {author} {\bibfnamefont {C.}~\bibnamefont {Sun}}, \bibinfo {author} {\bibfnamefont {K.}~\bibnamefont {M{\ae}land}}, \bibinfo {author} {\bibfnamefont {E.}~\bibnamefont {Thingstad}},\ and\ \bibinfo {author} {\bibfnamefont {A.}~\bibnamefont {Sudb{\o}}},\ }\bibfield  {title} {\bibinfo {title} {{Strong-coupling approach to temperature dependence of competing orders of superconductivity: Possible time-reversal symmetry breaking and nontrivial topology}},\ }\href {https://doi.org/10.1103/PhysRevB.109.174520} {\bibfield  {journal} {\bibinfo  {journal} {Phys. Rev. B}\ }\textbf {\bibinfo {volume} {109}},\ \bibinfo {pages} {174520} (\bibinfo {year} {2024})}\BibitemShut {NoStop}%
\bibitem [{\citenamefont {M{\ae}land}\ \emph {et~al.}(2021)\citenamefont {M{\ae}land}, \citenamefont {R{\o}st}, \citenamefont {Wells},\ and\ \citenamefont {Sudb{\o}}}]{Maeland2021Sep}%
  \BibitemOpen
  \bibfield  {author} {\bibinfo {author} {\bibfnamefont {K.}~\bibnamefont {M{\ae}land}}, \bibinfo {author} {\bibfnamefont {H.~I.}\ \bibnamefont {R{\o}st}}, \bibinfo {author} {\bibfnamefont {J.~W.}\ \bibnamefont {Wells}},\ and\ \bibinfo {author} {\bibfnamefont {A.}~\bibnamefont {Sudb{\o}}},\ }\bibfield  {title} {\bibinfo {title} {{Electron-magnon coupling and quasiparticle lifetimes on the surface of a topological insulator}},\ }\href {https://doi.org/10.1103/PhysRevB.104.125125} {\bibfield  {journal} {\bibinfo  {journal} {Phys. Rev. B}\ }\textbf {\bibinfo {volume} {104}},\ \bibinfo {pages} {125125} (\bibinfo {year} {2021})}\BibitemShut {NoStop}%
\end{thebibliography}%

\end{document}